\documentclass[manuscript]{acmart}
\AtBeginDocument{%
  }

\setcopyright{acmlicensed}
\copyrightyear{2026}
\acmYear{2026}
\acmDOI{XXXXXXX.XXXXXXX}
\acmConference[Conference acronym 'XX]{Make sure to enter the correct
  conference title from your rights confirmation email}{June 03--05,
  2018}{Woodstock, NY}
\acmISBN{978-1-4503-XXXX-X/2018/06}

\usepackage{float}

\usepackage{amsmath,amsfonts}
\usepackage{algorithmic}
\usepackage{graphicx}
\usepackage{textcomp}
\usepackage{xcolor}
\usepackage{multirow}
\usepackage{booktabs}
\usepackage{comment}
\usepackage{amsmath}
\usepackage{xspace}
\usepackage{balance}
\usepackage{todonotes}
\usepackage{subfigure}
\usepackage{caption} 
\usepackage{hyperref}
\usepackage{tcolorbox}

\newcommand\cbox[1]{\begin{tcolorbox}#1\end{tcolorbox}}

\begin{document}

\title{Adversarial Robustness of Camera–Radar Sensor Fusion for Autonomous Driving}

\author{Ce Zhou}
\affiliation{%
  \institution{Missouri University of Science and Technology}
  \city{Rolla}
  \country{USA}}
\email{cezhou@mst.edu}

\author{Yubo Wu}
\affiliation{%
  \institution{Virginia Tech}
  \city{Blacksburg}
  \country{USA}}
\email{wuyubo@vt.edu}

\author{Khang Nguyen}
\affiliation{%
  \institution{Michigan State University}
  \city{East Lansing}
  \country{USA}}
\email{nguy1055@msu.edu}

\author{Liyang Xiao}
\affiliation{%
  \institution{Michigan State University}
  \city{East Lansing}
  \country{USA}}
\email{xiaoliya@msu.edu}

\author{Thomas Hou}
\affiliation{%
  \institution{Virginia Tech}
  \city{Blacksburg}
  \country{USA}}
\email{thou@vt.edu}

\author{Qiben Yan}
\correspondingauthor
\affiliation{%
  \institution{Michigan State University}
  \city{East Lansing}
  \country{USA}}
\email{qyan@msu.edu}

\renewcommand{\shortauthors}{XX et al.}


\begin{abstract}
Autonomous driving (AD) relies on multi-sensor perception to
accurately understand and interpret the surrounding environment.
However, the robustness of camera--radar fusion under adversarial
conditions remains largely unexplored. This paper presents a
systematic measurement study of the robustness of camera--radar
sensor fusion in AD. We construct a unified evaluation pipeline on
the nuScenes dataset that simulates attacks on both sensing
modalities, including eight camera-image perturbations and three
radar point-cloud perturbations. We evaluate five representative
camera--radar fusion algorithms across two core perception tasks,
3D object detection and depth estimation, and quantify their
performance under clean and attacked inputs using standard metrics.
Our evaluation reveals a clear asymmetry in the robustness of
current fusion systems. Camera perturbations substantially degrade
both object detection and depth estimation, and moderate attack
strengths are already sufficient to cause severe performance
degradation. In contrast, the evaluated radar perturbations produce
smaller and task-dependent effects, with object-detection models
showing particularly limited degradation. Our analysis attributes
this difference to the strong dependence of existing fusion
architectures on dense camera features, together with the filtering,
association, and confidence-weighting mechanisms commonly applied
to sparse radar measurements. These results demonstrate that sensor
fusion does not inherently guarantee robustness and that the
contribution and reliability of each sensing modality must be
explicitly considered under adversarial conditions.
Based on four empirical findings, we further derive
measurement-guided design implications for future robust
camera--radar fusion, including modality-specific reliability
estimation, cross-modal consistency checking, adaptive fusion, and
graceful failover under severe sensor degradation.
\end{abstract}

\ccsdesc[500]{Security and privacy~Embedded systems security}
\ccsdesc[300]{Computer systems organization~Embedded and cyber-physical systems}
\ccsdesc[300]{Computing methodologies~Computer vision}

\keywords{Camera-radar sensor fusion, autonomous driving, sensor perception, adversarial attack, robustness}


\maketitle

\section{Introduction}

Autonomous driving (AD) relies on multiple sensors, such as cameras, radars, LiDARs, and ultrasonic sensors, to perceive and understand the surrounding environment. Sensor fusion is vital for AD because no single sensor is reliable enough in all scenarios. Many previous studies have demonstrated potential attacks on a single sensor~\cite{nassi2020phantom, hunt2023madradar, cao2019adversarial, zhang2017dolphinattack}. By combining complementary modalities, perception accuracy, robustness, and redundancy can be improved in the decision-making process. 

In particular, radar–camera fusion leverages radar’s precise range and Doppler velocity measurements, which are resilient to darkness, glare, rain, and fog, together with the camera's high-resolution appearance cues such as texture, color, lanes, and semantics \cite{patole2017automotive,sheeny2021radiate,barnes2019radarrobotcar}. 
By aligning radar returns with image features, the perception stack reduces false positives/negatives, resolves occlusions, and stabilizes multi-object tracking (e.g., using Doppler effect to disambiguate movers from static clutter) \cite{Nabati2021:WACV:OD,Lin2024:CVPR:OD,baumann2024cr3dt}. 
Radar depth priors also help regularize image-based depth and 3D detection, enhancing performance at long range and under adverse weather \cite{Nabati2021:WACV:OD,Lin2024:CVPR:OD,sheeny2021radiate}.

Depending on requirements, sensor fusion can be performed early (feature-level), mid-level (region/feature association), or late (decision-level), often with uncertainty-aware weighting to handle disagreements \cite{fayyad2020deepfusion,huang2022survey}. 
Beyond accuracy, cross-modal consistency checks also improve safety by detecting sensor faults or adversarial anomalies~\cite{matos2024sensorfailures,chen2021crossmodal,you2021tc2}.

However, an open question remains: ``\emph{are radar-camera sensor fusion systems robust under adversarial driving scenarios?}" To the best of our knowledge, no prior work has systematically evaluated the robustness of radar–camera fusion algorithms. In this paper, we conduct a systematic study of five radar–camera fusion algorithms across two tasks, object detection and depth estimation, quantifying their robustness under a suite of adversarial perturbations.

To construct the robustness-evaluation pipeline, we first simulate adversarial attacks on camera images and radar point clouds in the nuScenes Dataset~\cite{nuscenes2019}. We implement eight camera-based attacks and three radar-based attacks, treating each perturbed modality as a potentially malicious input to the sensor-fusion algorithms for object detection and depth estimation. We then assess robustness by measuring perception performance under attacks using standard evaluation metrics.


Drawing on our evaluation, we report four principal findings, provide in-depth analyses that reveal vulnerabilities in state-of-the-art sensor-fusion algorithm design, and outline future research directions to enhance the robustness of camera–radar fusion.

We summarize our contributions as follows. 

\begin{itemize}
  \item \textbf{First systematic robustness evaluation of camera-radar fusion under adversarial attacks.} To the best of our knowledge, this is the first comprehensive evaluation of five representative camera-radar sensor-fusion algorithms across two core tasks, 3D object detection and depth estimation, explicitly under adversarial conditions in autonomous driving.

  \item \textbf{Unified robustness-evaluation pipeline and attack suite.} We build a reproducible pipeline on the nuScenes dataset that simulates eight camera image attacks and three radar point cloud attacks, feeds perturbed single-modality inputs into fusion stacks, and quantifies degradation using standard metrics.

  \item \textbf{Empirical findings and design guidance.} We report four key findings exposing vulnerabilities in state-of-the-art fusion (e.g., brittleness to modality-specific and cross-modal inconsistencies, trade-offs across mid/late fusion, and limits of Doppler/depth priors in adverse weather/attack), and we translate them into actionable recommendations, such as uncertainty calibration, cross-modal consistency checks, robust association/alignment, and attack-aware training, to improve the robustness of camera-radar fusion. We will release our implementation upon acceptance.
\end{itemize}
\section{Related Work}

In the field of sensor fusion algorithm design for automotive systems, existing research primarily falls into two categories: sensor fusion for depth estimation \cite{lo2021depth, Lo2023:ICASSP:DE, long2021radar, singh2023depth, Li2020:WACV:DE, Ma2018:ICRA:DE} and sensor fusion for object detection \cite{long2023radiant, kim2023crn, Zhao2024:CVPR:OD, Nabati2021:WACV:OD, Lin2024:CVPR:OD, Kim2023:AAAI:OD}. 
This section provides a review of both categories, establishing a foundation for the subsequent discussions in this paper.
\subsection{Depth Estimation}
Sensor fusion algorithms for depth estimation typically integrate camera and radar sensor inputs to determine distances between the vehicle and surrounding objects. These algorithms leverage inputs from both sensors to produce accurate depth estimations, using LiDAR data as ground truth during the training phase. The typical pipeline involves two main stages: data association and depth completion.
Lo et al. \cite{lo2021depth} introduced a method that initially converts sparse 2D radar points into extended-height 3D measurements, subsequently incorporating these measurements into the network via a late fusion strategy.
Building upon similar ideas, Lo et al. \cite{Lo2023:ICASSP:DE} integrated radar data into a dense prediction transformer network, merging radar representations with camera features. Rather than using traditional readout tokens, radar data provides supplementary depth information, which could enhance monocular depth estimation performance.
Additionally, Long et al. \cite{long2021radar} proposed a fusion algorithm specifically designed to handle the sparse radar point cloud association problem. Singh et al. 
\cite{singh2023depth} addressed challenges posed by sparse and noisy radar inputs differently from prior approaches. Instead of treating the radar point cloud as a faulty depth map, their approach independently queries each raw radar point to associate it accurately with corresponding image pixels, producing a semi-dense radar depth map.
Li et al. \cite{Li2020:WACV:DE} tackled variability in depth maps using a multi-scale guided cascade hourglass network, efficiently and effectively managing diverse depth map patterns. Furthermore, Ma et al. \cite{Ma2018:ICRA:DE} explored direct learning from RGB-D raw data through a single deep regression network, and examined how the quantity of depth samples could influence prediction accuracy.

\subsection{Object Detection}
Sensor fusion algorithms for object detection leverage radar input to enhance detection accuracy, completeness, and robustness when identifying targets on the road.
Long et al. \cite{long2023radiant} introduced a method that combines feature-level and detection-level fusion, effectively utilizing state-of-the-art monocular detection techniques without requiring retraining.
Kim et al. \cite{kim2023crn} proposed a radar-camera fusion method that initially associates image proposals with radar points in the polar coordinate system to address spatial discrepancies. 
Moreover, the proposed adaptive cross-attention layers facilitate robust and attentive spatial-contextual feature exchange between camera and radar modalities.
Zhao et al. \cite{Zhao2024:CVPR:OD} aimed to bridge the performance gap between LiDAR-camera and camera-radar detectors using a cross-modality knowledge distillation framework.
This approach leverages a Bird’s-Eye-View (BEV) representation as a unified feature space to effectively transfer knowledge between modalities.
Nabati et al. \cite{Nabati2021:WACV:OD} developed a method employing a center-point detection network to identify object centers directly from images. 
A frustum-based data association approach subsequently associates radar detections accurately with these identified center points.
Lin et al. \cite{Lin2024:CVPR:OD} presented an efficient radar feature extraction method coupled with a deformable cross-attention fusion module, enabling robust and precise radar-camera feature alignment and integration.

Although the sensor fusion algorithms discussed above have demonstrated strong performance in typical scenarios, none explicitly considered the impact of adversarial attacks targeting road-based sensors. 
Existing adversarial defense research in autonomous driving primarily focuses on unimodal architectures: single-camera or single-radar sensors.  

On the camera side, several studies use detection-based techniques to identify anomalies and perturbed regions in the input image and suppress noise~\cite{zheng2018robust, chiang2021adversarial, naseer2019local, xiang2021detectorguard}. Other studies attempt to mitigate and filter the adversarial perturbations by intercepting the preprocessing pipeline of the vehicle system before they reach the model's perception module \cite{deng2021deep, kyrkou2020towards, krishna2020defending}. 

A commonly adopted technique is adversarial training, where models are trained with a combination of clean data and adversarially perturbed inputs. Accordingly, the network can improve the generalization and robustness against many different adversarial manipulations \cite{liu2021training, mahmood2019deep, yan2018deep}. However, adversarial training is computationally expensive, and the victim model can still fail when exposed to novel attack methods.
At the training stage, Lee et al. \cite{lee_random_sample} proposed a defense mechanism based on random layer sampling, training ensemble architectures by using the weights of different sampled networks, reducing the effectiveness of gradient-based attacks.

On the radar side, existing security studies mainly focus on utilizing sensor fusion algorithms that combine radar with camera and LiDAR data \cite{oluwaferanmi2025secure, long2023radiant, lo2021depth, li2022modality, gohring2011radar, wang2022interfusion, wang2023bi}. Given the inherently sparse and ambiguous nature of radar inputs, directly employing radar for downstream tasks such as depth estimation often leads to noisy or erroneous outcomes \cite{singh2023depth}. Therefore, integrating radar with other modalities such as camera and LiDAR has become a common strategy to enhance the accuracy and reliability of vehicle perception systems, including depth estimation and object detection. However, despite the recent advancements of multimodal fusion, both camera and LiDAR sensors have shown to be highly vulnerable to  natural perturbations and physical adversarial attacks, posing significant threats to the robustness of autonomous driving systems \cite{zhu2021can, zhu2021adversarial, jia2020fooling}.

In this paper, we evaluate the robustness and effectiveness of  camera-radar sensor fusion algorithms when subjected to various types of adversarial attacks on both camera and radar sensors.




\section{System Design}

In this section, we provide an overview of robustness evaluation of camera-radar sensor fusion algorithms. 

\subsection{System Overview}
\label{sec:system_overview}

\begin{figure*}
    \centering
    \includegraphics[width=0.75\linewidth]{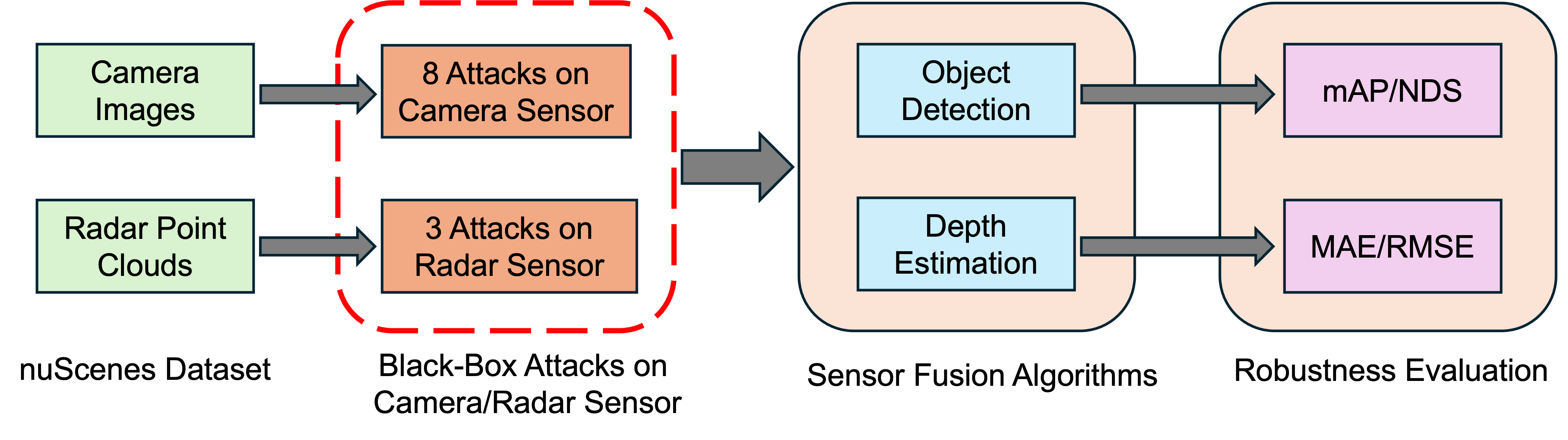}
    \caption{Overview of robustness evaluation on camera-radar sensor fusion algorithms under different black-box attacks.}
    \label{fig:system_design}
\end{figure*}

Fig.~\ref{fig:system_design} presents our robustness-evaluation pipeline
for camera--radar sensor fusion under different black-box attacks.
We use the nuScenes dataset~\cite{nuscenes2019}, a large-scale
autonomous-driving dataset containing synchronized camera, radar,
and LiDAR measurements collected in real-world driving scenarios.

We denote the camera image and radar point cloud at time $t$ as
$I_t$ and $P_t$, respectively. A camera attack is represented by
$\mathcal{A}_c(\cdot;\lambda_c)$, where $\lambda_c$ denotes the
attack-specific strength. The resulting attacked image is

\begin{equation}
    \widetilde{I}_t
    =
    \mathcal{A}_c(I_t;\lambda_c).
    \label{eq:camera_attack_general}
\end{equation}

Similarly, a radar attack is represented by
$\mathcal{A}_r(\cdot;\lambda_r)$, producing

\begin{equation}
    \widetilde{P}_t
    =
    \mathcal{A}_r(P_t;\lambda_r),
    \label{eq:radar_attack_general}
\end{equation}
where $\lambda_r$ controls the severity of the radar corruption.
The specific definitions of $\lambda_c$ and $\lambda_r$ depend on
the corresponding attack and are described in
Sections~\ref{sec:camera_attacks} and~\ref{sec:radar_attacks}.

For a given fusion architecture, we denote the camera and radar
encoders as $\mathcal{E}_c$ and $\mathcal{E}_r$. Their intermediate
representations are

\begin{equation}
    F_t^c = \mathcal{E}_c(I_t),
    \qquad
    F_t^r = \mathcal{E}_r(P_t),
    \label{eq:modality_features}
\end{equation}
for clean inputs. Under attack, the corresponding encoder receives
$\widetilde{I}_t$ or $\widetilde{P}_t$. Although the five evaluated
models implement fusion at different stages, we use
$\mathcal{F}(\cdot)$ as a general notation for their fusion
operation,

\begin{equation}
    F_t^f
    =
    \mathcal{F}(F_t^c,F_t^r),
    \label{eq:general_fusion}
\end{equation}
where $F_t^f$ denotes the resulting fused representation. Depending
on the evaluated task, the downstream prediction is either an
object-detection output $\hat{Y}_t^{\mathrm{OD}}$ or a depth map
$\hat{D}_t$.

In our evaluation, only one sensing modality is attacked at a time.
For camera-side attacks, the model receives
$(\widetilde{I}_t,P_t)$, whereas for radar-side attacks it receives
$(I_t,\widetilde{P}_t)$. We implement eight attacks on the camera
sensor and three attacks on the radar sensor and quantify the
resulting degradation using standard object-detection and
depth-estimation metrics.


Fig.~\ref{fig:system_design} presents the system overview of robustness evaluation pipeline on camera-radar sensor fusion algorithms under different black-box attacks. Here, we use the nuScence dataset~\cite{nuscenes2019}, a large-scale public dataset widely used in autonomous driving research. The dataset includes camera, radar, and LiDAR data collected from real-world driving scenarios. We focus on simulating adversarial attacks targeting camera images and radar point clouds. Specifically, we design and implement eight attacks on the camera sensor and three attacks on the radar sensor. Each attacked modality serves as a potentially malicious input to the object detection (OD) or depth estimation (DE) fusion algorithms. Finally, we assess the robustness of these algorithms by measuring their performance under attacks using a variety of standard evaluation metrics.

\subsection{Threat Model and Evaluation Scope}
\label{sec:threat_model}

We consider adversarial perturbations targeting the sensing inputs
of camera--radar fusion systems. The adversary aims to degrade the
downstream perception performance of object detection (OD) or depth
estimation (DE) by manipulating either the camera observation
$I_t$ or the radar observation $P_t$ before they are processed by
the corresponding perception pipeline.

\textbf{Attacker Capability.}
We consider a black-box setting in which the adversary does not
require access to the internal model parameters or gradients of the
target camera--radar fusion algorithm. Instead, the attacker
perturbs the sensor input through camera-side or radar-side
manipulations. For camera attacks, the resulting observation is

\begin{equation}
    \widetilde{I}_t
    =
    \mathcal{A}_c(I_t;\lambda_c),
\end{equation}
while the radar observation $P_t$ remains unchanged. For radar
attacks,

\begin{equation}
    \widetilde{P}_t
    =
    \mathcal{A}_r(P_t;\lambda_r),
\end{equation}
while the camera observation $I_t$ remains unchanged.

\textbf{Single-Modality Attack Assumption.}
We focus on attacks against one sensing modality at a time. Thus,
the fusion model receives either
$(\widetilde{I}_t,P_t)$ under a camera attack or
$(I_t,\widetilde{P}_t)$ under a radar attack. This setting allows
us to isolate how perturbations to each modality propagate through
the fusion pipeline and affect downstream perception.

\textbf{Attack Goal.}
We consider untargeted attacks whose objective is to degrade the
overall perception performance rather than force a specific object,
class, depth, or bounding-box prediction. Accordingly, attack
effectiveness is quantified using the standard task metrics adopted
in our evaluation, including mAP/NDS or MATE/MAP for object
detection and MAE/RMSE for depth estimation.

\textbf{Evaluation Scope.}
Our study focuses on the robustness of camera--radar perception
under sensor-input perturbations. We do not consider simultaneous
attacks on both modalities, attacks on the communication links or
fusion software, or attacks on other sensing modalities such as
LiDAR. We also do not assume that the adversary modifies the
parameters of the evaluated fusion models. These cases represent
important directions for future study.

\subsection{Sensor Fusion Algorithm Selection}

We conduct a comprehensive study of camera-radar sensor fusion algorithms developed between 2019 and the present. Fusion strategies between camera and radar modalities are typically grouped into three categories:  
\begin{itemize}
  \item \textbf{Early fusion}, which concatenates raw inputs or low-level features before joint processing.  
  \item \textbf{Mid-level fusion}, which processes each modality separately before merging their feature representations through mechanisms such as  concatenation, attention, or association modules.  
  \item \textbf{Late fusion}, which continues to separate modality processing until near the output stage, often combining predictions or high-level features.  
\end{itemize}

For consistency, we use $F_t^c$ and $F_t^r$ to denote the
modality-specific representations immediately before the principal
camera--radar interaction of each architecture, and $F_t^f$ to
denote the representation after this interaction. These symbols
provide a unified description of the evaluated models and do not
imply that the architectures perform fusion at identical network
layers.

From this, we select five algorithms on the object detection or depth estimation tasks that are open-source and reproducible for our study in both mid-level fusion and late fusion. 
There are several early fusion works available~\cite{siddiqui2020multi, lim2019radar}. Siddiqui et al. \cite{siddiqui2020multi} integrated radar data as an additional channel of the image input to enhance the overall depth estimation result through a series of Convolutional Neural Networks (CNNs). However, these models are not open source and therefore cannot be reproduced in our experiments. The selected algorithms include the object detection approaches RADIANT~\cite{long2023radiant} and CRN~\cite{kim2023crn}, as well as the depth estimation approaches RC-PDA~\cite{long2021radar}, Cam\&mmWave~\cite{singh2023depth}, and DORN~\cite{lo2021depth}.

RADIANT~\cite{long2023radiant} employs a mid-level fusion design, where radar and camera features are extracted independently and then associated through a feature interaction module for robust 3D object detection. 
Similarly, CRN~\cite{kim2023crn} applies mid-level fusion by lifting image features into bird’s-eye view with radar guidance and aggregating multi-modal features through deformable attention. 
RC-PDA~\cite{long2021radar} also follows a mid-level paradigm: it first learns radar-to-pixel associations to densify radar depth maps, and then fuses them with image features in the completion network. 
In contrast, Cam\&mmWave~\cite{singh2023depth} integrates semi-dense radar depth with image embeddings through a gated fusion module. 
Finally, DORN with radar extension~\cite{lo2021depth} adopts a late-fusion strategy, concatenating radar and camera features after separate encoders, which simplifies integration but may lose fine-grained cross-modal interactions. 
Overall, most recent models favor mid-level fusion for balancing complementary information exchange and modality-specific feature learning, while late fusion provides modularity and ease of extension at the cost of weaker cross-modal alignment.


\section{Robustness Evaluation Under Camera Attacks}

In this section, we evaluate the impact of the camera sensor
attacks on the performance of the state-of-the-art (SOTA) camera-radar sensor
fusion algorithms.

\subsection{Attacks on Camera Sensors}
\label{sec:camera_attacks}

We evaluate eight black-box camera attacks belonging to three
categories: three conventional blur attacks, four motion blur attacks~\cite{ji2021poltergeist} and LensAttack~\cite{zhou2024optical}.
Following Eq.~\eqref{eq:camera_attack_general}, each camera attack
is represented as

\begin{equation}
    \widetilde{I}_t
    =
    \mathcal{A}_c(I_t;\lambda_c),
\end{equation}
where $\lambda_c$ denotes the attack-specific perturbation strength.

The conventional blur attacks include Gaussian, median, and
bilateral blur, which emulate image degradation caused by
conditions such as dust or contamination on the camera lens. For
Gaussian and median blur, $\lambda_c$ is controlled by the kernel
size, while the bilateral-blur configuration additionally specifies
the color- and spatial-filter parameters.

The motion-blur attacks are motivated by Poltergeist~\cite{ji2021poltergeist},
which demonstrated that acoustic interference with a camera
stabilization system can induce motion blur. We evaluate vertical,
horizontal, radial, and rotational motion blur. Their attack
strengths are controlled by the corresponding kernel size or
geometric transformation parameters.

Finally, LensAttack~\cite{zhou2024optical} introduces an optical
distortion by placing an additional lens in front of the camera,
causing local image magnification or shrinkage. For LensAttack,
$\lambda_c$ is controlled by the focal-length setting used to
generate the distortion.

Unless otherwise specified, only the camera stream is perturbed in
these experiments. Thus, the fusion model receives

\begin{equation}
    (\widetilde{I}_t,P_t)
    =
    \left(
    \mathcal{A}_c(I_t;\lambda_c),P_t
    \right),
    \label{eq:camera_attack_input}
\end{equation}
while the radar observation remains unchanged.



\subsection{Evaluation Results}

The experiments were conducted on the nuScenes dataset~\cite{nuscenes2019}, a large-scale public dataset widely used in autonomous driving research. Since all the selected fusion algorithms were originally evaluated on nuScenes, we applied adversarial modifications to the ``CAMERA FRONT'' images in the dataset to ensure compatibility. We restricted our experiments to the test dataset.

For evaluating depth estimation performance, we use kernel size to control how much blurring is applied to the dataset. We adopt two commonly used regression metrics: Mean Absolute Error (MAE) and Root Mean Square Error (RMSE). For object detection, we use two standard detection metrics: Mean Average Precision (mAP) and NuScenes Detection Score (NDS). 

For all camera experiments, $\lambda_c=0$ denotes the clean
condition, while increasing $\lambda_c$ corresponds to stronger
attack-specific perturbations. Because different attacks use
different physical or image-processing parameters, the numerical
value of $\lambda_c$ is not compared directly across attack
families. Instead, we evaluate low, moderate, and high settings for
each attack according to its corresponding parameterization.

Given the similarity among attacks within each category, we focus on model performance under three representative perturbations: Gaussian blur, vertical motion blur, and LensAttack. Tables~\ref{tab:gaussian_blur}, \ref{tab:vertical_blur}, and \ref{tab:lensattack_blur} present the results for the selected five sensor fusion models under these three attacks. We define adversarial strength based on parameter values in different attacks, e.g., kernel size and focal length. It can be observed that for all target models, both mAP and NDS decrease as the adversarial strength increases, indicating a decline in object detection accuracy. Additionally, MAE and RMSE increase significantly with stronger adversarial perturbations for all models, which shows a deterioration in depth estimation performance.

\begin{table}[]
\centering
\caption{The performance of different camera-radar sensor fusion models under the Gaussian blur attack with various adversarial strengths. OD Metrics: mAP/NDS. DE Metrics: MAE/RMSE.}
\label{tab:gaussian_blur}
\resizebox{0.6\columnwidth}{!}{%
\begin{tabular}{|l|c|llll|}
\hline
\multicolumn{1}{|c|}{\multirow{2}{*}{Model}} & \multirow{2}{*}{Type} & \multicolumn{4}{c|}{Adversarial Strength (Kernel Size)}                                                                                                                                                                                                                                                                                      \\ \cline{3-6} 
\multicolumn{1}{|c|}{}                       &                       & \multicolumn{1}{c|}{No Attack}                                                         & \multicolumn{1}{c|}{15}                                                                & \multicolumn{1}{c|}{45}                                                                & \multicolumn{1}{c|}{75}                                           \\ \hline
RADIANT                                      & OD                    & \multicolumn{1}{l|}{0.363/0.340}                                                       & \multicolumn{1}{l|}{0.336/0.322}                                                       & \multicolumn{1}{l|}{0.277/0.294}                                                       & 0.272/0.290                                                       \\ \hline
CRN                                          & OD                    & \multicolumn{1}{l|}{0.473/0.562}                                                       & \multicolumn{1}{l|}{0.442/0.540}                                                       & \multicolumn{1}{l|}{0.362/0.494}                                                       & 0.3526/0.490                                                      \\ \hline
RC-PDA                                       & DE                    & \multicolumn{1}{l|}{\begin{tabular}[c]{@{}l@{}}0.831/2.018\\ 1.279/2.682\end{tabular}} & \multicolumn{1}{l|}{\begin{tabular}[c]{@{}l@{}}1.012/2.355\\ 1.518/3.130\end{tabular}} & \multicolumn{1}{l|}{\begin{tabular}[c]{@{}l@{}}1.433/3.097\\ 1.952/3.728\end{tabular}} & \begin{tabular}[c]{@{}l@{}}1.910/3.793\\ 2.450/4.265\end{tabular} \\ \hline
Cam\&mmWave                                & DE                    & \multicolumn{1}{l|}{2.037/4.548}                                                       & \multicolumn{1}{l|}{4.004/7.257}                                                       & \multicolumn{1}{l|}{10.146/ 16.286}                                                    & 10.533/16.677                                                     \\ \hline
DORN                                 & DE                    & \multicolumn{1}{l|}{2.384/5.264}                                                       & \multicolumn{1}{l|}{3.061/6.263}                                                       & \multicolumn{1}{l|}{3.992/7.789}                                                       & 4.68/9.138                                                        \\ \hline
\end{tabular}%
}
\end{table}

\begin{table}[]
\centering
\caption{The performance of different camera-radar sensor fusion models under the vertical motion blur attack with various adversarial strengths. OD Metrics: mAP/NDS. DE Metrics: MAE/RMSE.}
\label{tab:vertical_blur}
\resizebox{0.6\columnwidth}{!}{%
\begin{tabular}{|l|l|llll|}
\hline
\multirow{2}{*}{Model} & \multirow{2}{*}{Type} & \multicolumn{4}{c|}{Adversarial Strength (Kernel Size)}                                                                                                                                                                                                                                                                                           \\ \cline{3-6} 
                                &                                & \multicolumn{1}{l|}{No Attack}                                                         & \multicolumn{1}{c|}{20}                                                       & \multicolumn{1}{c|}{50}                                                       & \multicolumn{1}{c|}{80}                                   \\ \hline
RADIANT                         & OD                             & \multicolumn{1}{l|}{0.363/0.340}                                                       & \multicolumn{1}{l|}{0.290/0.301}                                                       & \multicolumn{1}{l|}{0.272/0.293}                                                       & 0.272/0.292                                                       \\ \hline
CRN                             & OD                             & \multicolumn{1}{l|}{0.473/0.562}                                                       & \multicolumn{1}{l|}{0.4066/0.519}                                                      & \multicolumn{1}{l|}{0.361/0.496}                                                       & 0.3541/0.491                                                      \\ \hline
RC-PDA                          & DE                             & \multicolumn{1}{l|}{\begin{tabular}[c]{@{}l@{}}0.831/2.018\\ 1.279/2.682\end{tabular}} & \multicolumn{1}{l|}{\begin{tabular}[c]{@{}l@{}}1.226/2.691\\ 1.755/3.362\end{tabular}} & \multicolumn{1}{l|}{\begin{tabular}[c]{@{}l@{}}1.750/3.538\\ 2.273/3.825\end{tabular}} & \begin{tabular}[c]{@{}l@{}}2.282/4.357\\ 2.760/4.558\end{tabular} \\ \hline
Cam\&mmWave                     & DE                             & \multicolumn{1}{l|}{2.037/4.548}                                                       & \multicolumn{1}{l|}{7.156/12.467}                                                      & \multicolumn{1}{l|}{10.029/16.329}                                                     & 10.421/16.809                                                     \\ \hline
DORN                     & DE                             & \multicolumn{1}{l|}{2.384/5.264}                                                       & \multicolumn{1}{l|}{4.87/9.931}                                                        & \multicolumn{1}{l|}{8.261/14.902}                                                      & 10.325/17.305                                                     \\ \hline
\end{tabular}%
}
\end{table}

\begin{figure*}[t]
\centering
\subfigure[No attack]{\includegraphics[width=0.23\textwidth]{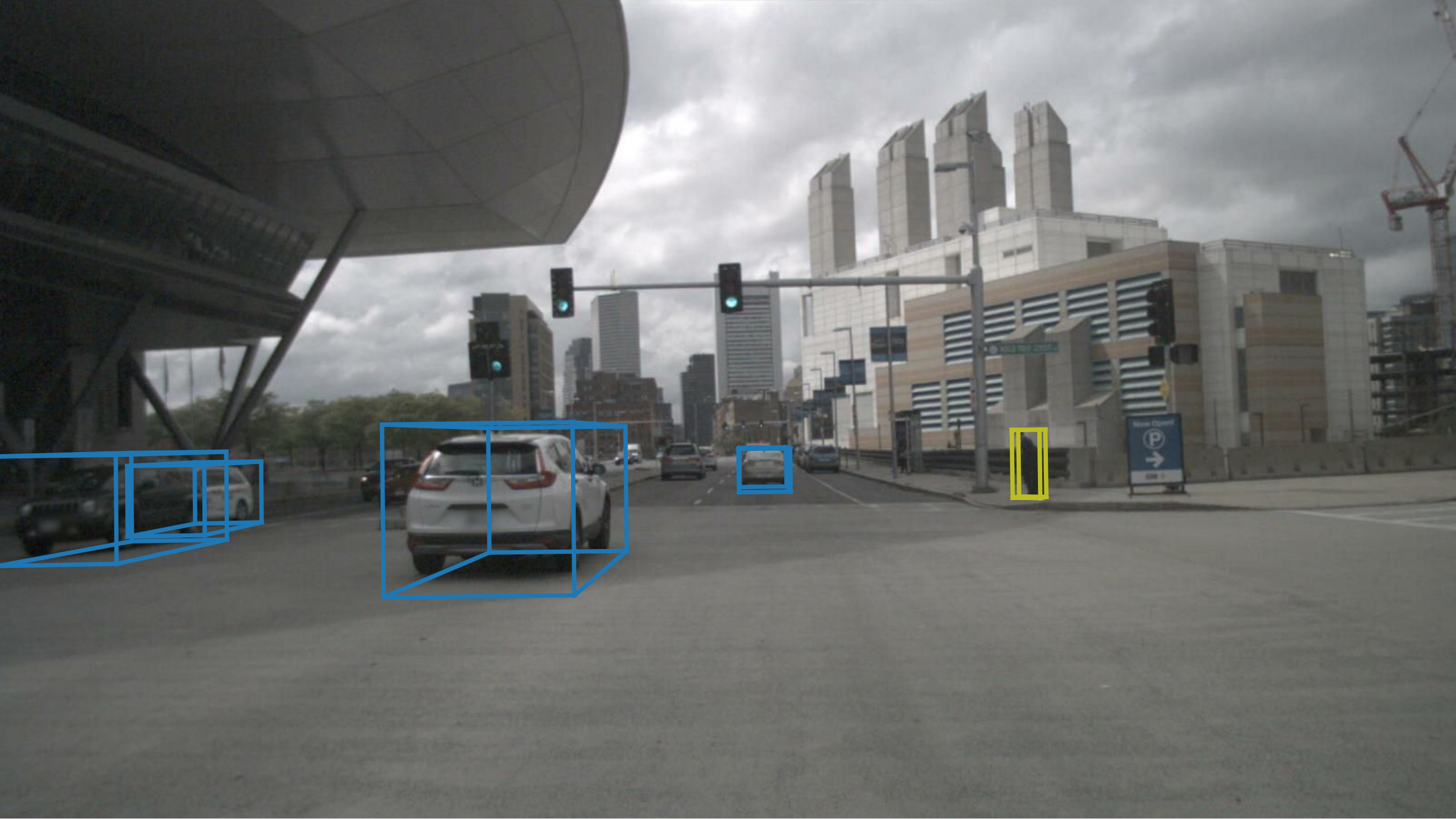}\label{fig:gaussian_NoAttack}}
\subfigure[Kernel=15]{\includegraphics[width=0.23\textwidth]{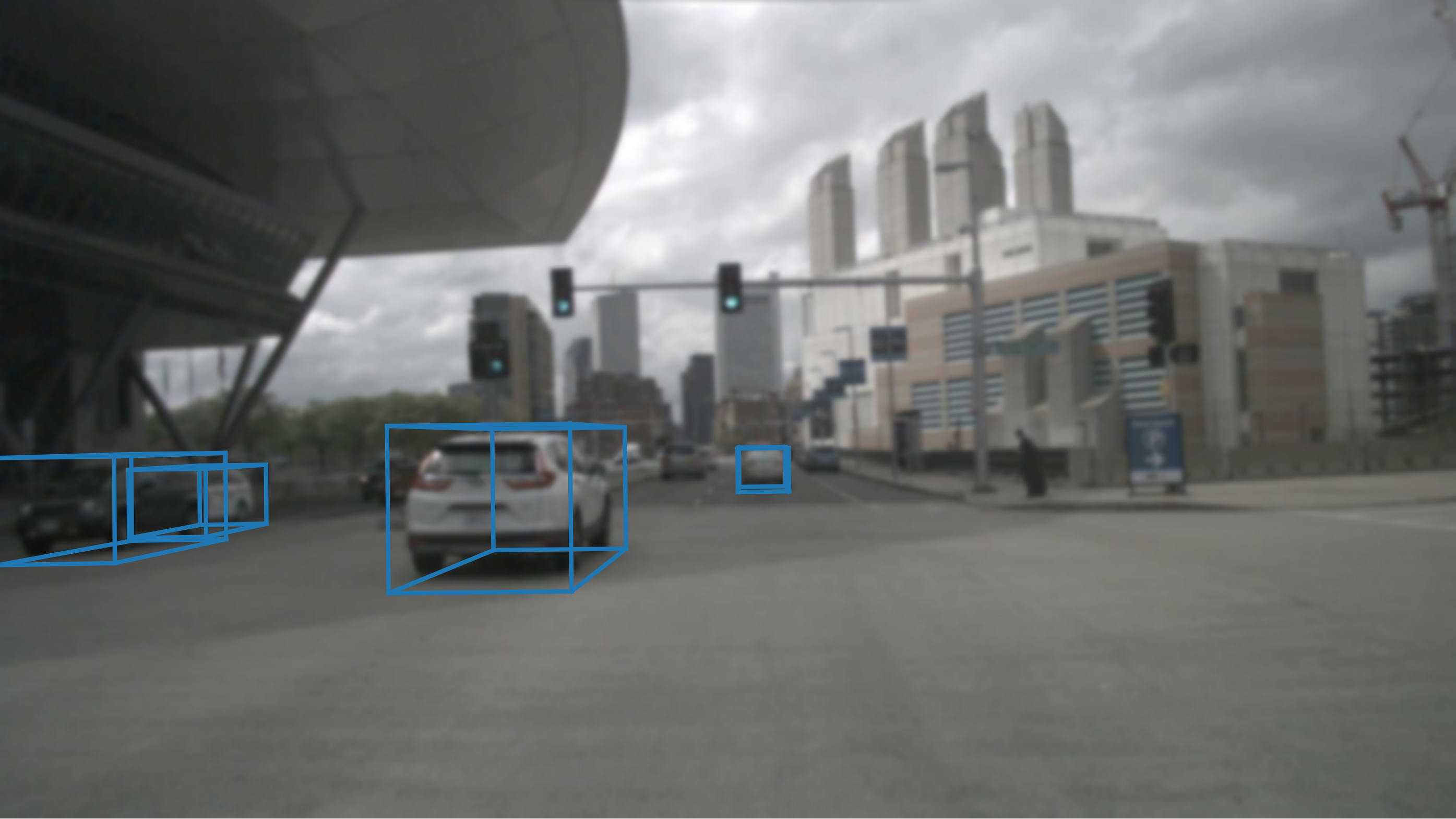}\label{fig:gaussian1}}
\subfigure[Kernel=45]{\includegraphics[width=0.23\textwidth]{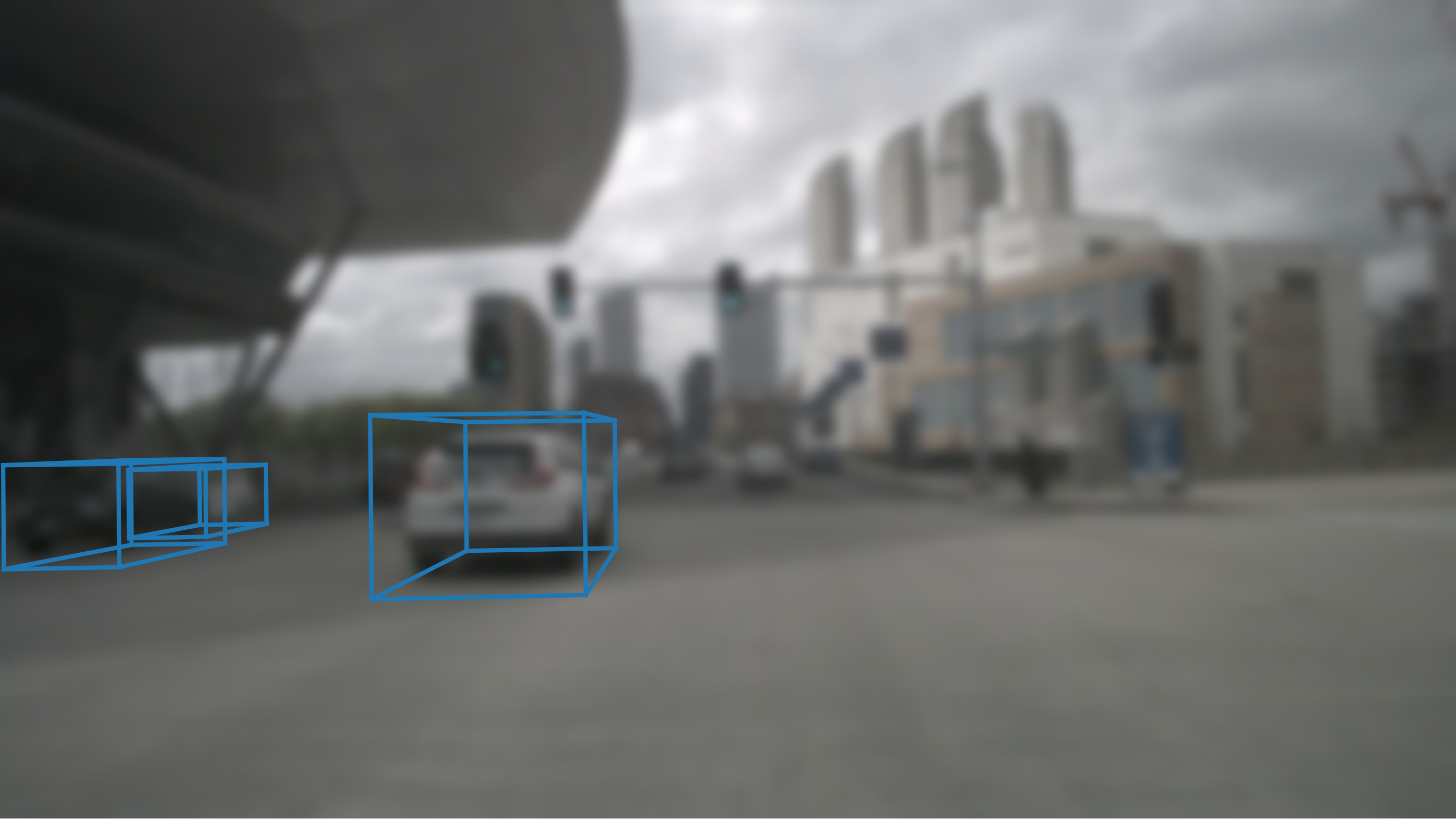}\label{fig:gaussian2}}
\subfigure[Kernel=75]{\includegraphics[width=0.23\textwidth]{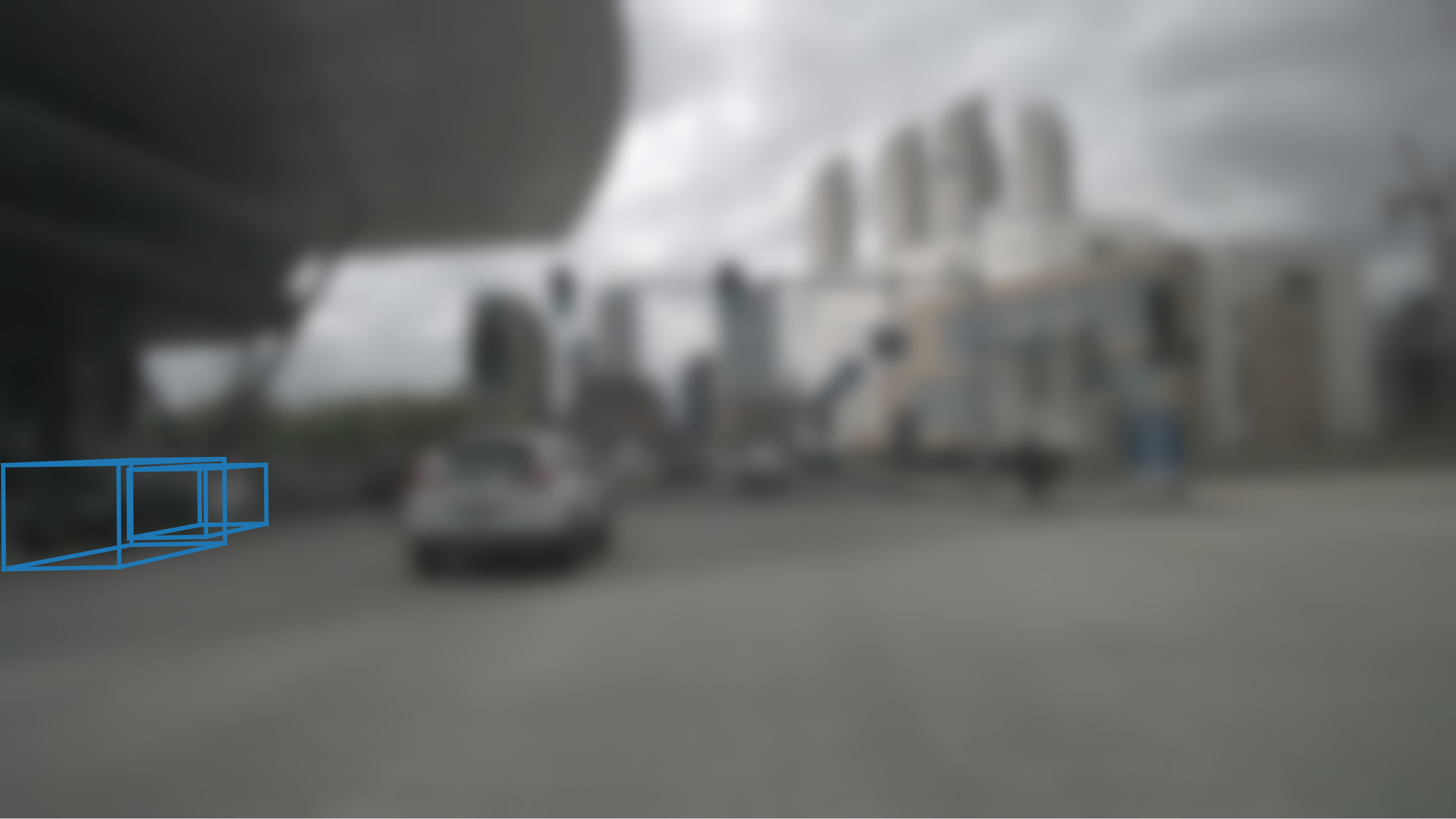}\label{fig:gaussian3}}

\subfigure[No attack]{\includegraphics[width=0.23\textwidth]{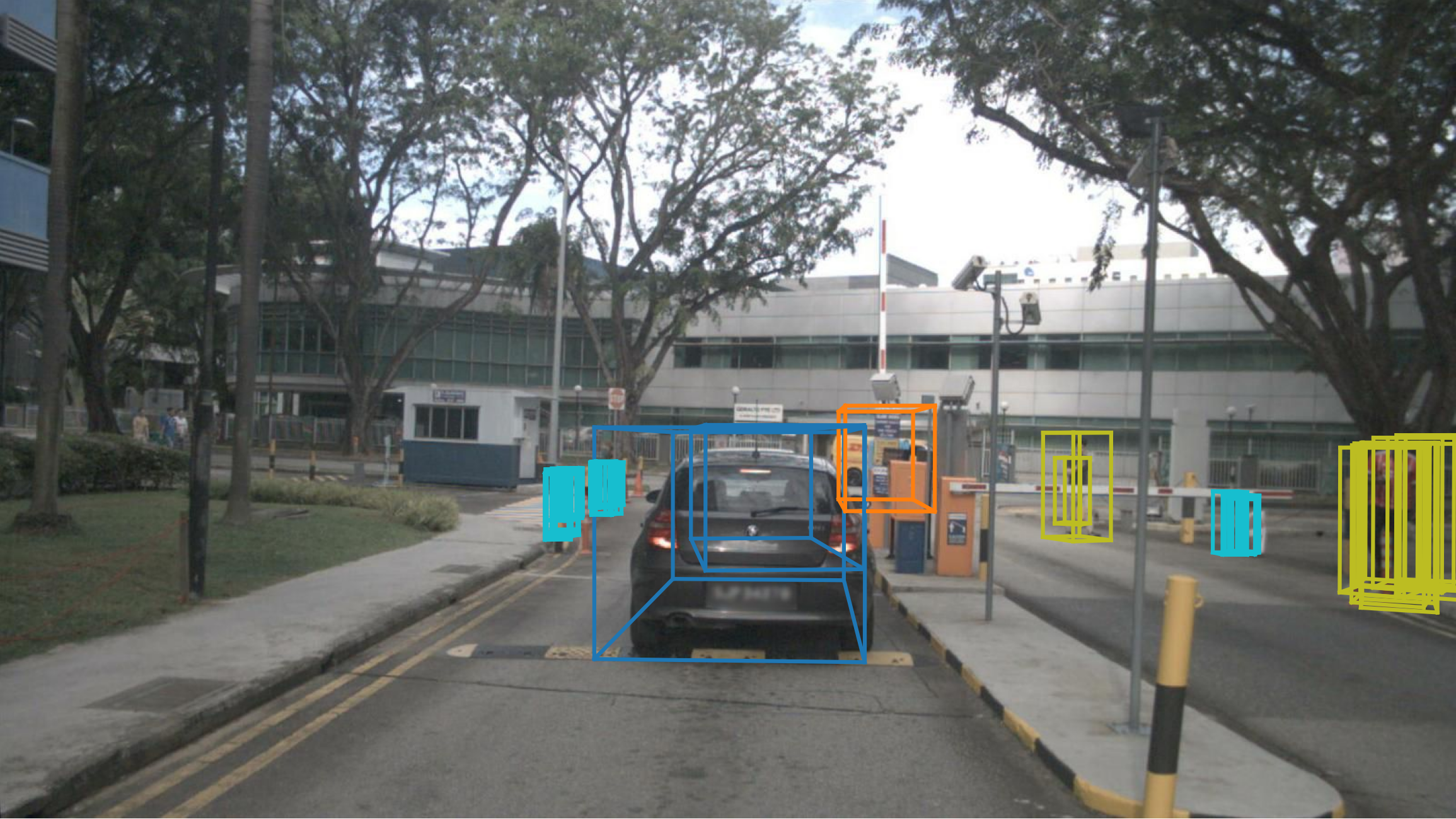}\label{fig:serial_design}}
\subfigure[Kernel=15]{\includegraphics[width=0.23\textwidth]{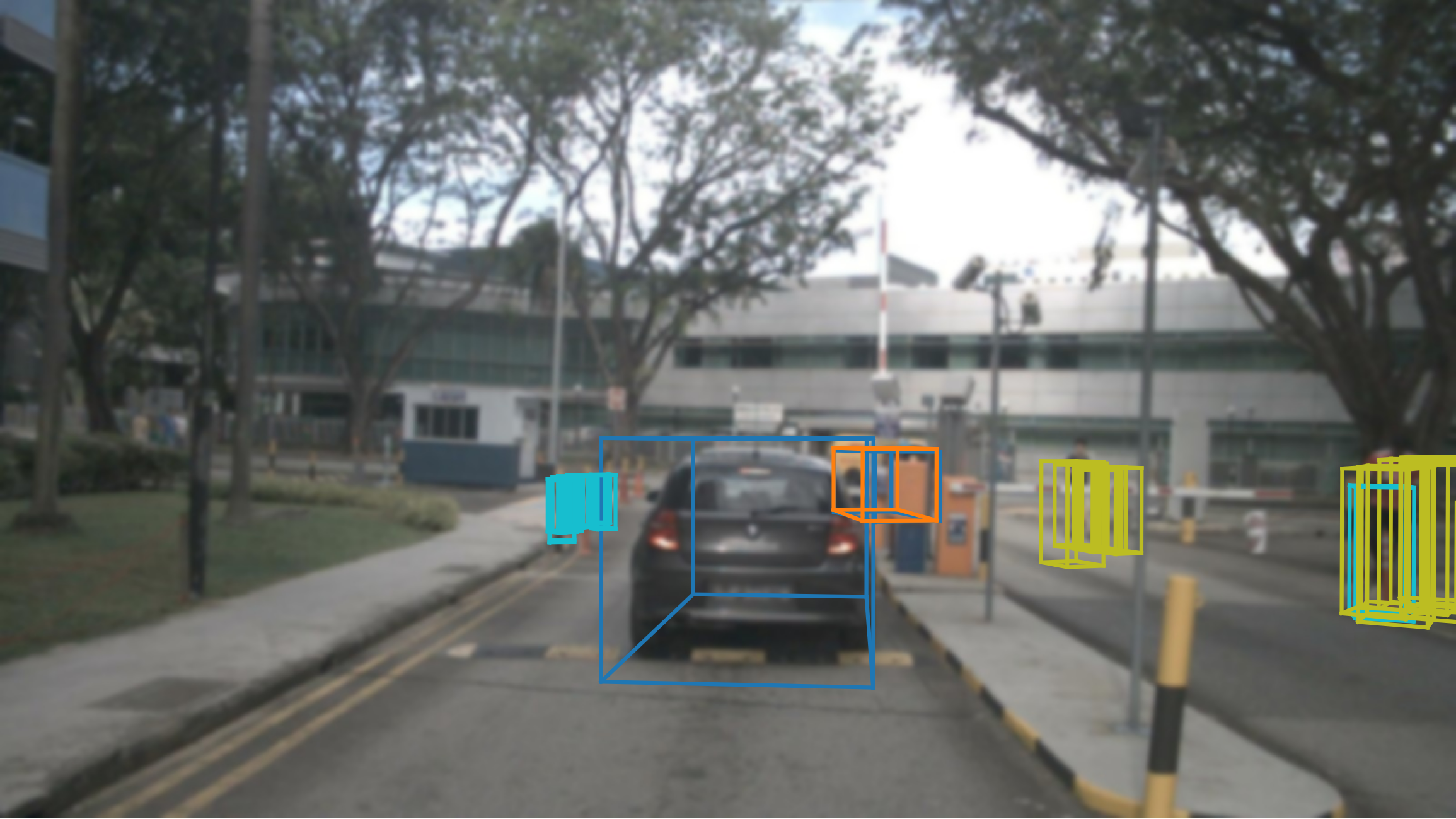}\label{fig:parallel_design}}
\subfigure[Kernel=45]{\includegraphics[width=0.23\textwidth]{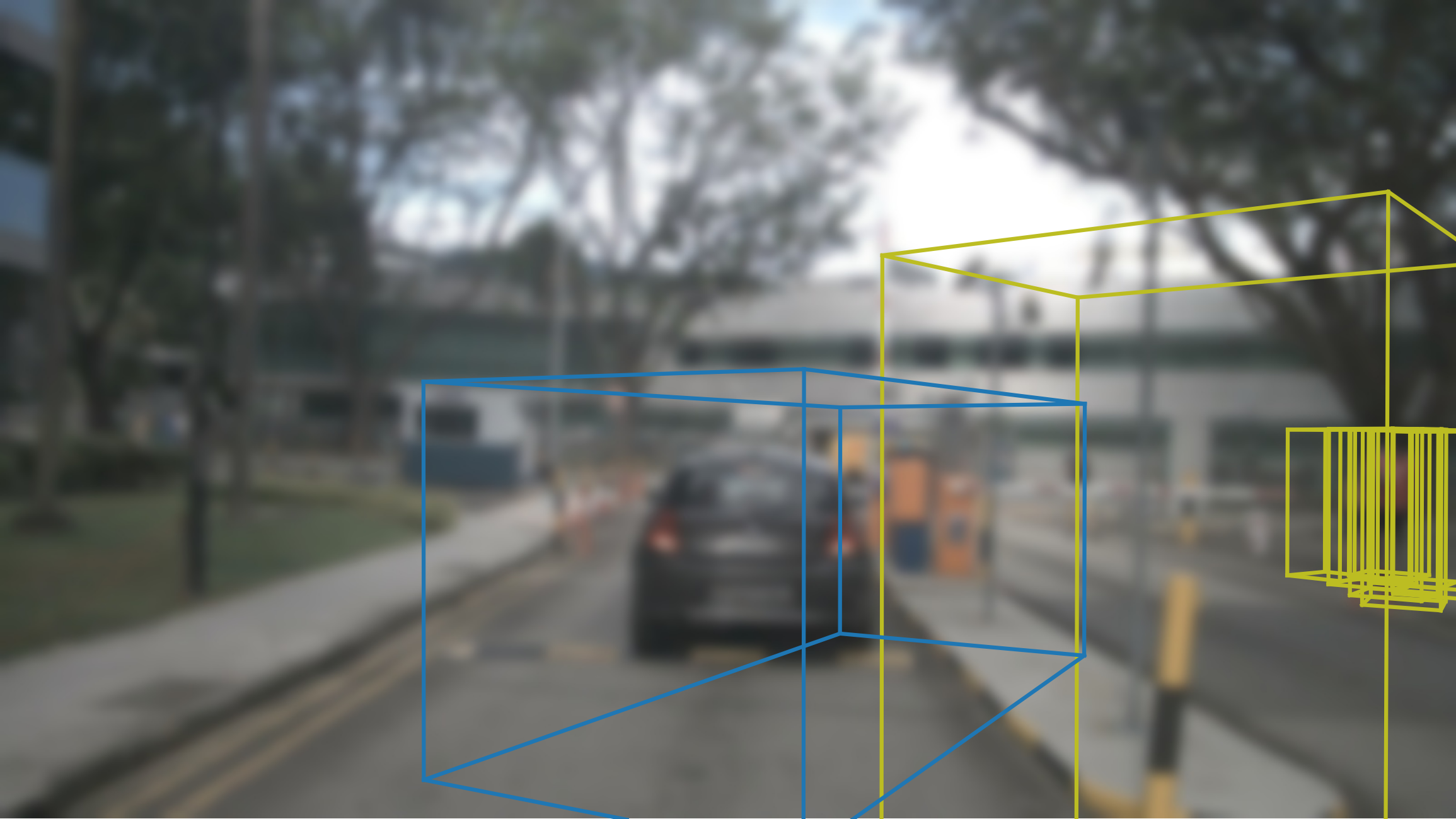}\label{fig:parallel_design}}
\subfigure[Kernel=75]{\includegraphics[width=0.23\textwidth]{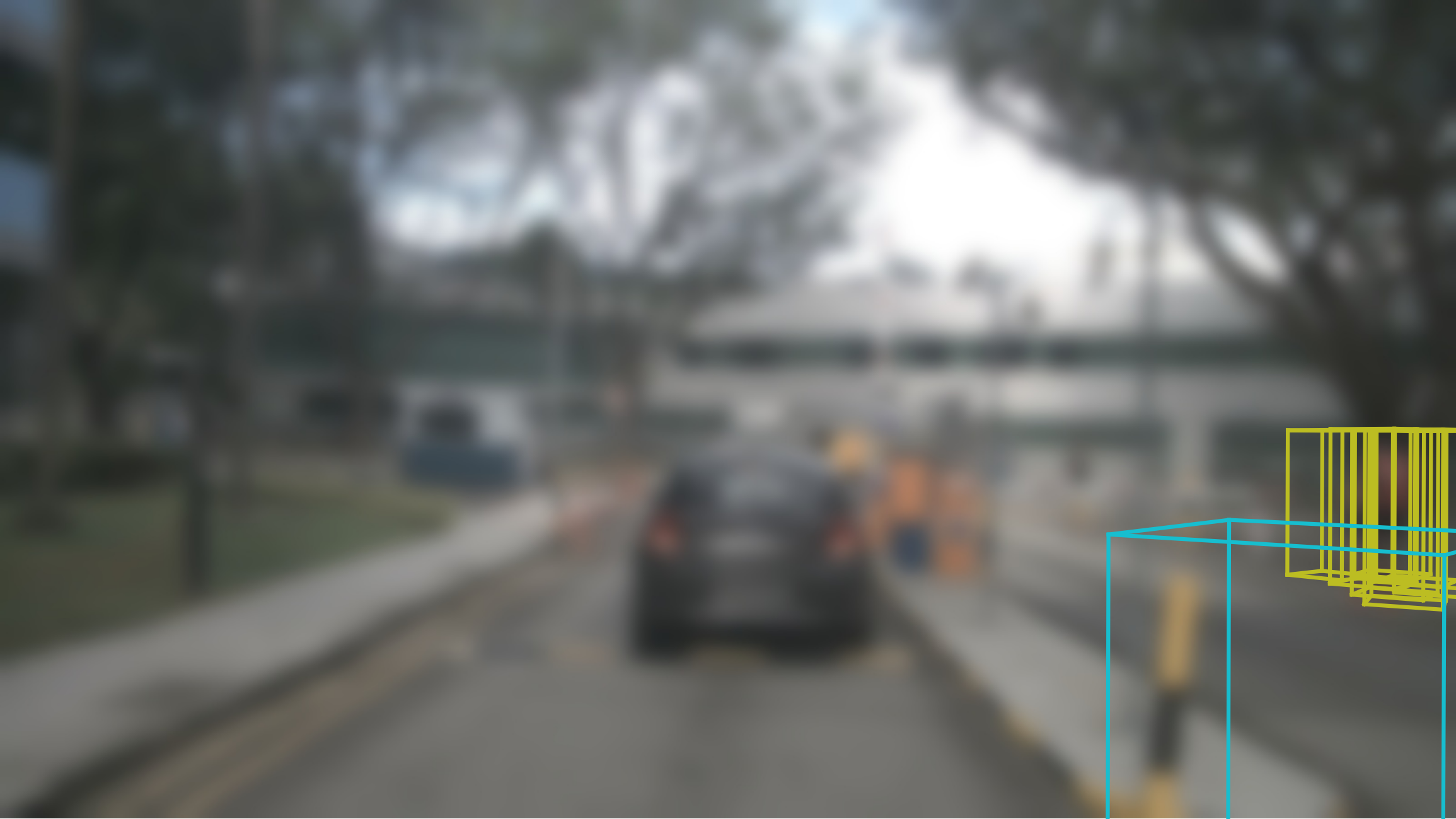}\label{fig:parallel_design}}

\subfigure[No attack]{\includegraphics[width=0.23\textwidth,trim=5 20 80 238, clip]{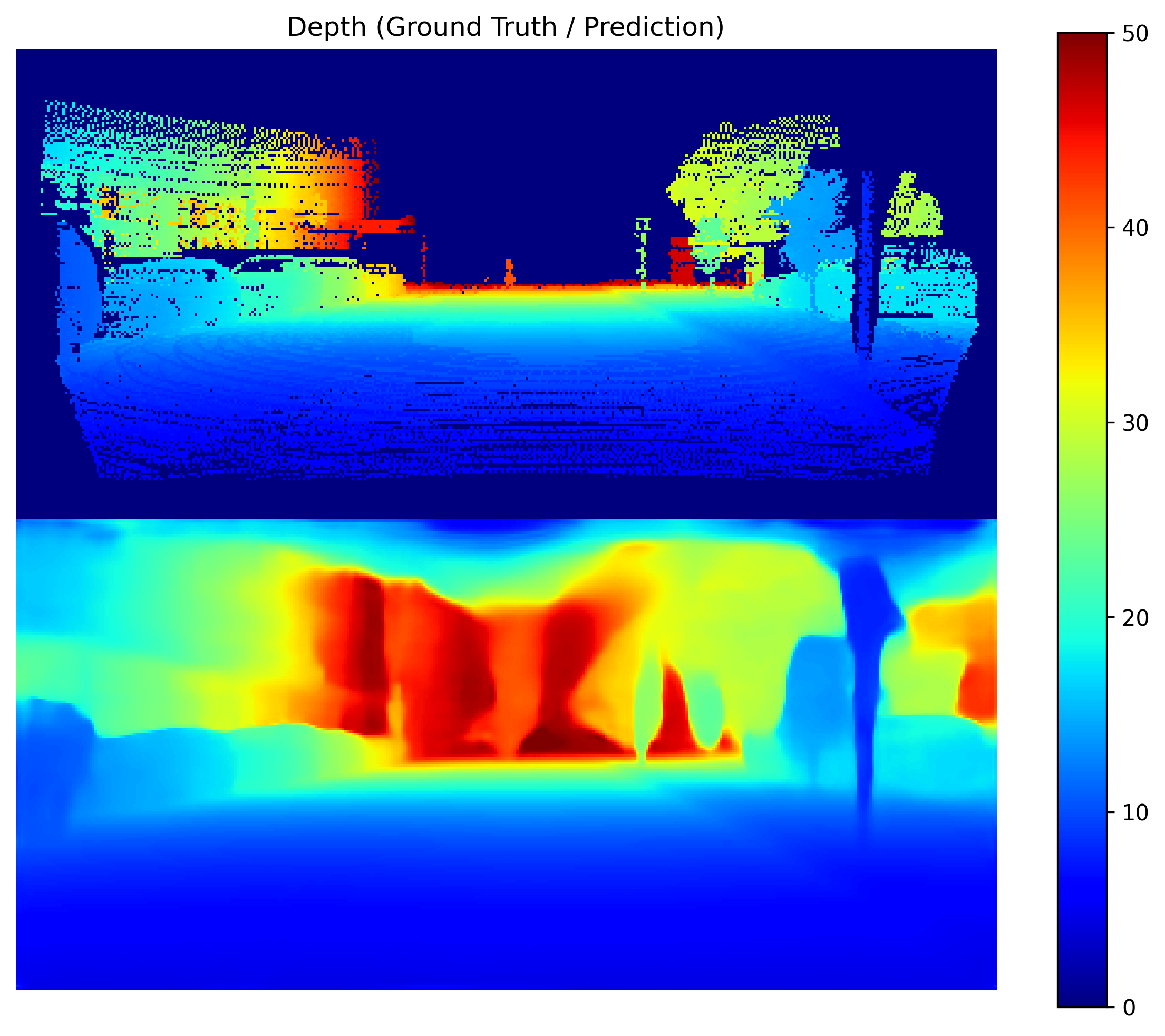}\label{fig:serial_design}}
\subfigure[Kernel=15]{\includegraphics[width=0.23\textwidth,trim=5 20 80 238, clip]{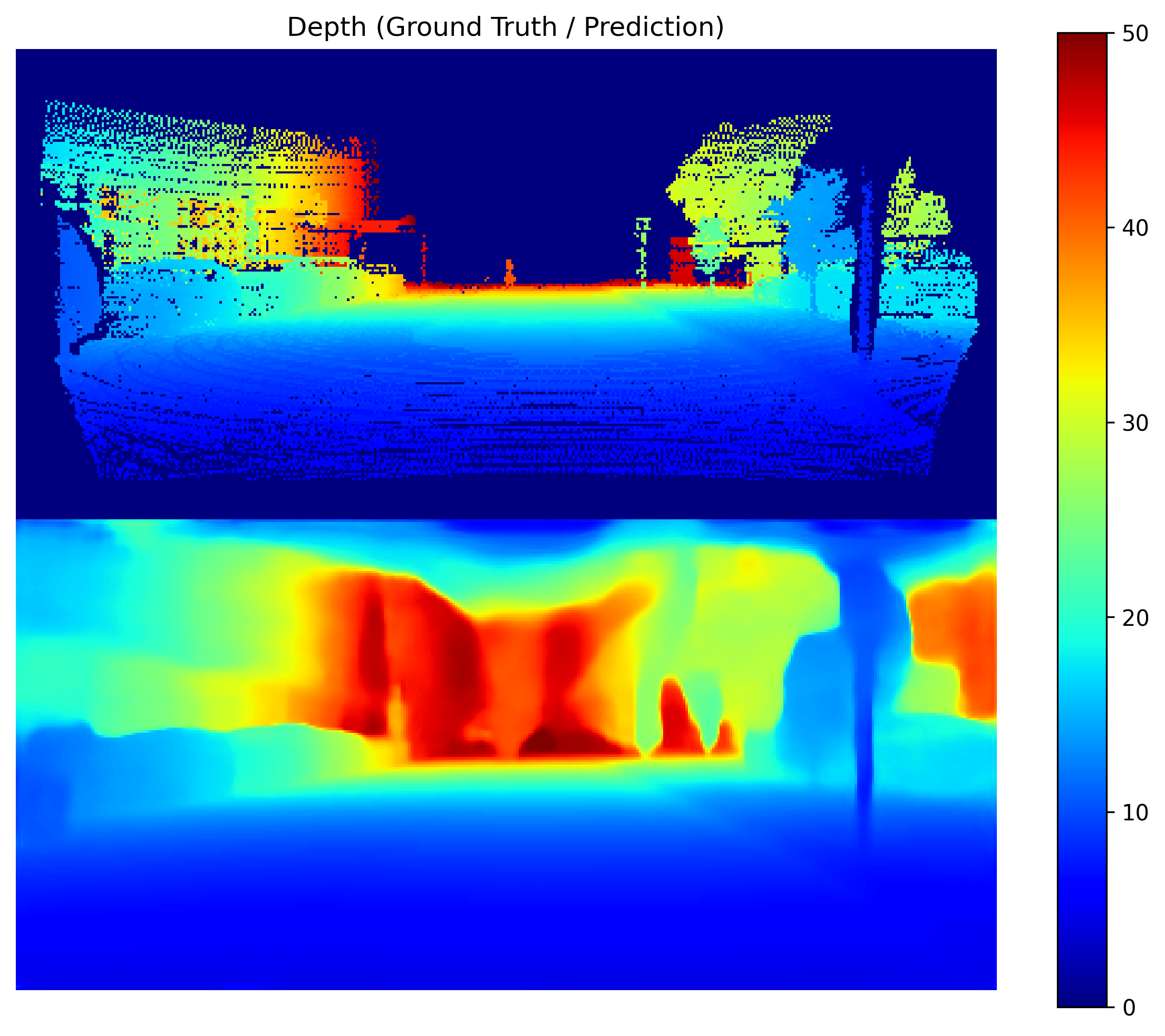}\label{fig:parallel_design}}
\subfigure[Kernel=45]{\includegraphics[width=0.23\textwidth,trim=5 20 80 238, clip]{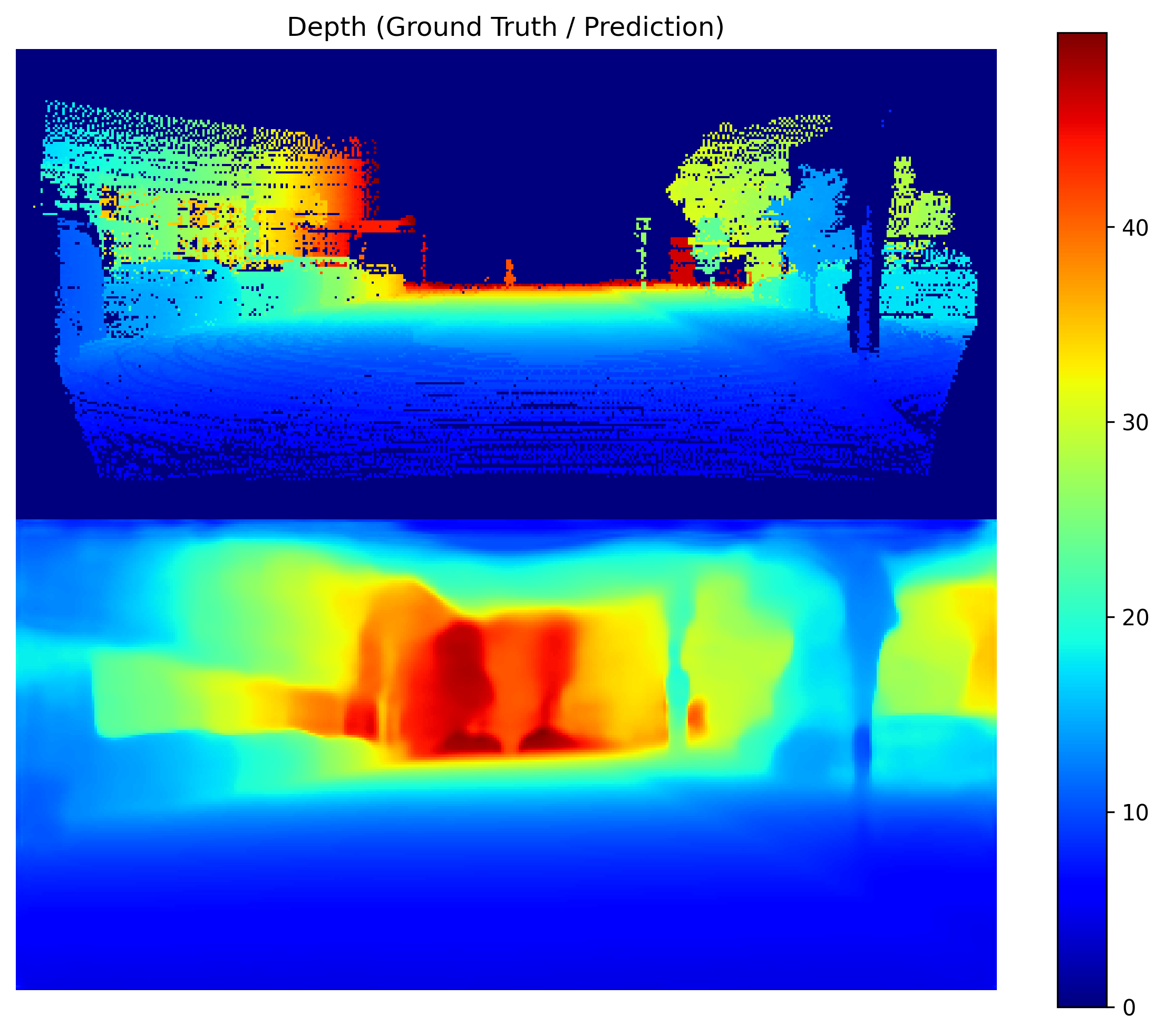}\label{fig:parallel_design}}
\subfigure[Kernel=75]{\includegraphics[width=0.23\textwidth,trim=5 20 80 238, clip]{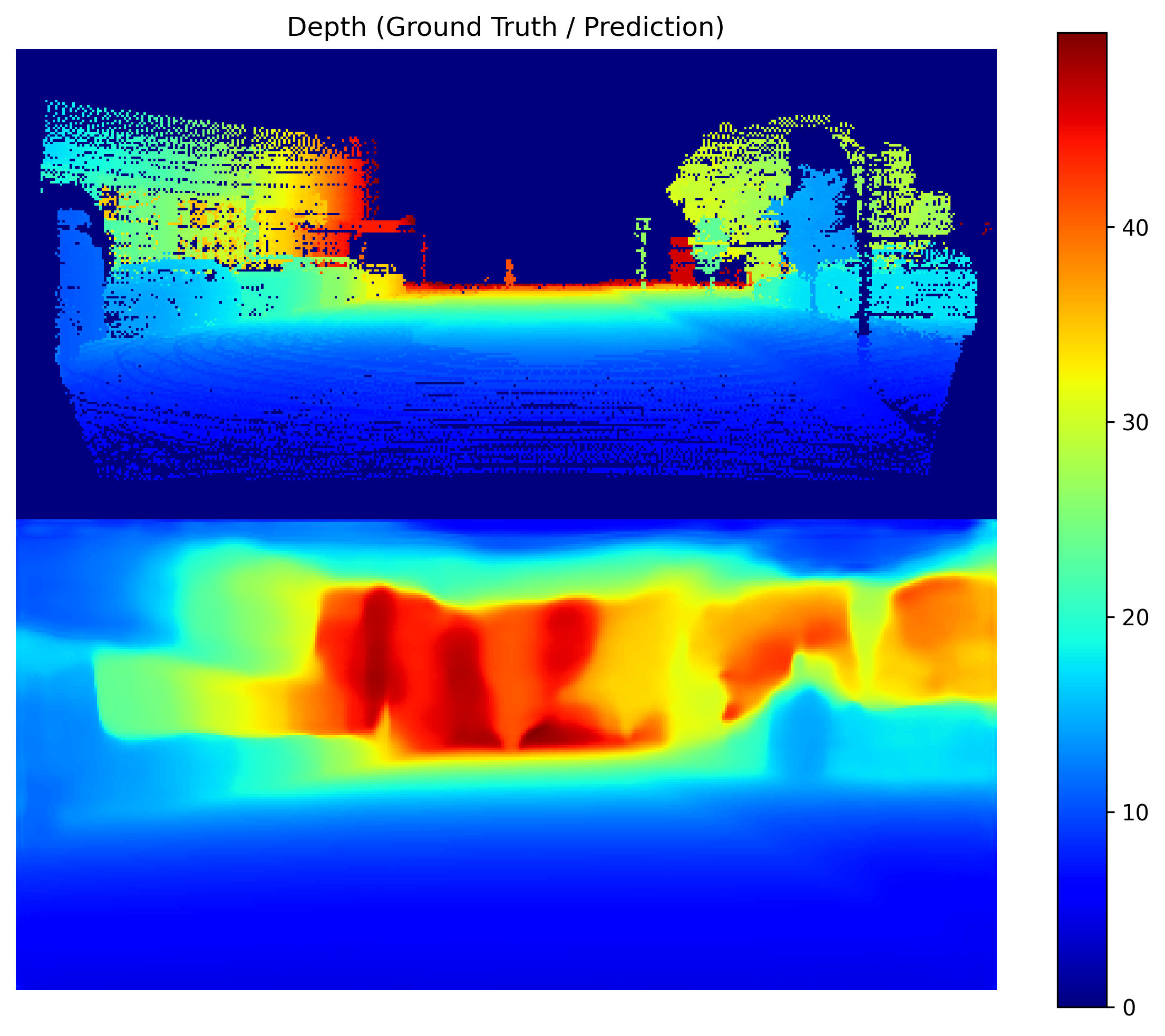}\label{fig:parallel_design}}

\subfigure[No attack]{\includegraphics[width=0.23\textwidth]{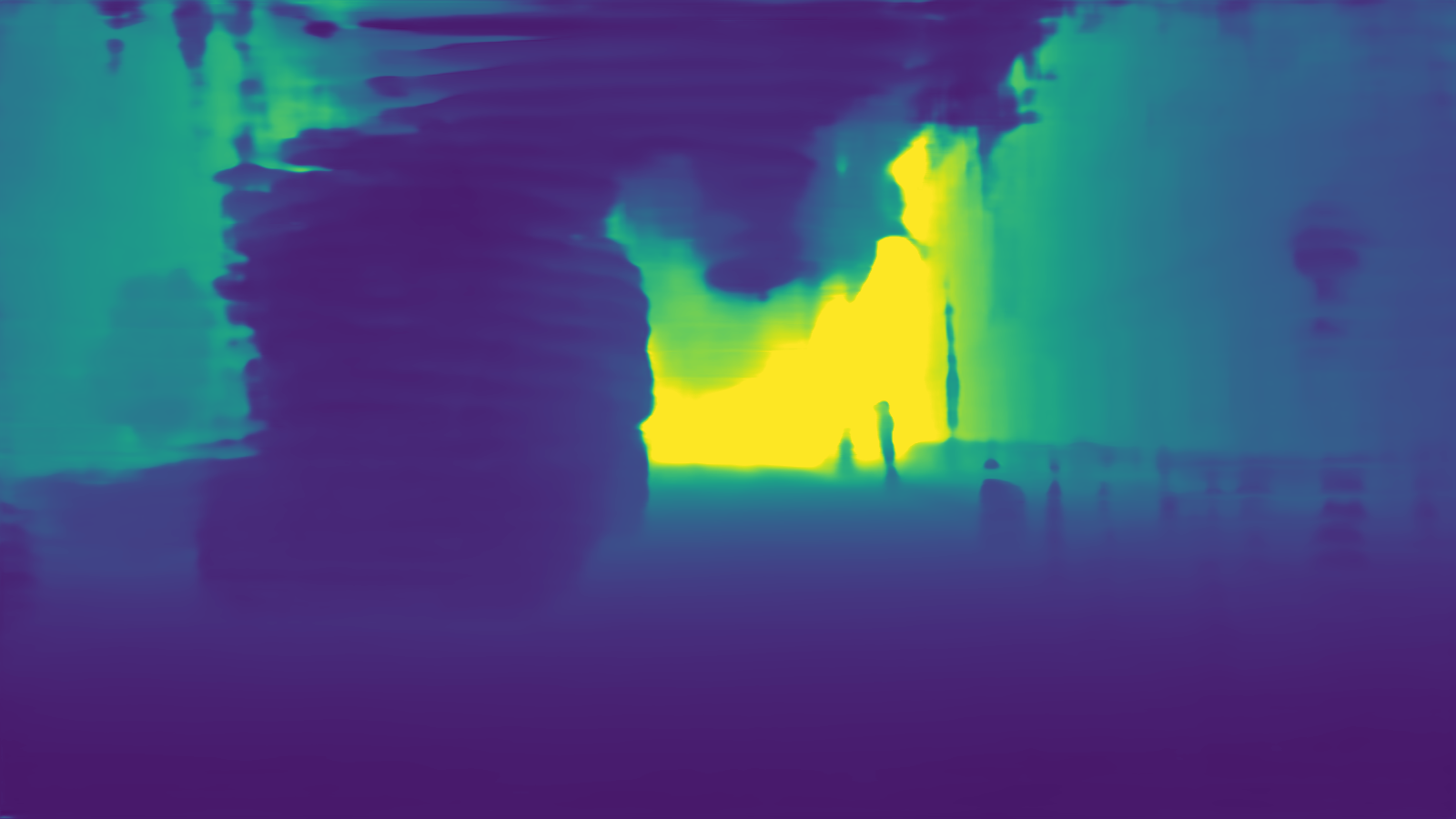}\label{fig:serial_design}}
\subfigure[Kernel=15]{\includegraphics[width=0.23\textwidth]{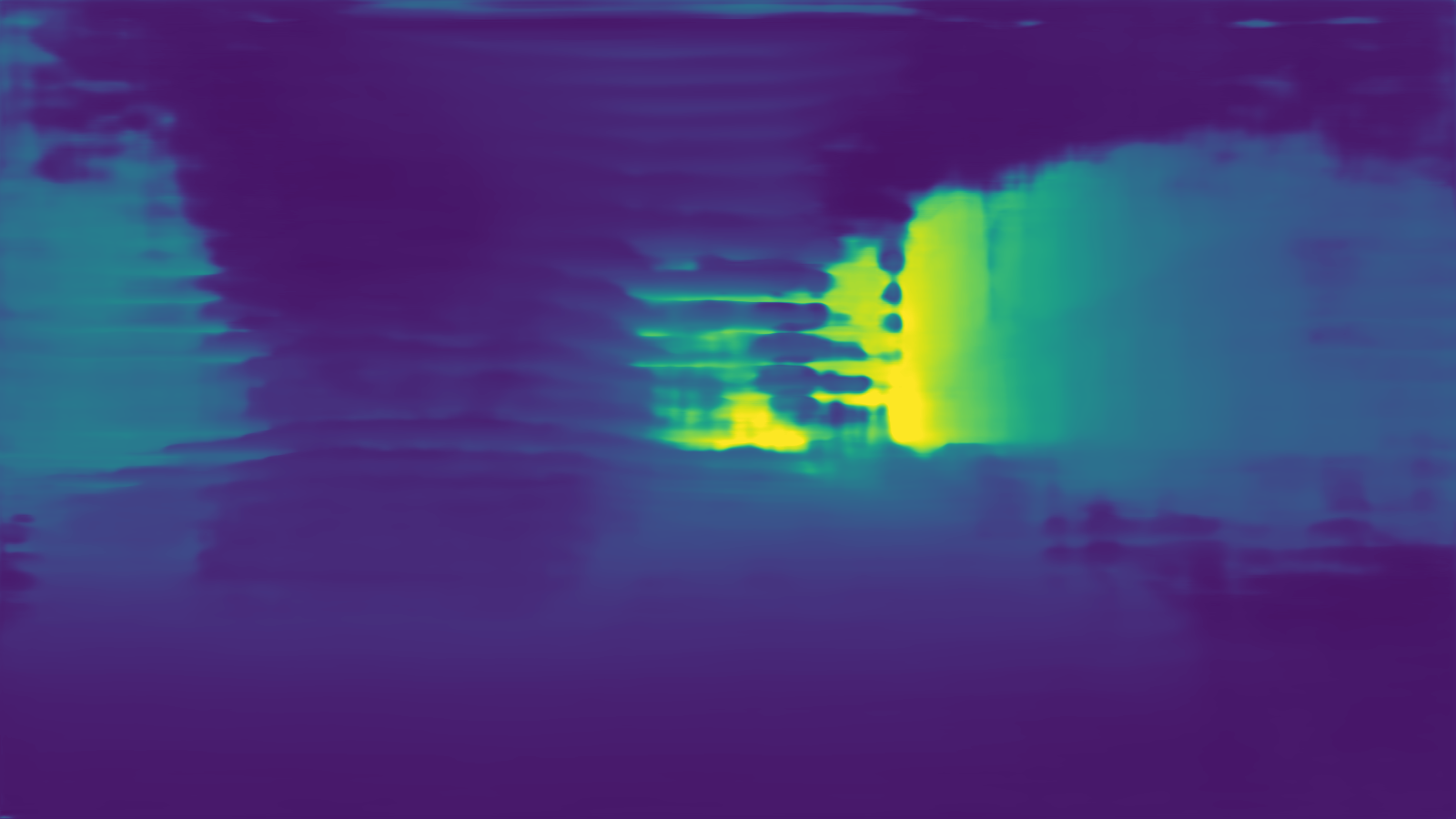}\label{fig:parallel_design}}
\subfigure[Kernel=45]{\includegraphics[width=0.23\textwidth]{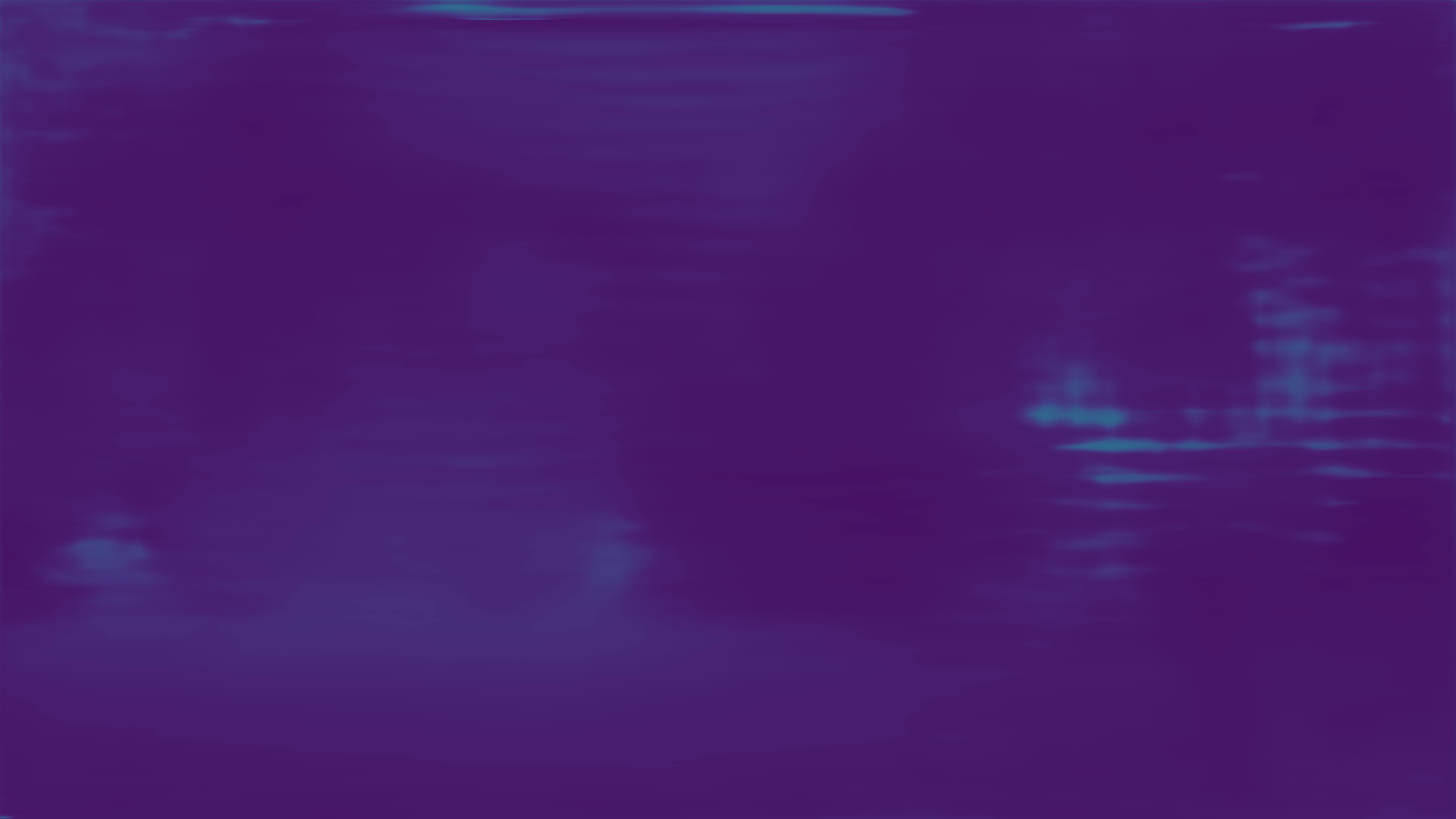}\label{fig:parallel_design}}
\subfigure[Kernel=75]{\includegraphics[width=0.23\textwidth]{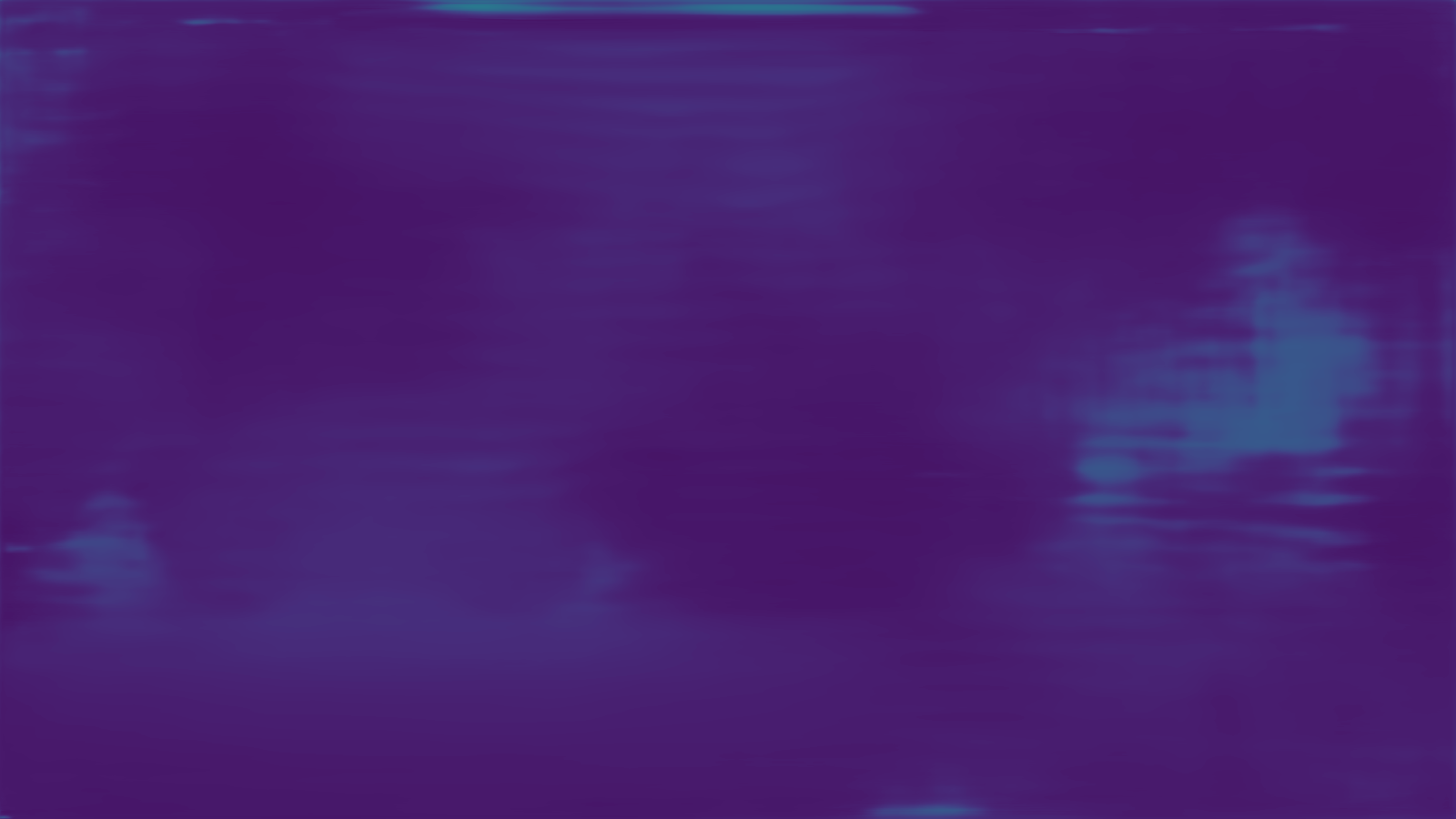}\label{fig:parallel_design}}

\subfigure[No attack]{\includegraphics[width=0.23\textwidth]{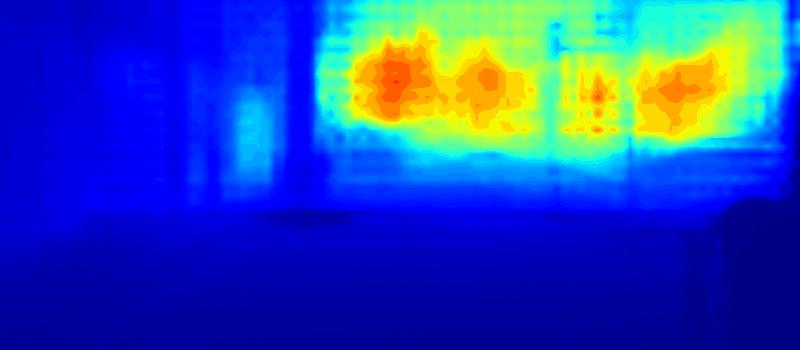}\label{fig:serial_design}}
\subfigure[Kernel=15]{\includegraphics[width=0.23\textwidth]{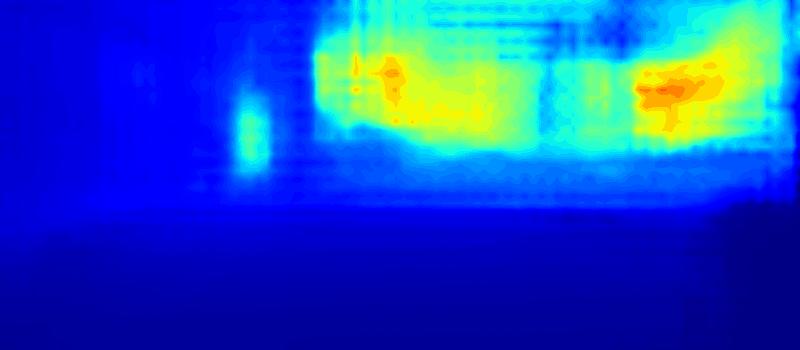}\label{fig:parallel_design}}
\subfigure[Kernel=45]{\includegraphics[width=0.23\textwidth]{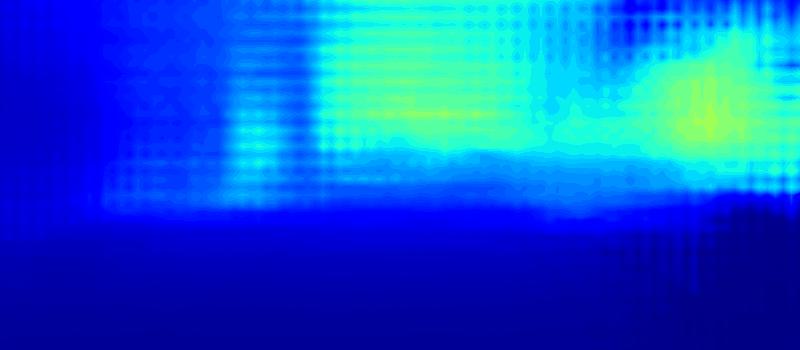}\label{fig:parallel_design}}
\subfigure[Kernel=75]{\includegraphics[width=0.23\textwidth]{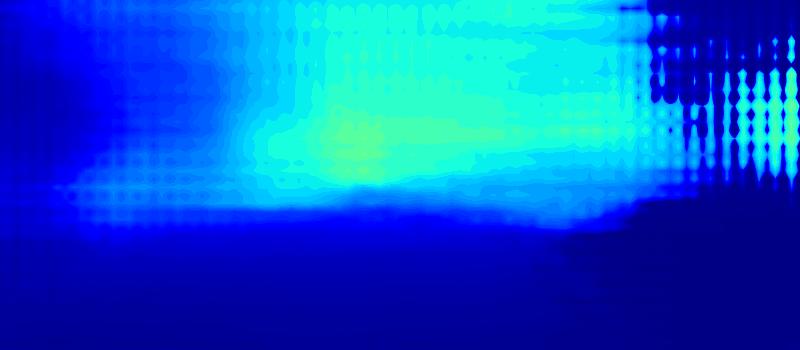}\label{fig:parallel_design}}

\caption{Examples of Gaussian Blur attack on various object detection and depth estimation sensor fusion algorithms. The first to the last rows correspond to the object detection algorithms, RADIANT~\cite{long2023radiant} and CRN~\cite{kim2023crn}, and the depth estimation algorithms, RC-PDA~\cite{long2021radar}, Cam\&mmWave~\cite{singh2023depth}, and DORN~\cite{lo2021depth}. }
\label{fig:gaussian_attack}
\end{figure*}

\begin{figure*}[t]
\centering
\subfigure[No attack]{\includegraphics[width=0.23\textwidth]{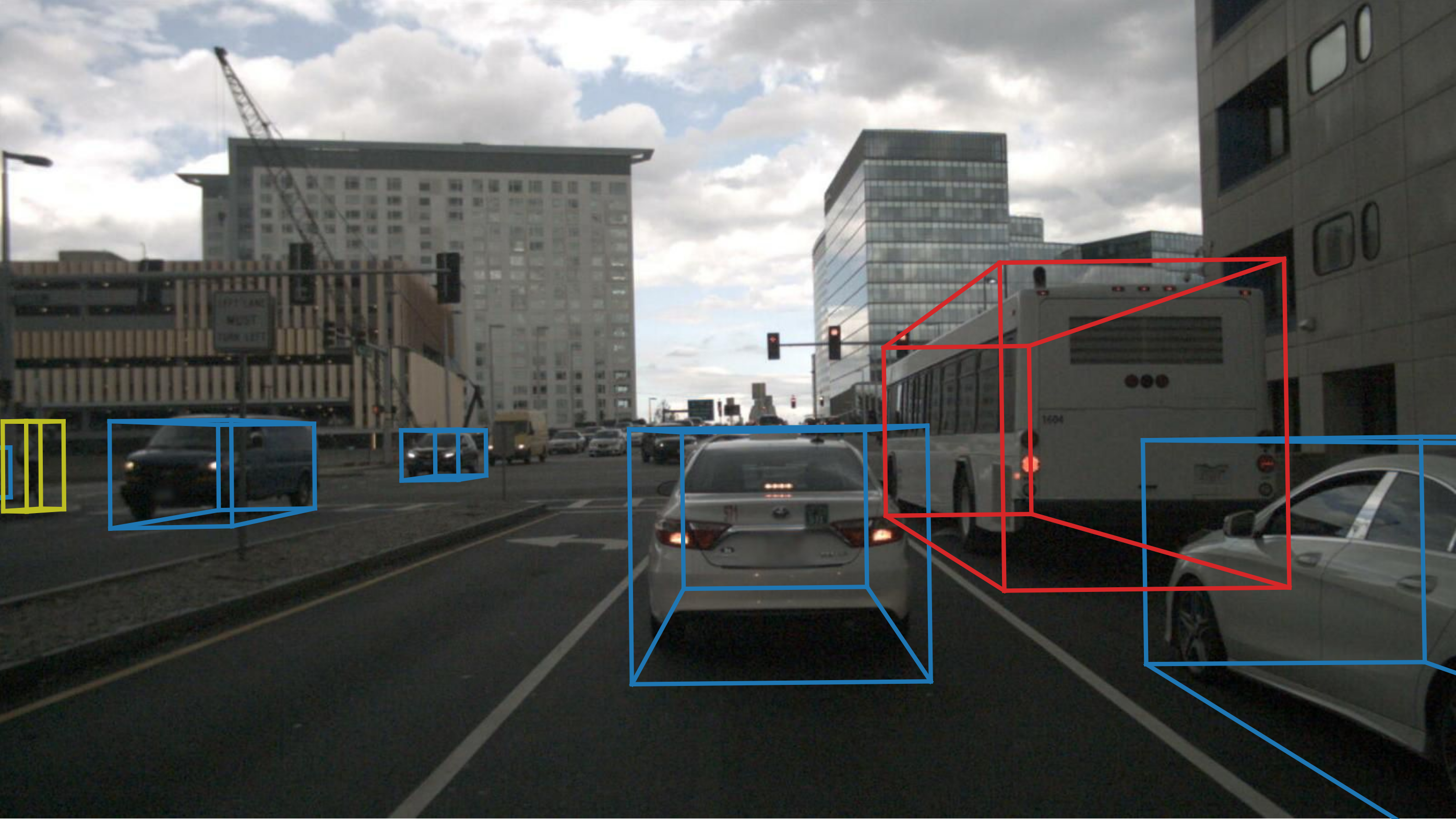}\label{fig:gaussian_NoAttack}}
\subfigure[Kernel=20]{\includegraphics[width=0.23\textwidth]{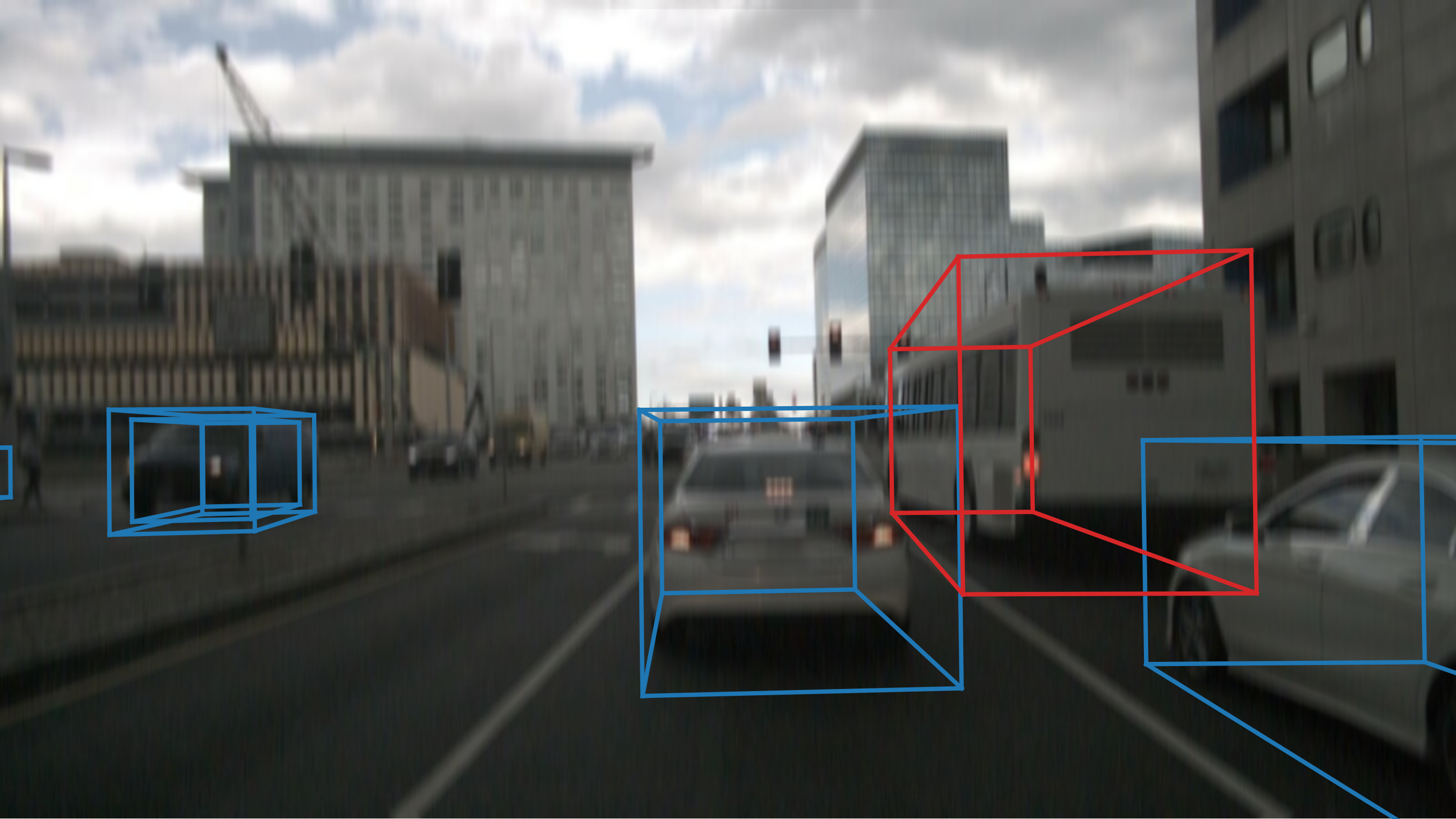}\label{fig:gaussian1}}
\subfigure[Kernel=50]{\includegraphics[width=0.23\textwidth]{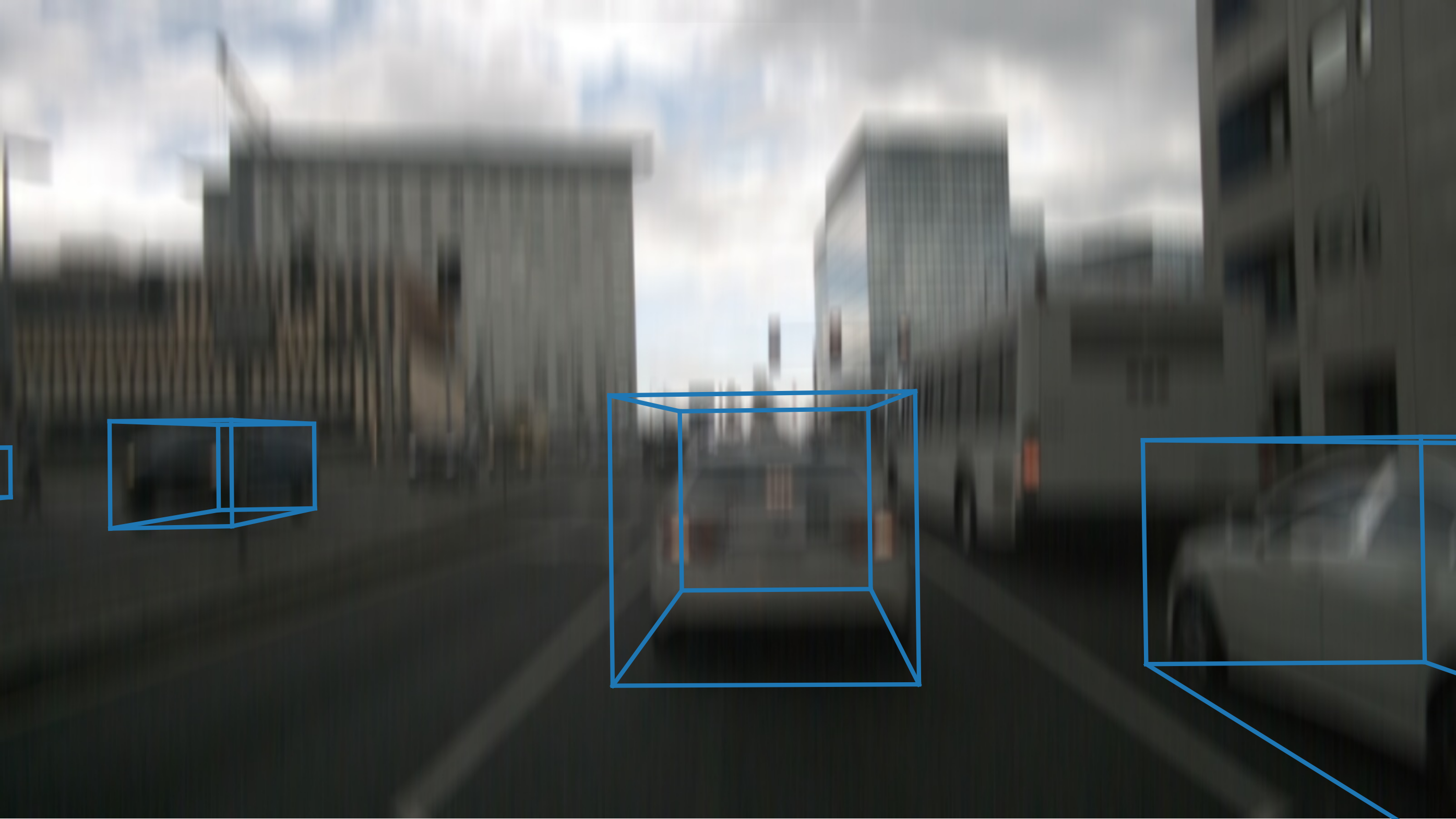}\label{fig:gaussian2}}
\subfigure[Kernel=80]{\includegraphics[width=0.23\textwidth]{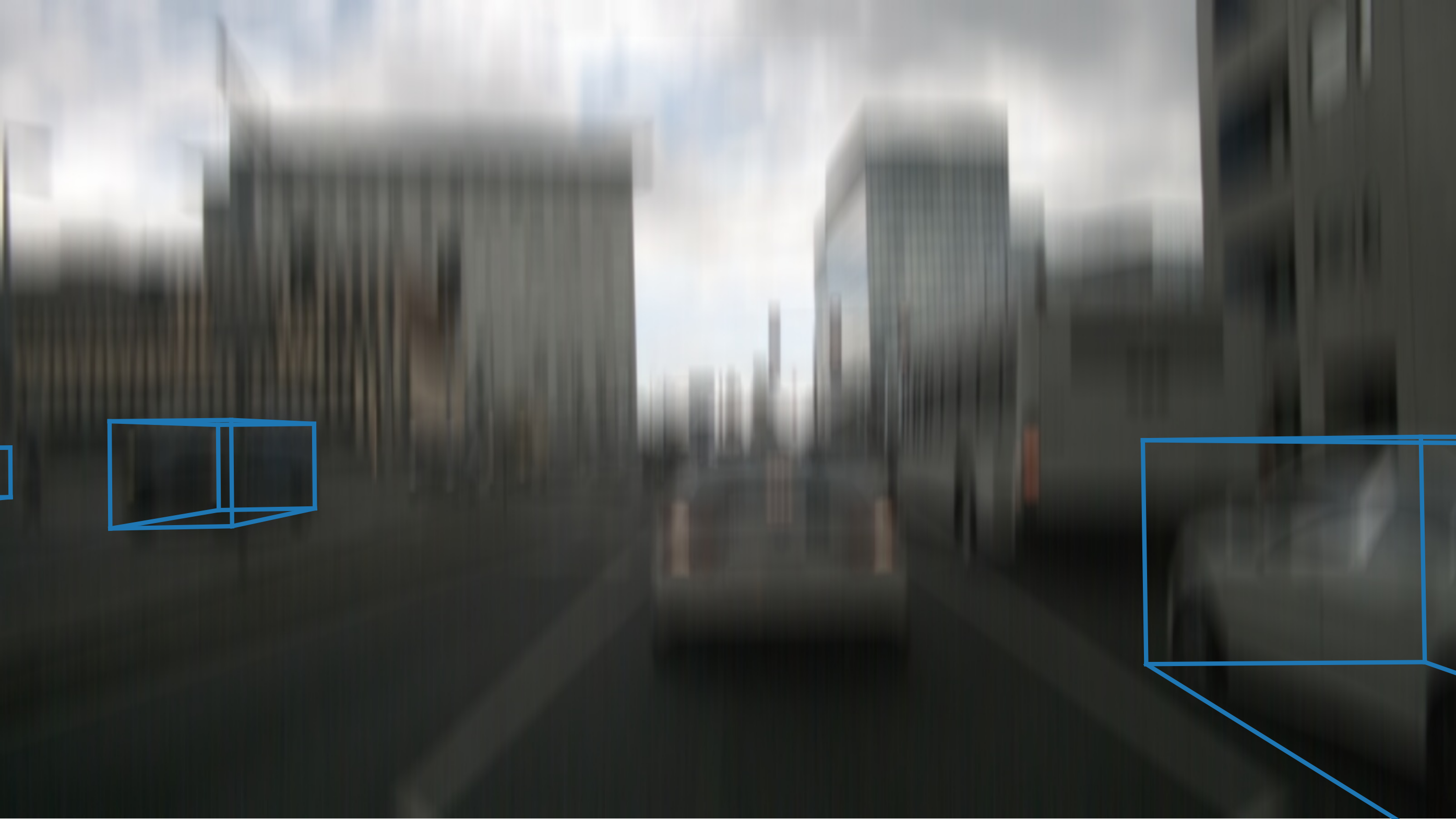}\label{fig:gaussian3}}

\subfigure[No attack]{\includegraphics[width=0.23\textwidth]{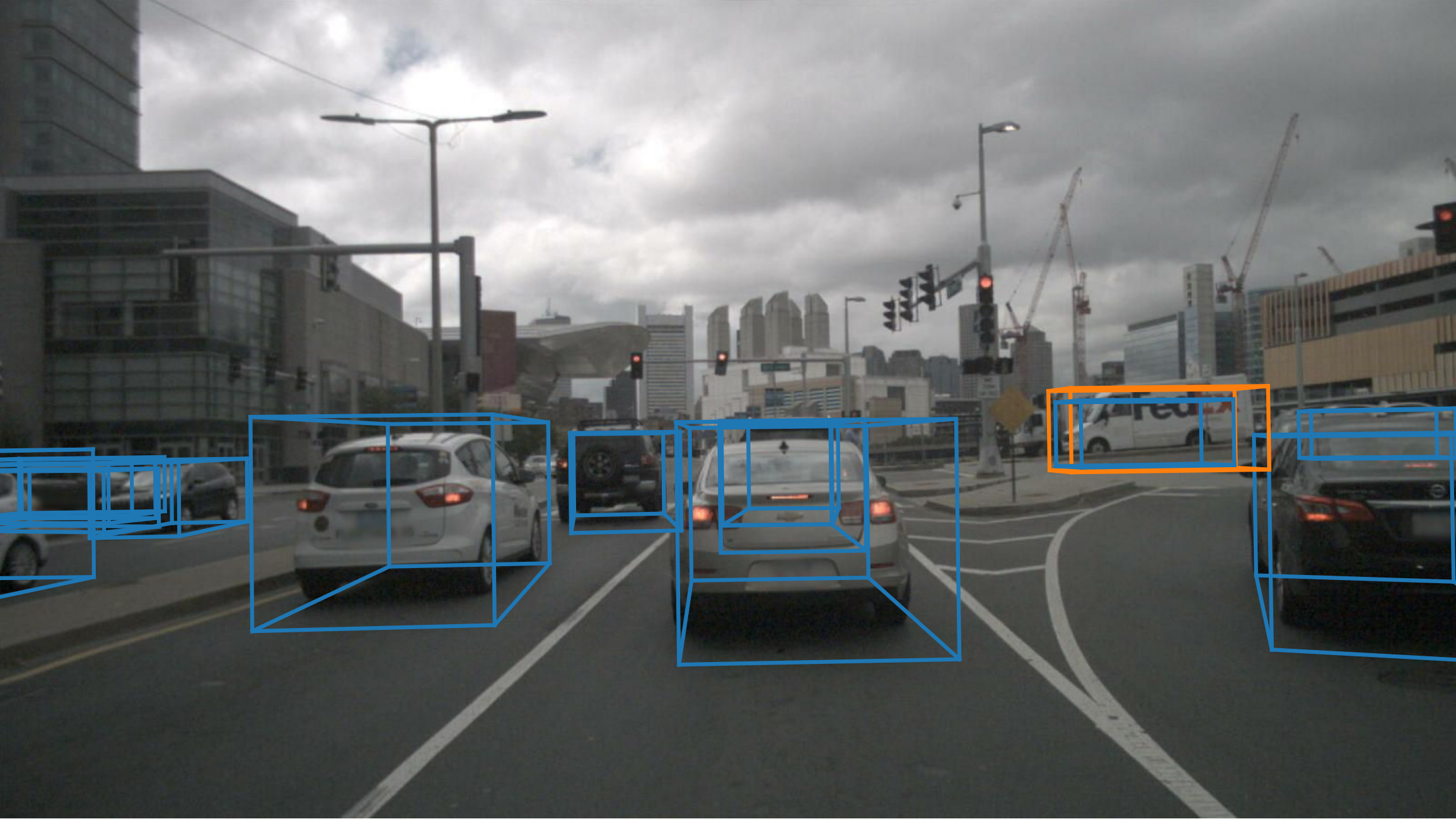}\label{fig:serial_design}}
\subfigure[Kernel=20]{\includegraphics[width=0.23\textwidth]{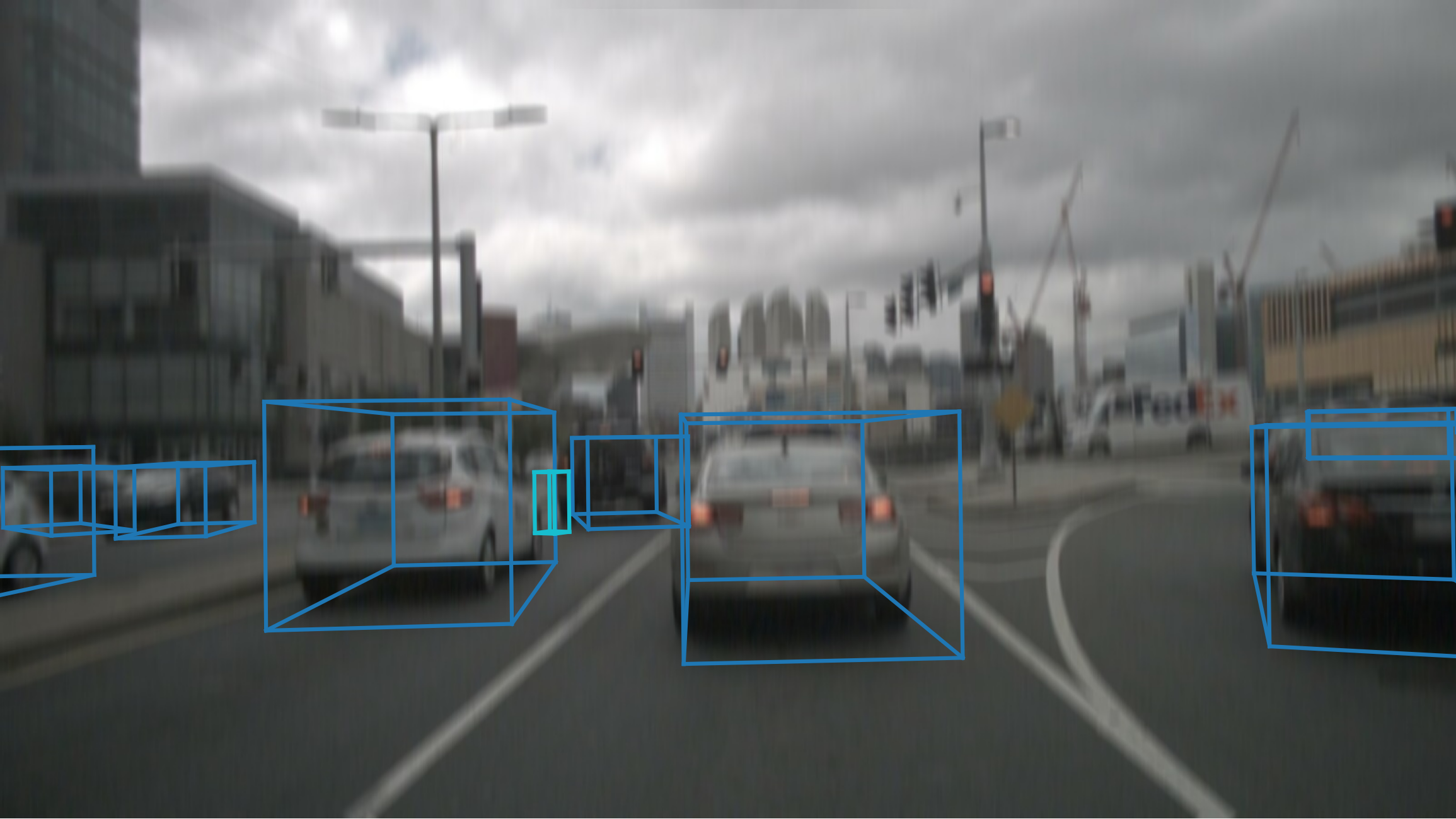}\label{fig:parallel_design}}
\subfigure[Kernel=50]{\includegraphics[width=0.23\textwidth]{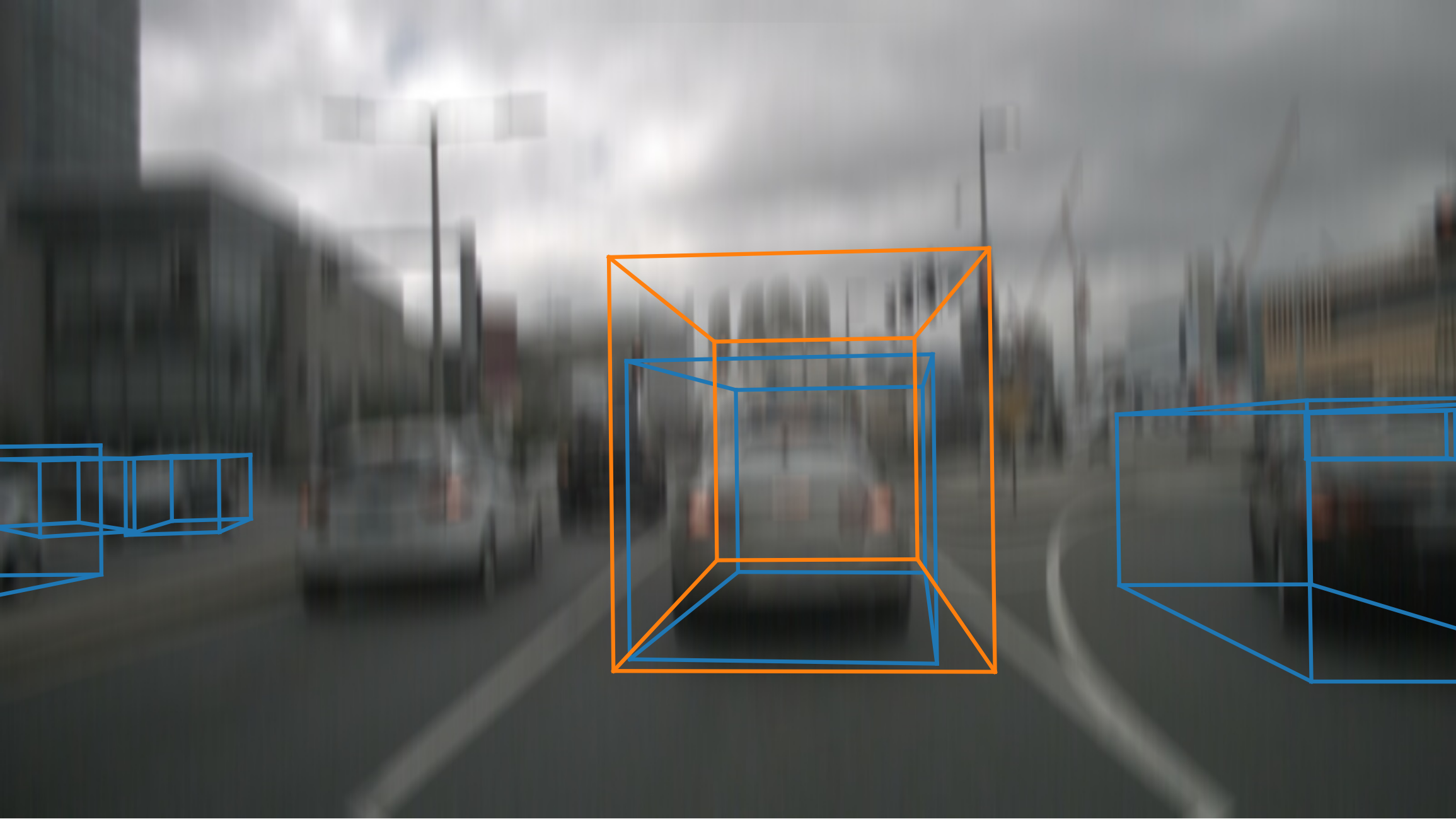}\label{fig:parallel_design}}
\subfigure[Kernel=80]{\includegraphics[width=0.23\textwidth]{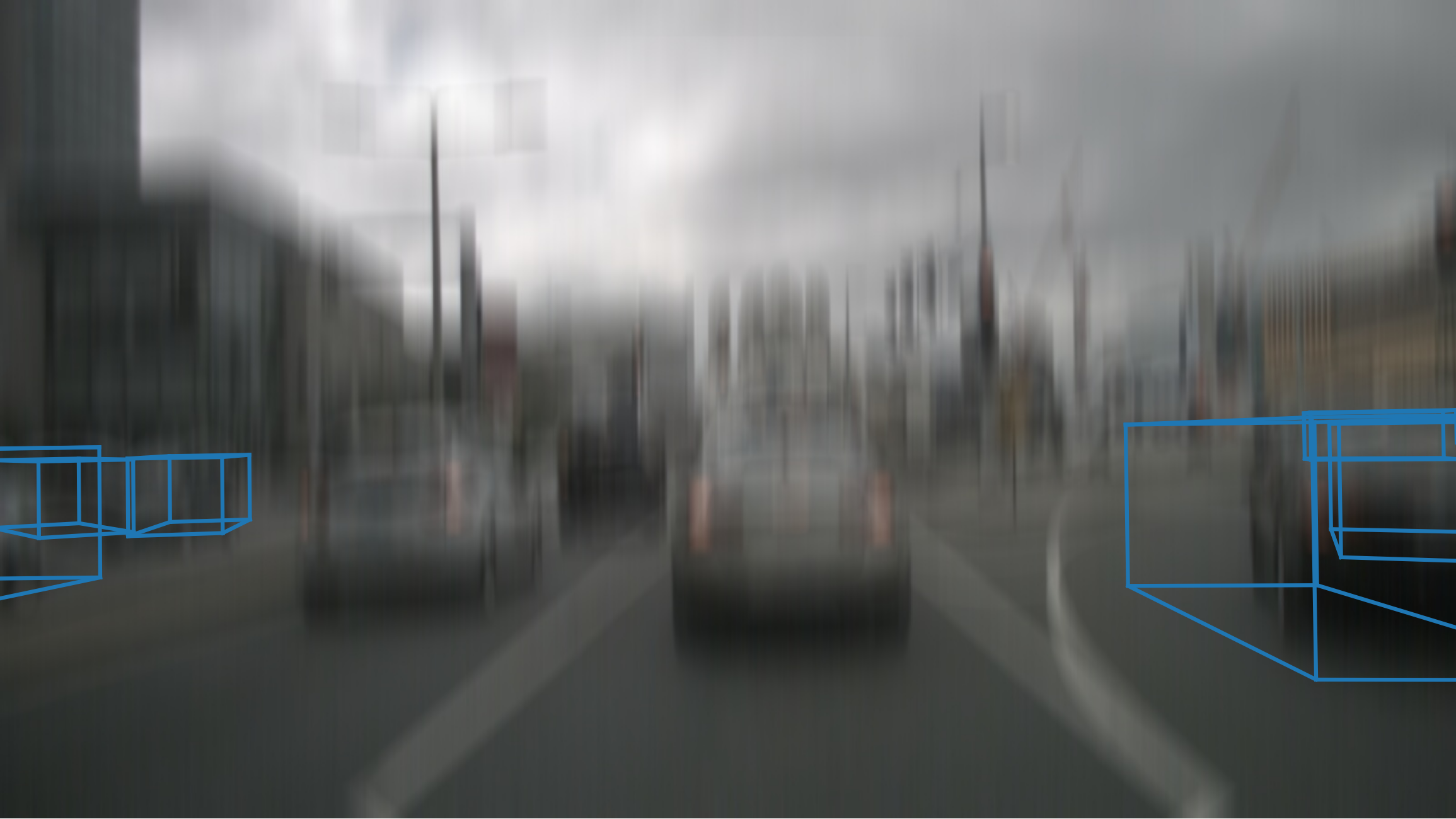}\label{fig:parallel_design}}

\subfigure[No attack]{\includegraphics[width=0.23\textwidth,trim=5 20 80 238, clip]{figure/3_rc-pda/NoAttack.png}\label{fig:serial_design}}
\subfigure[Kernel=20]{\includegraphics[width=0.23\textwidth,trim=5 20 80 238, clip]{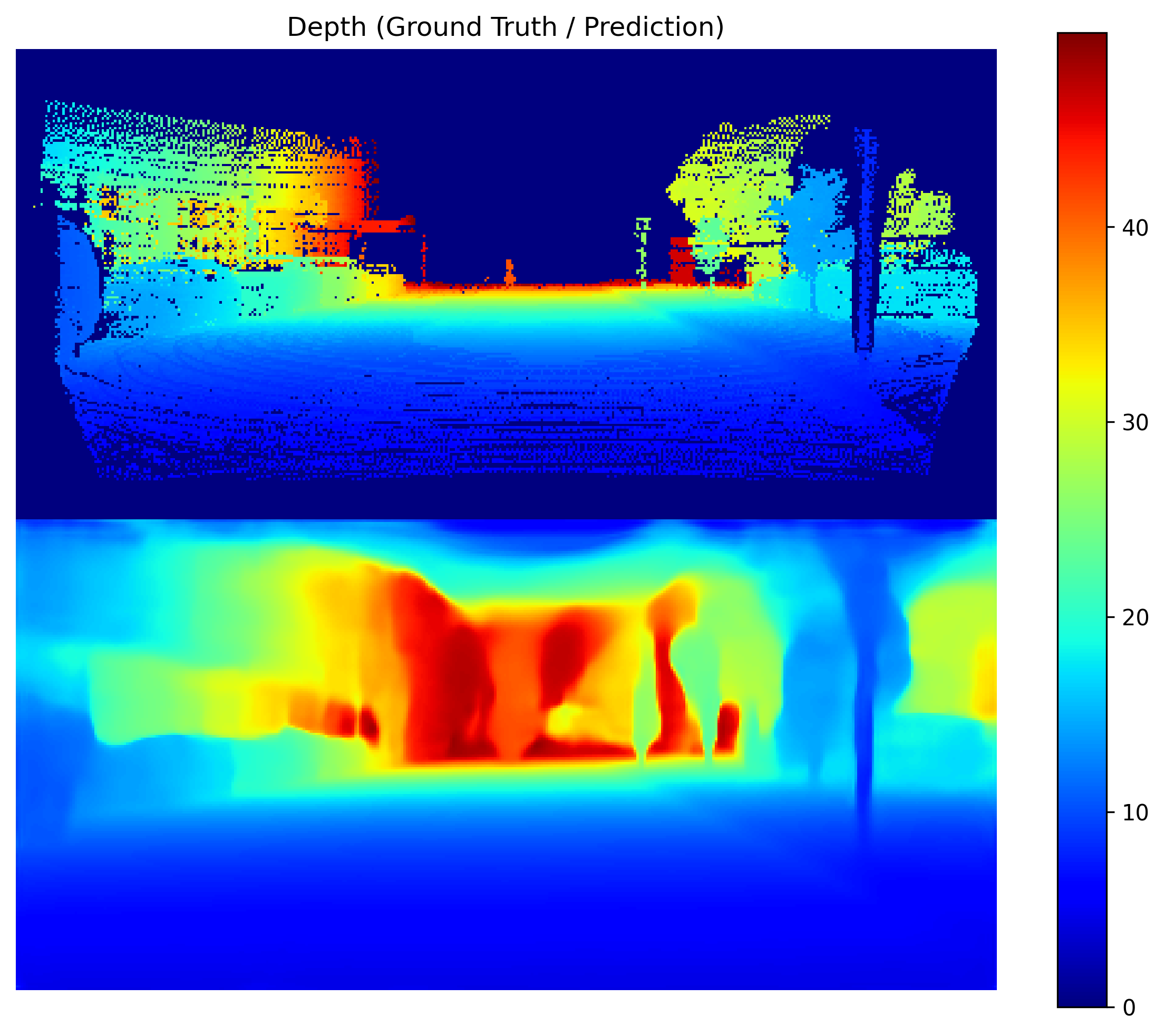}\label{fig:parallel_design}}
\subfigure[Kernel=50]{\includegraphics[width=0.23\textwidth,trim=5 20 80 238, clip]{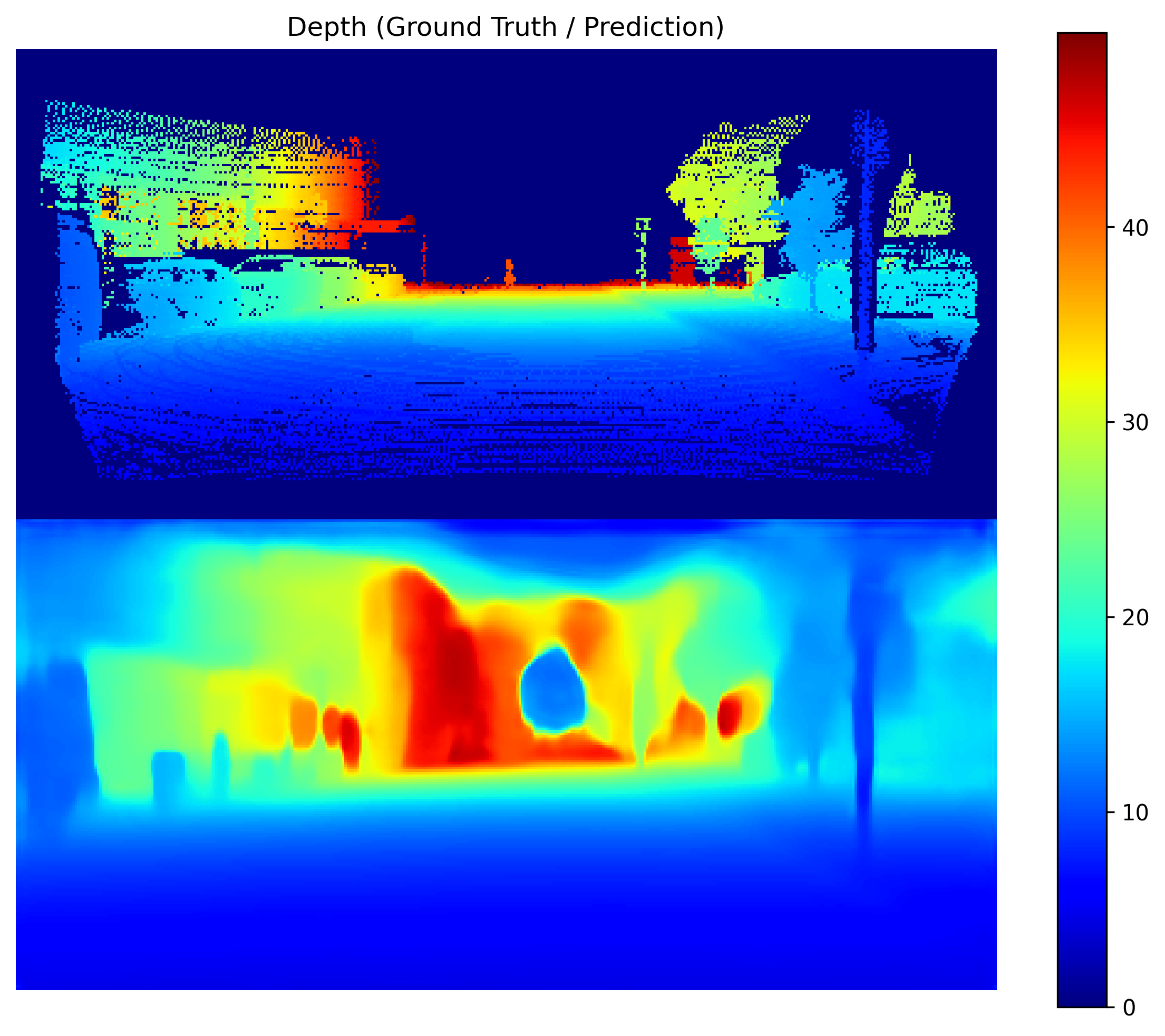}\label{fig:parallel_design}}
\subfigure[Kernel=80]{\includegraphics[width=0.23\textwidth,trim=5 20 80 238, clip]{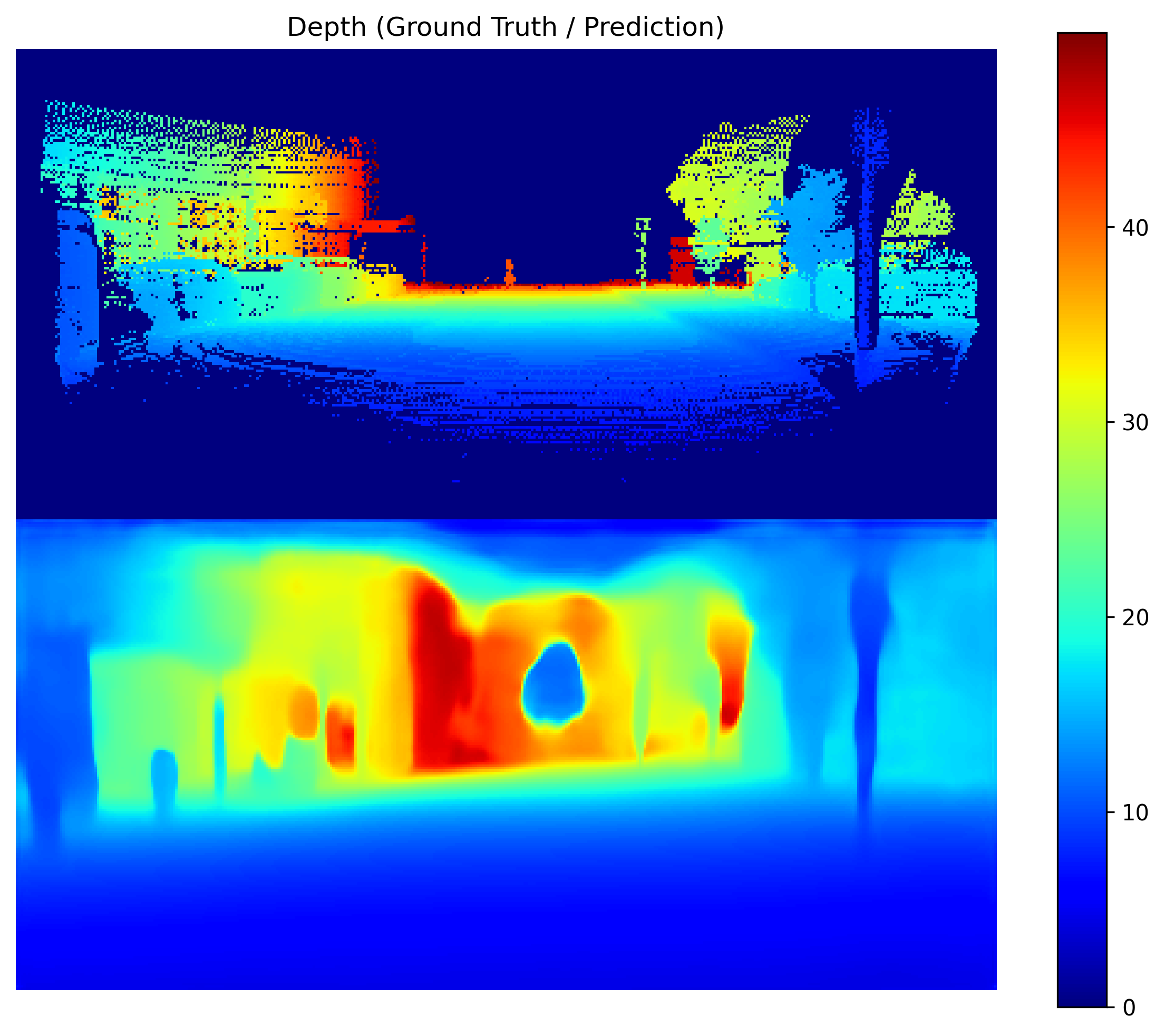}\label{fig:parallel_design}}

\subfigure[No attack]{\includegraphics[width=0.23\textwidth]{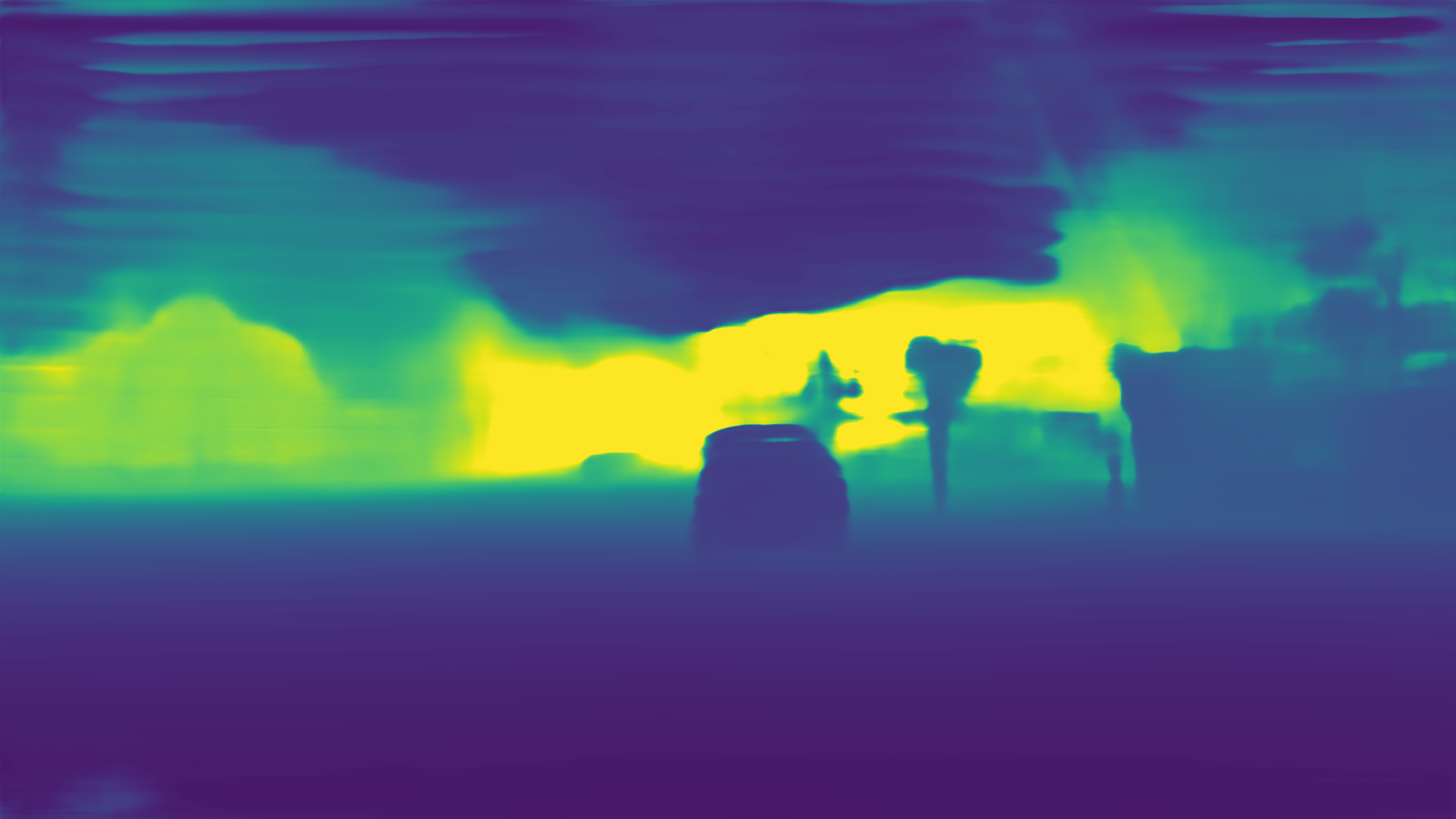}\label{fig:serial_design}}
\subfigure[Kernel=20]{\includegraphics[width=0.23\textwidth]{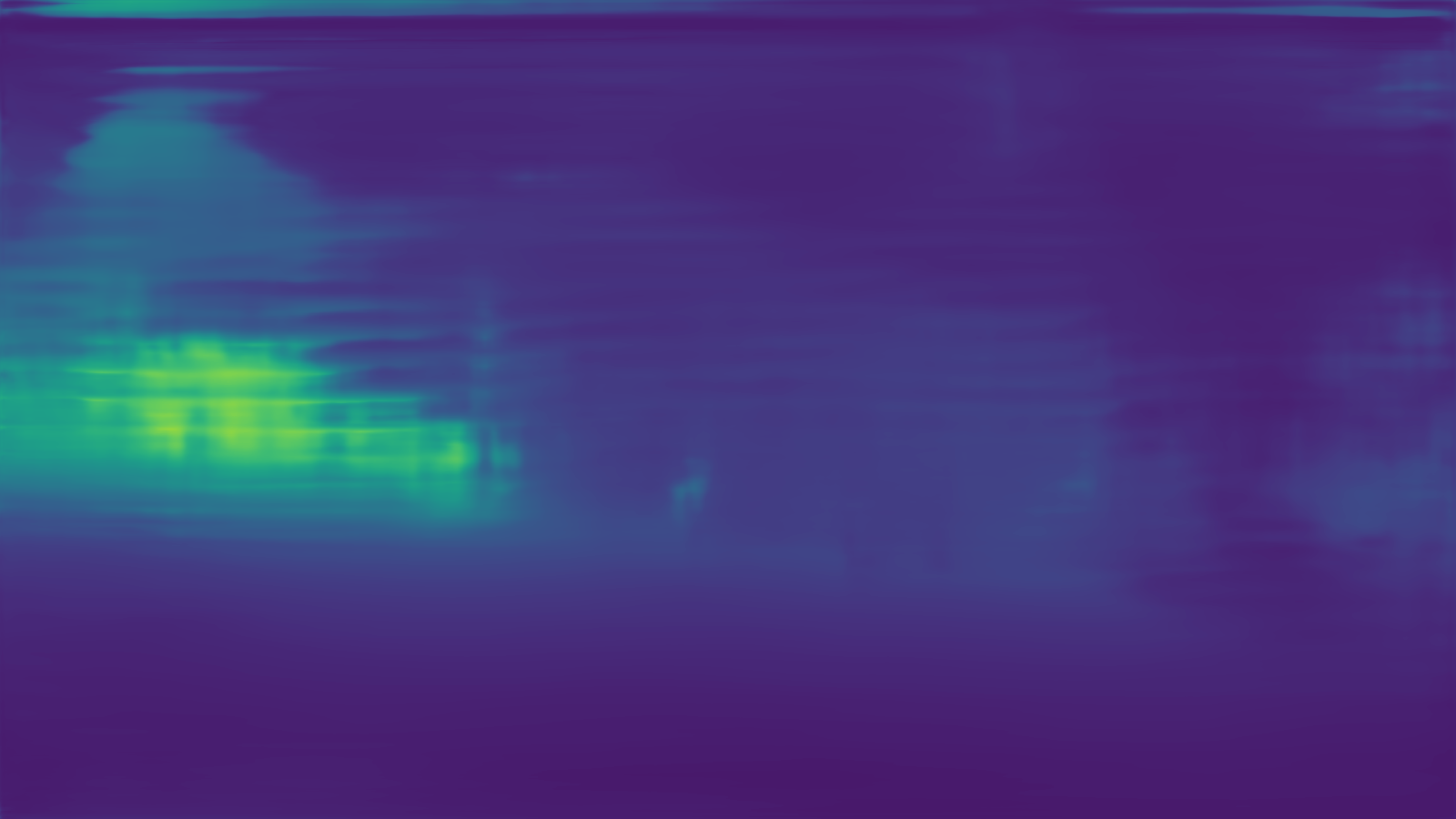}\label{fig:parallel_design}}
\subfigure[Kernel=50]{\includegraphics[width=0.23\textwidth]{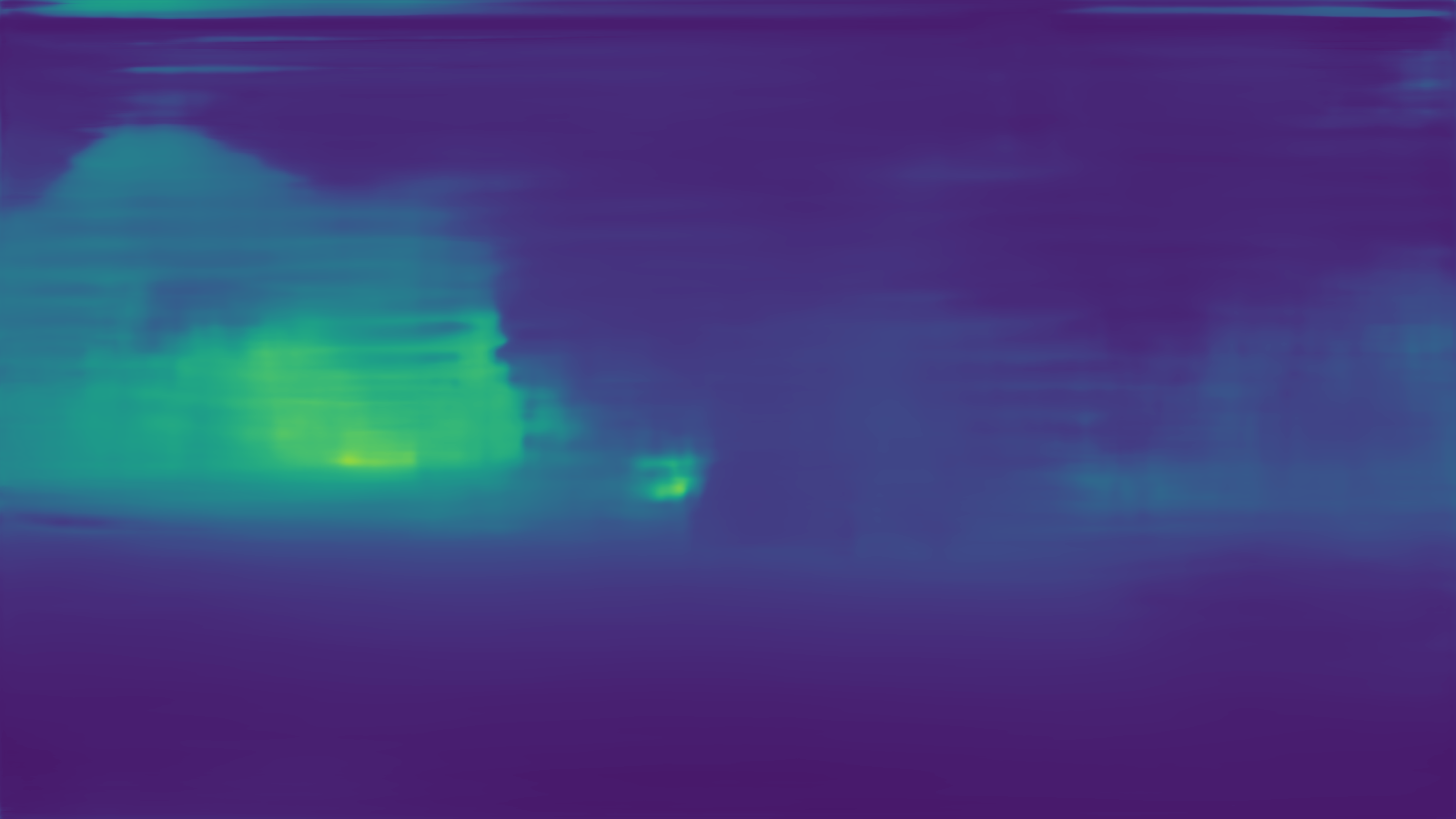}\label{fig:parallel_design}}
\subfigure[Kernel=80]{\includegraphics[width=0.23\textwidth]{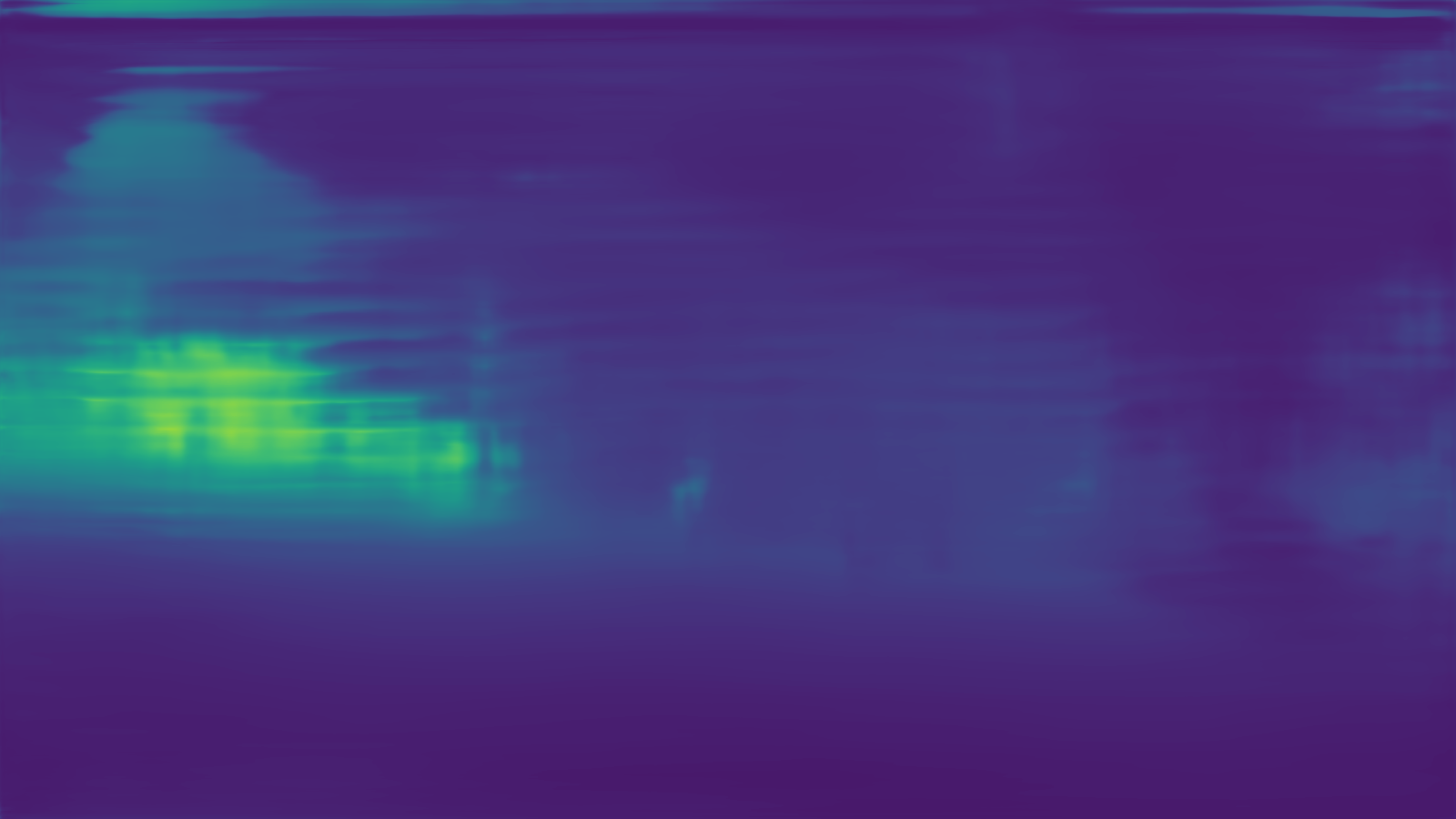}\label{fig:parallel_design}}

\subfigure[No attack]{\includegraphics[width=0.23\textwidth]{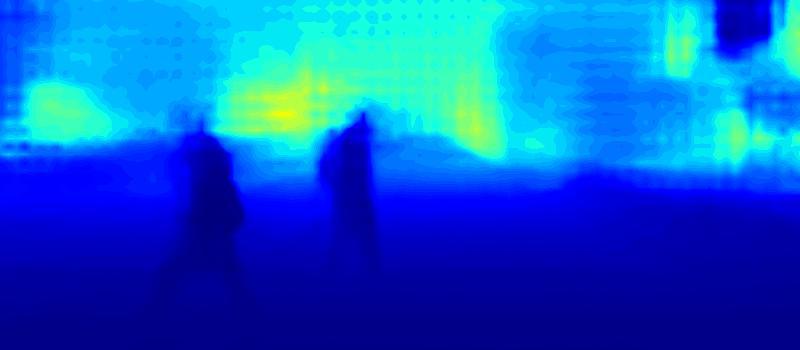}\label{fig:serial_design}}
\subfigure[Kernel=20]{\includegraphics[width=0.23\textwidth]{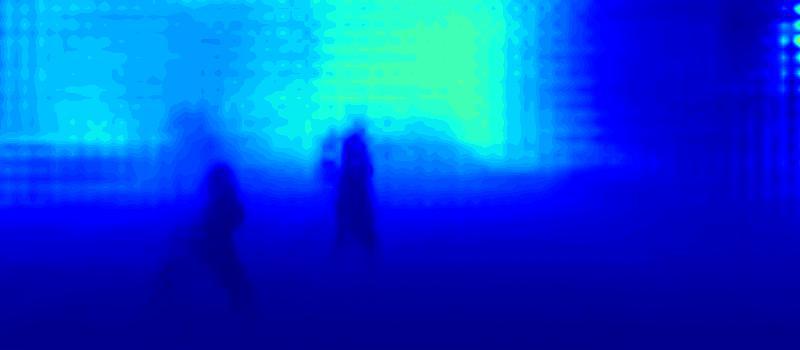}\label{fig:parallel_design}}
\subfigure[Kernel=50]{\includegraphics[width=0.23\textwidth]{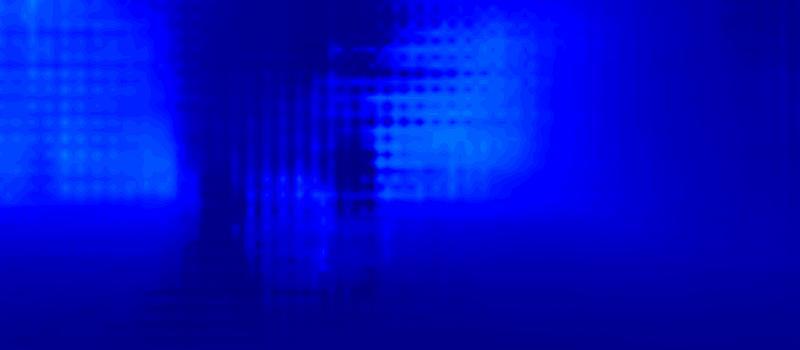}\label{fig:parallel_design}}
\subfigure[Kernel=80]{\includegraphics[width=0.23\textwidth]{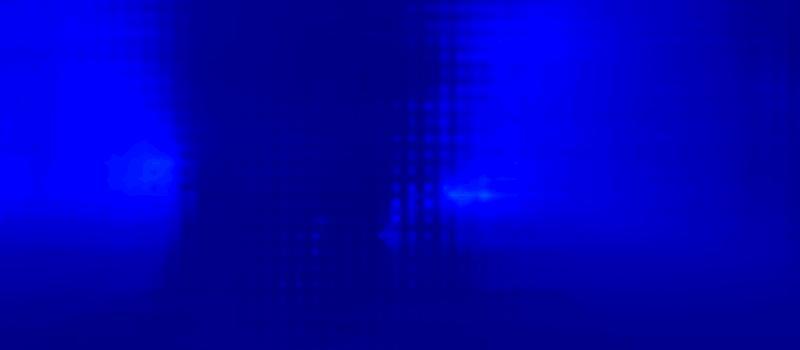}\label{fig:parallel_design}}

\caption{Examples of Vertical Blur attack on various object detection and depth estimation sensor fusion algorithms. The first to the last rows correspond to the object detection algorithms,  RADIANT~\cite{long2023radiant} and CRN~\cite{kim2023crn}, and the depth estimation algorithms, RC-PDA~\cite{long2021radar}, Cam\&mmWave~\cite{singh2023depth}, and DORN~\cite{lo2021depth}.}
\label{fig:Vertical_attack}
\end{figure*}

\begin{figure*}[t]
\centering
\subfigure[No attack]{\includegraphics[width=0.23\textwidth]{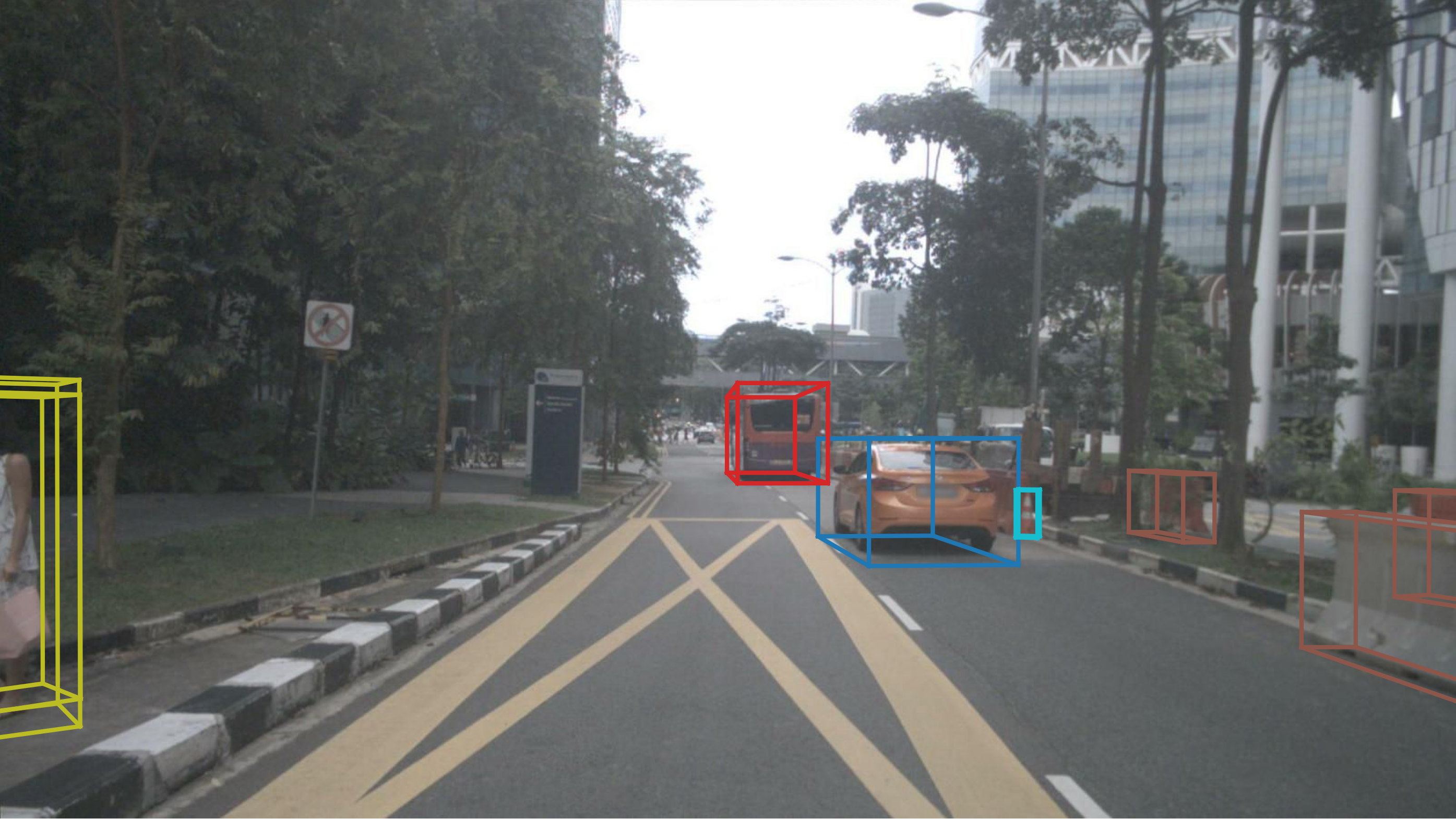}\label{fig:gaussian_NoAttack}}
\subfigure[Focal Length=5]{\includegraphics[width=0.23\textwidth]{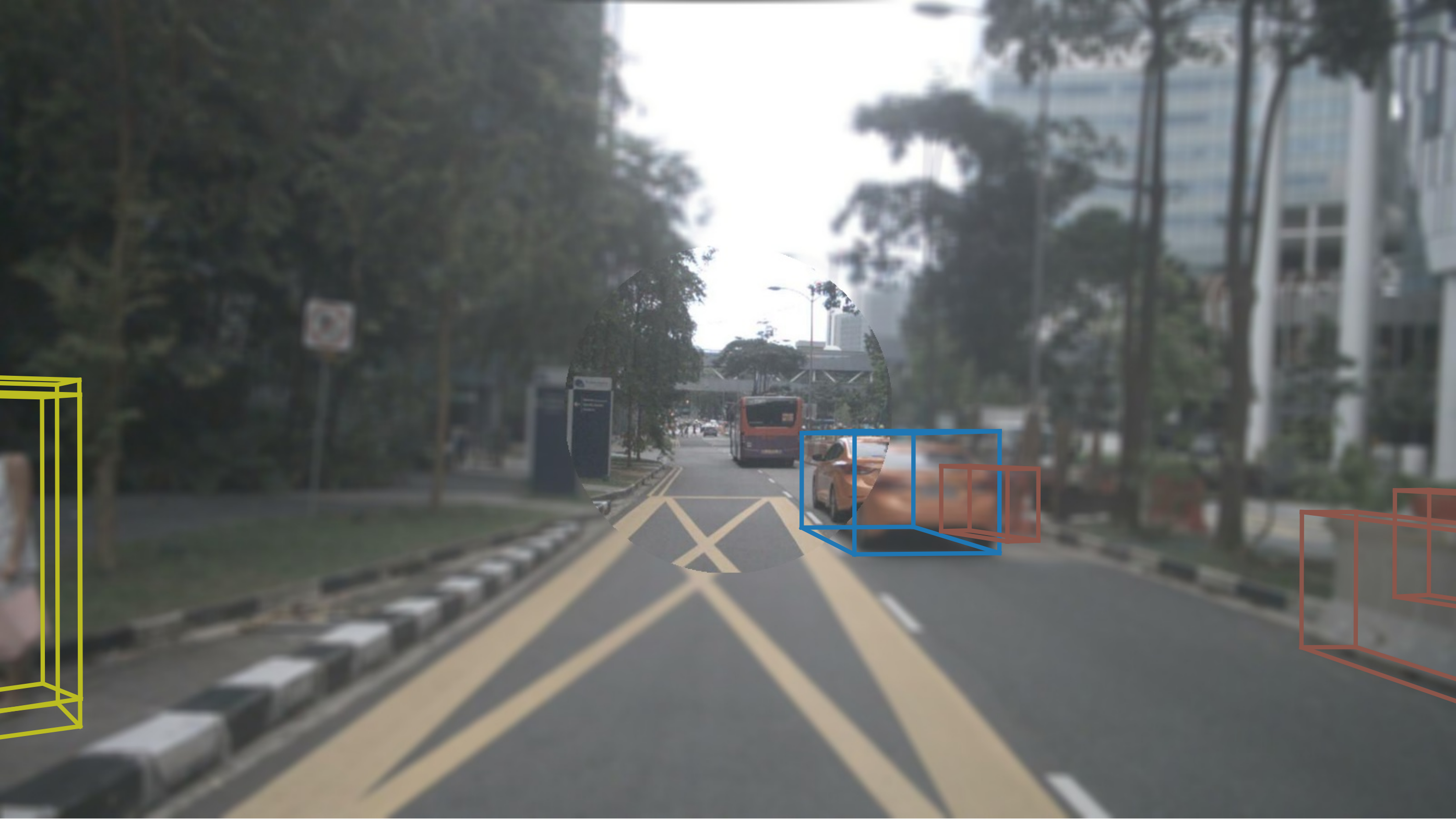}\label{fig:gaussian1}}
\subfigure[Focal Length=10]{\includegraphics[width=0.23\textwidth]{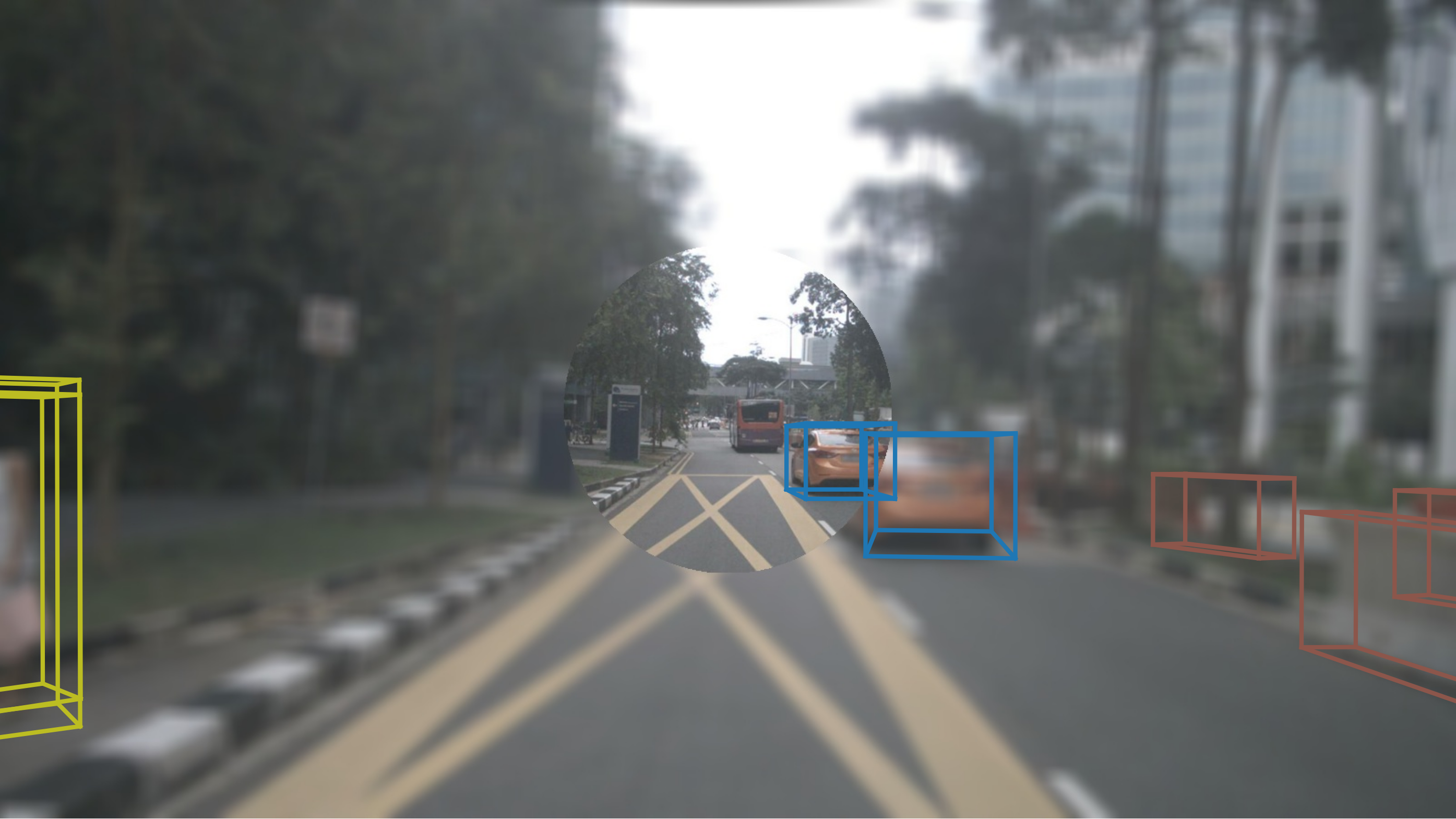}\label{fig:gaussian2}}
\subfigure[Focal Length=15]{\includegraphics[width=0.23\textwidth]{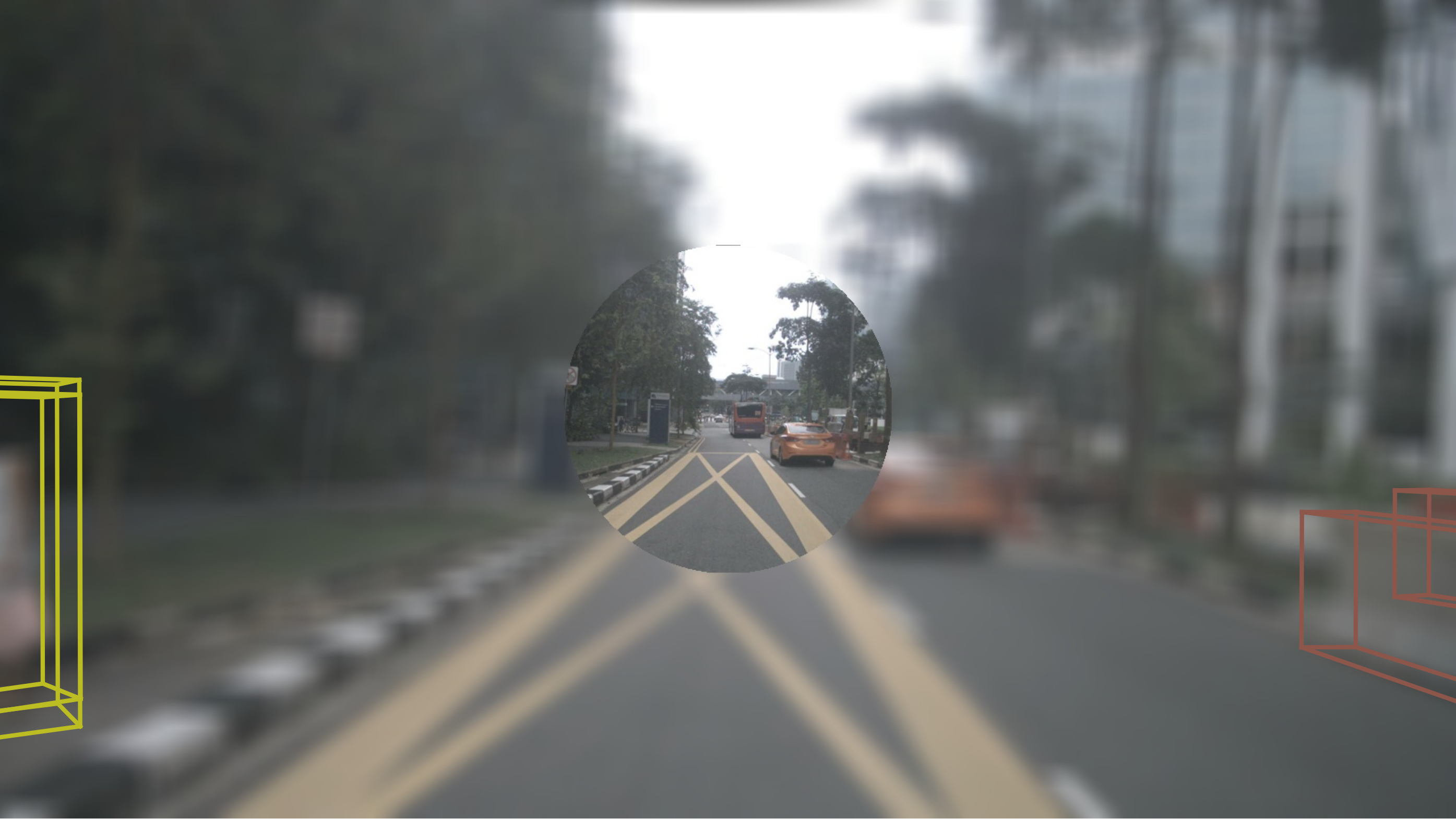}\label{fig:gaussian3}}

\subfigure[No attack]{\includegraphics[width=0.23\textwidth]{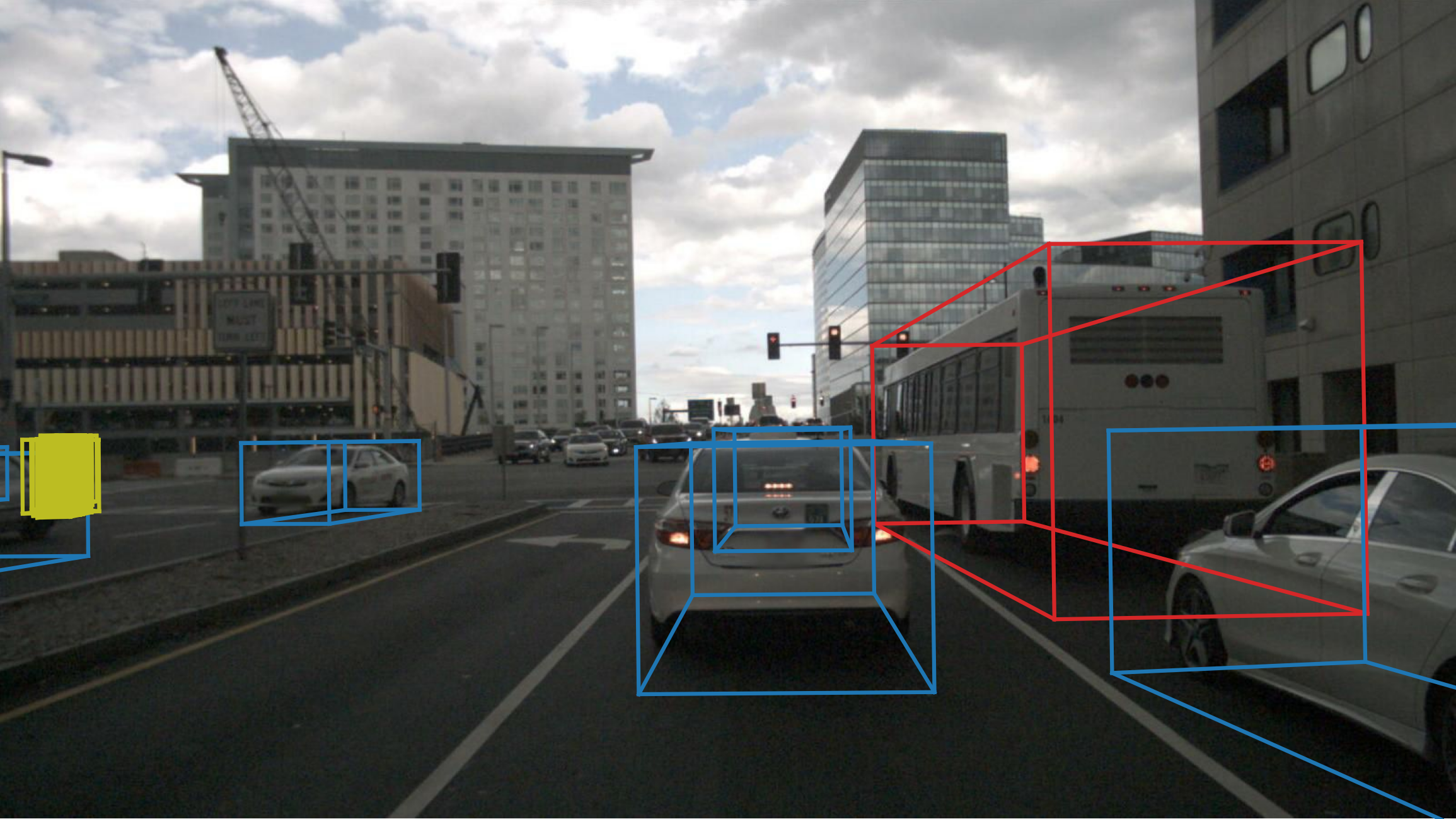}\label{fig:lens_NoAttack}}
\subfigure[Focal Length=5]{\includegraphics[width=0.23\textwidth]{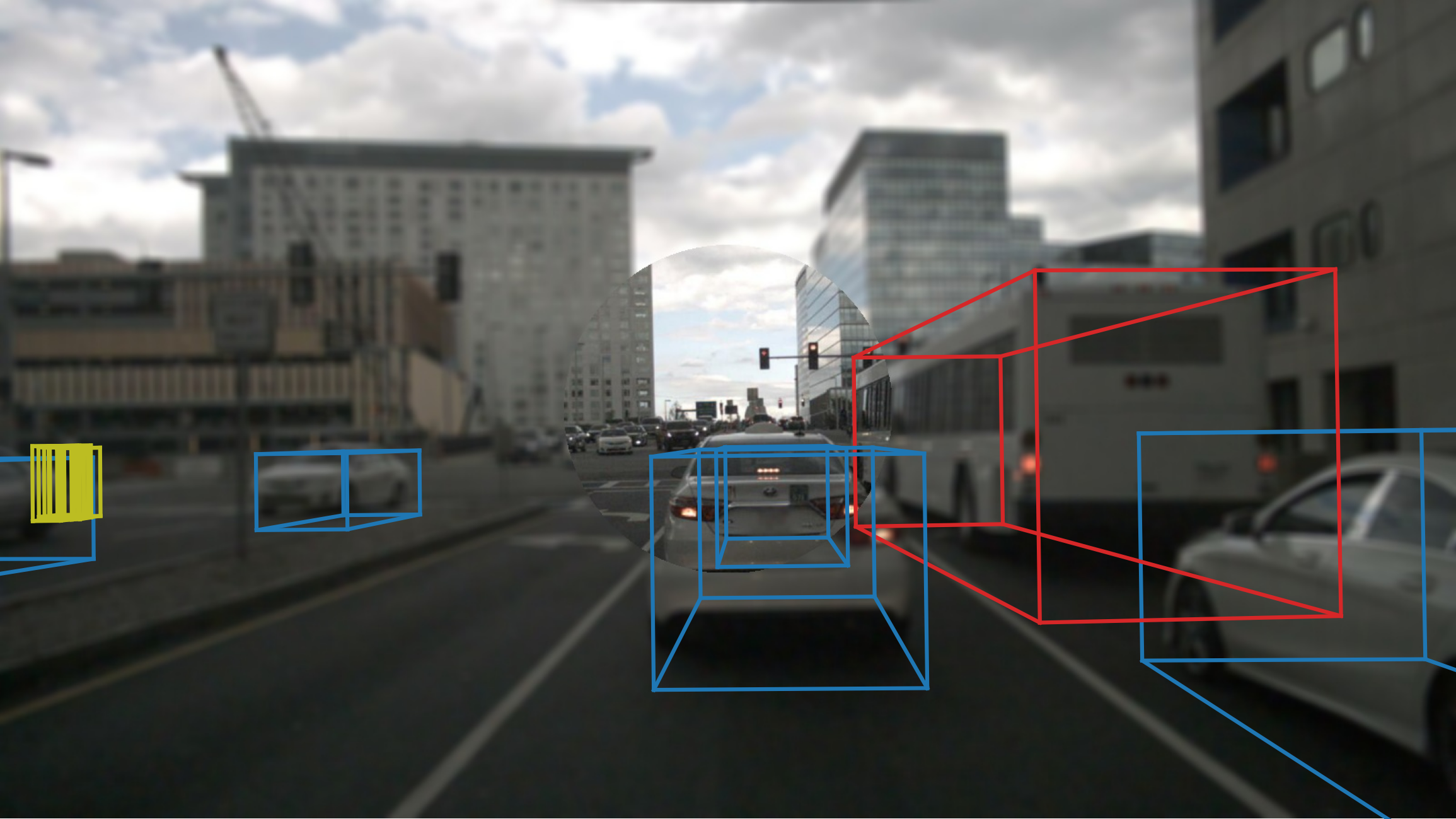}\label{fig:lens1}}
\subfigure[Focal Length=10]{\includegraphics[width=0.23\textwidth]{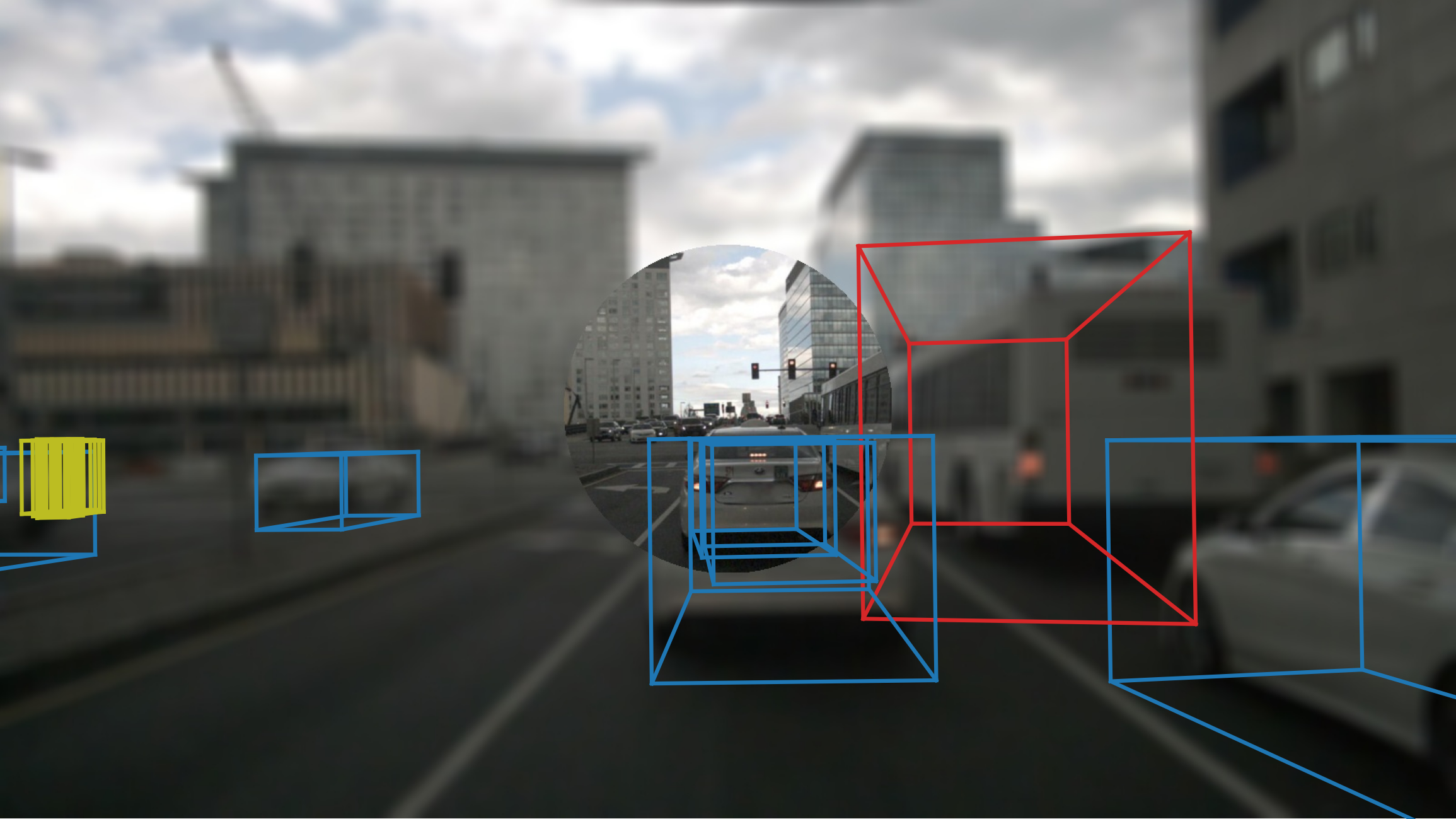}\label{fig:lens2}}
\subfigure[Focal Length=15]{\includegraphics[width=0.23\textwidth]{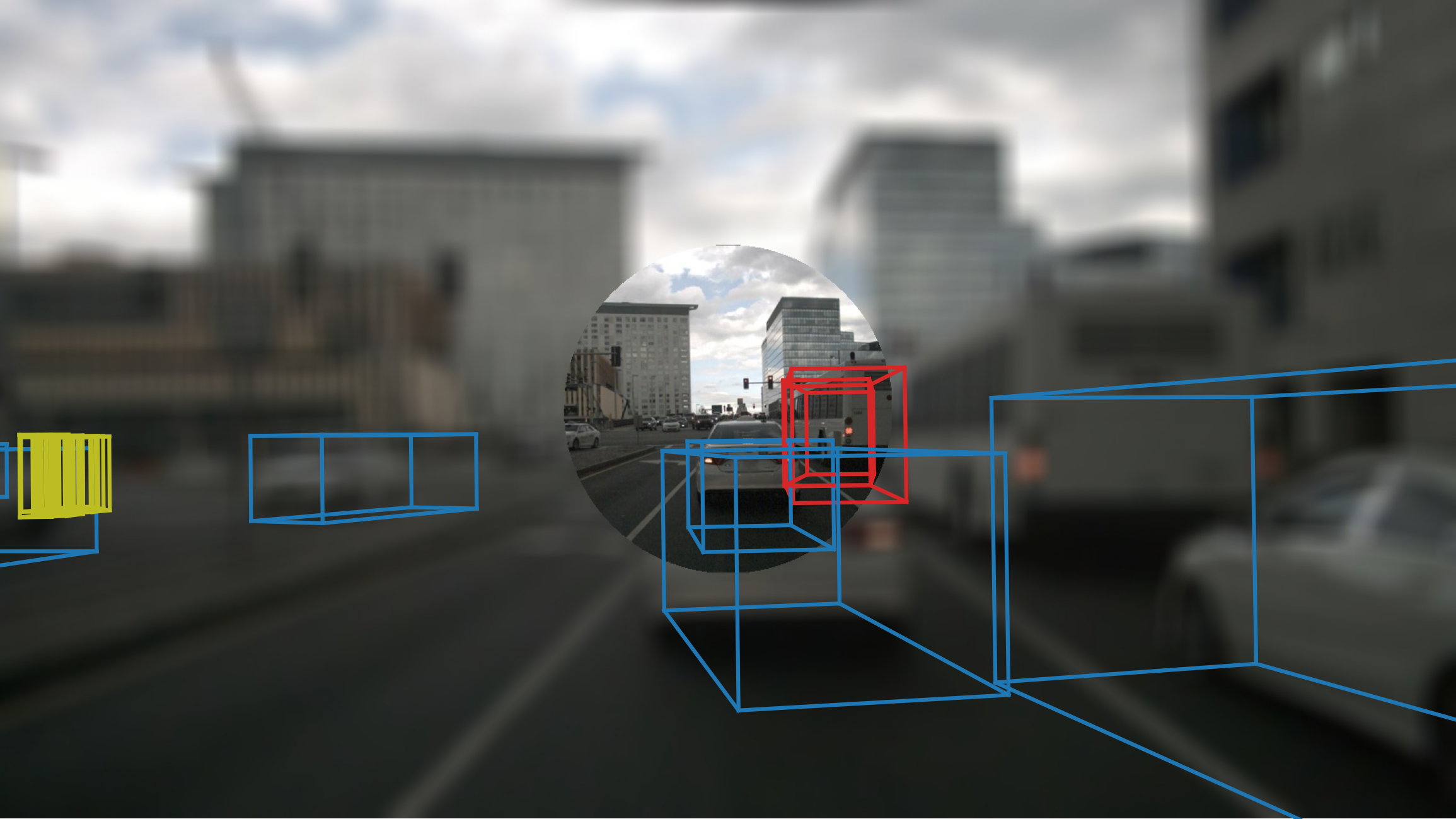}\label{fig:lens3}}

\subfigure[No attack]{\includegraphics[width=0.23\textwidth,trim=5 20 80 238, clip]{figure/3_rc-pda/NoAttack.png}\label{fig:serial_design}}
\subfigure[Focal Length=5]{\includegraphics[width=0.23\textwidth,trim=5 20 80 238, clip]{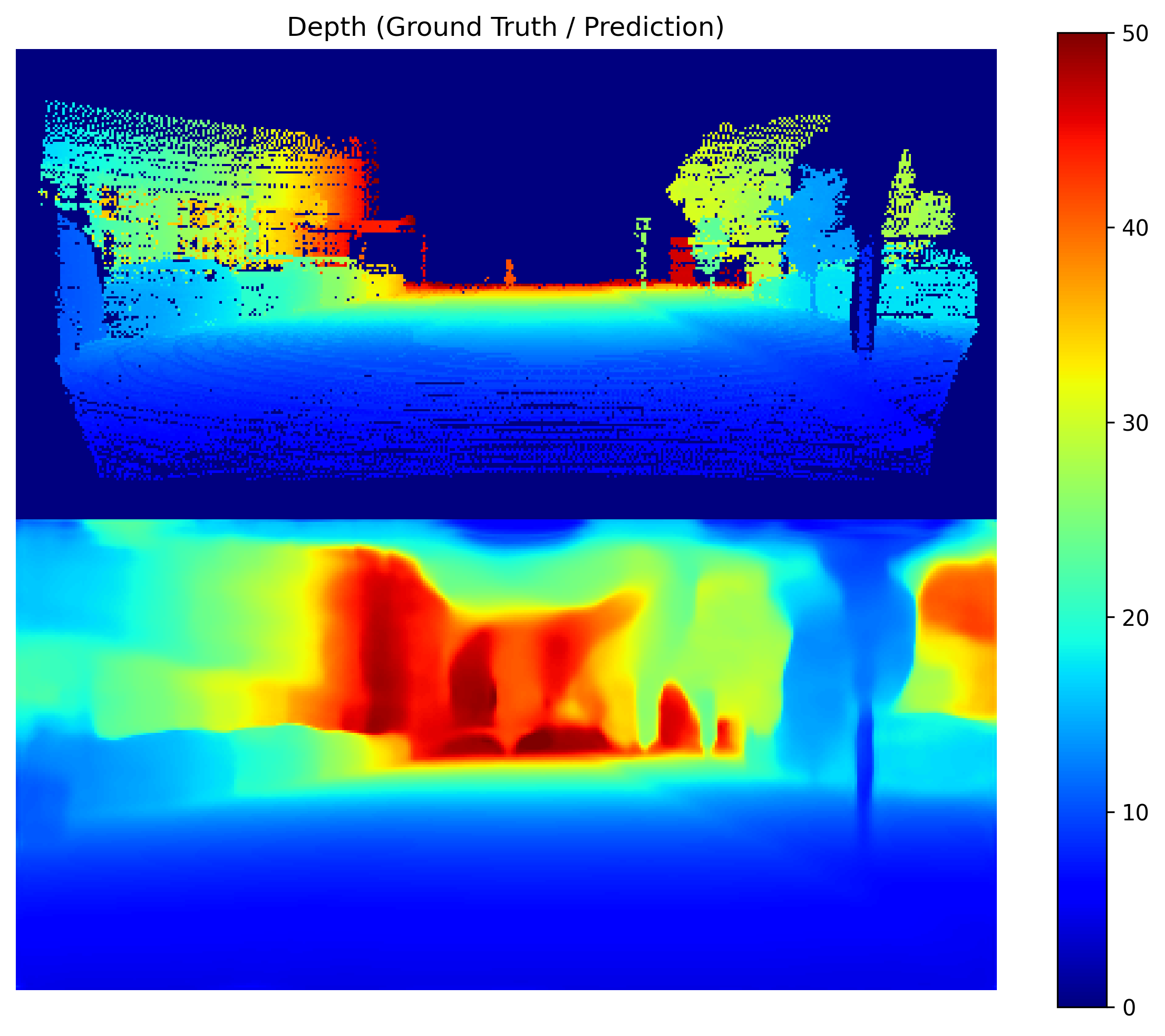}\label{fig:parallel_design}}
\subfigure[Focal Length=10]{\includegraphics[width=0.23\textwidth,trim=5 20 80 238, clip]{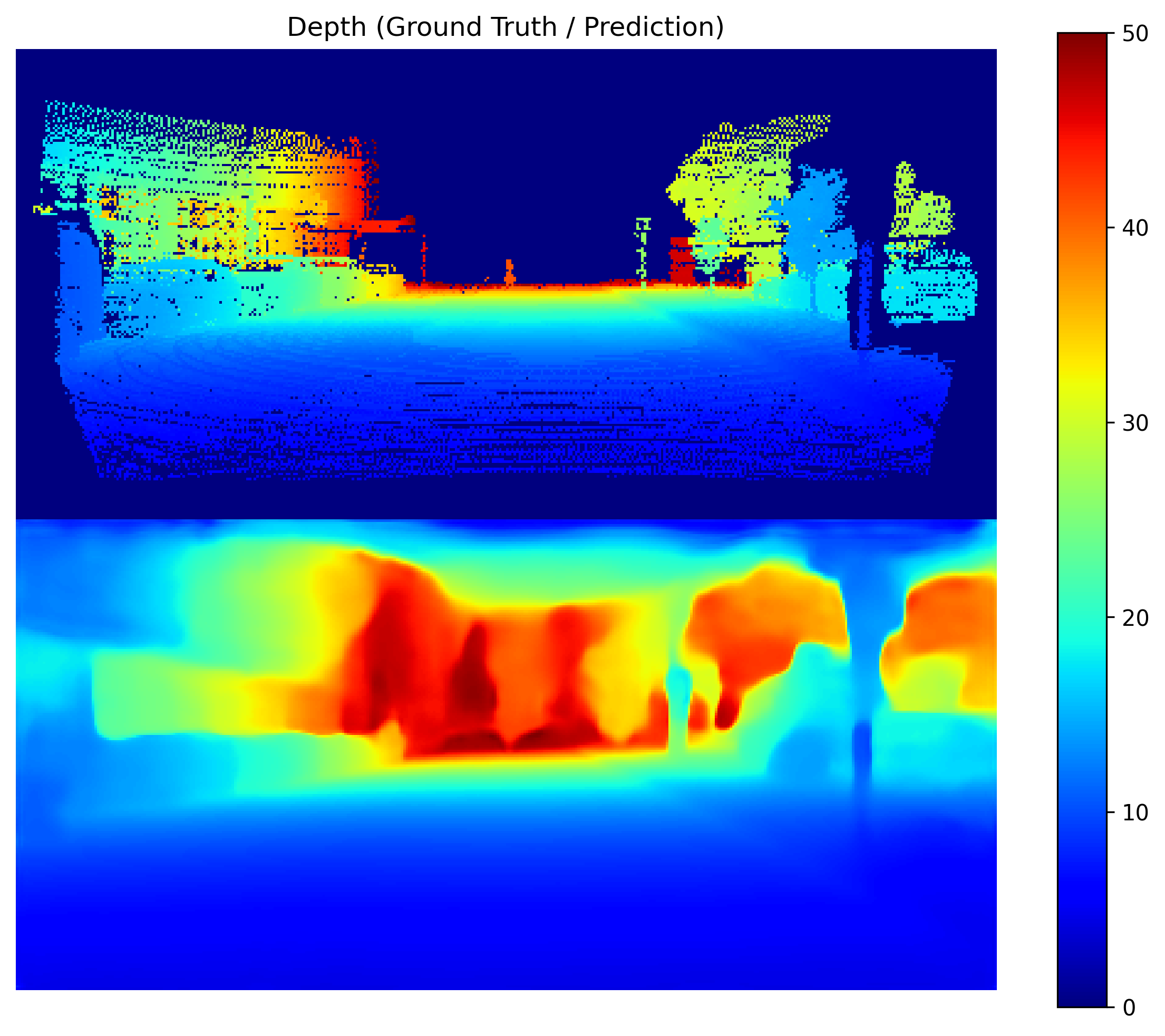}\label{fig:parallel_design}}
\subfigure[Focal Length=15]{\includegraphics[width=0.23\textwidth,trim=5 20 80 238, clip]{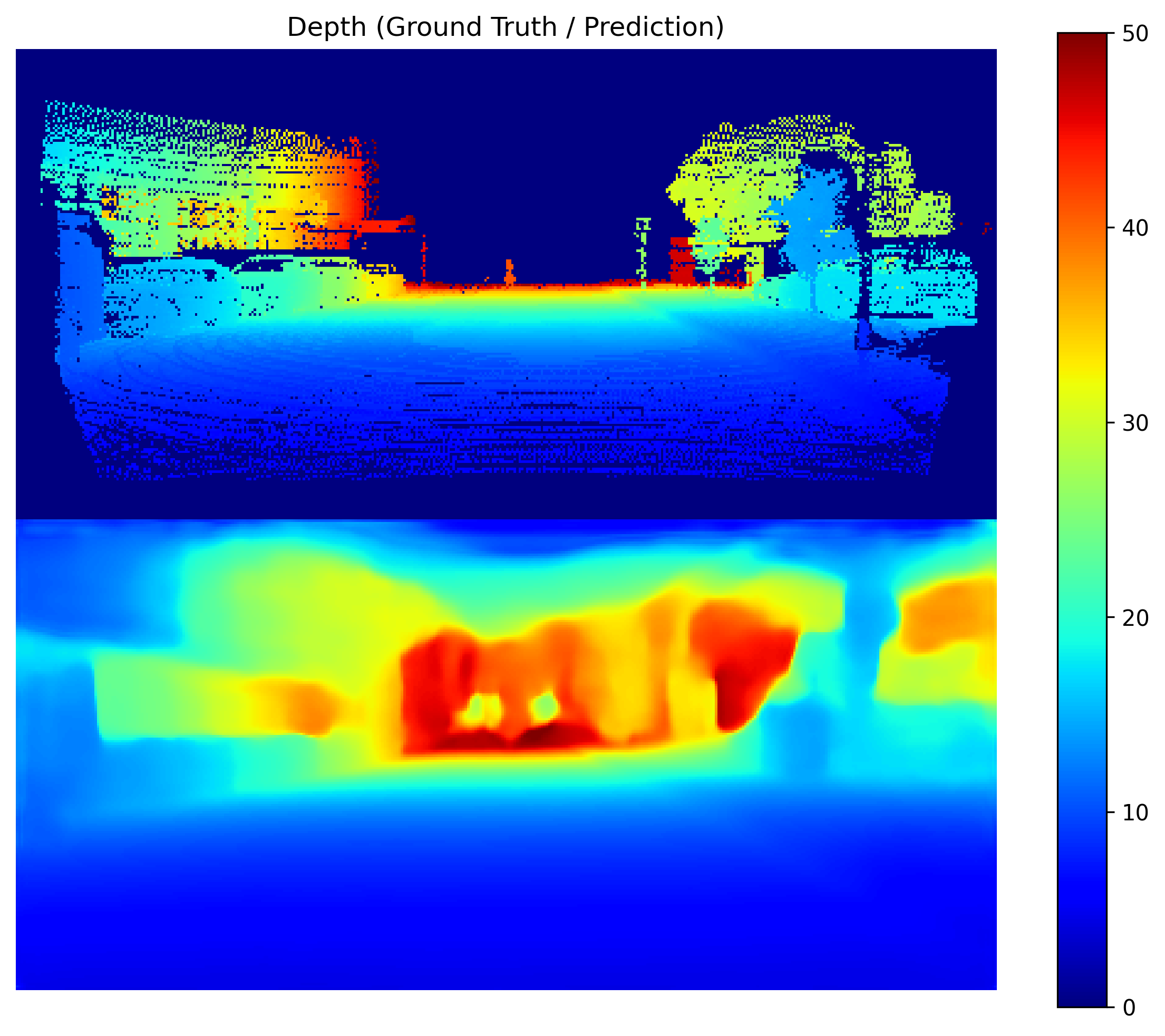}\label{fig:parallel_design}}

\subfigure[No attack]{\includegraphics[width=0.23\textwidth]{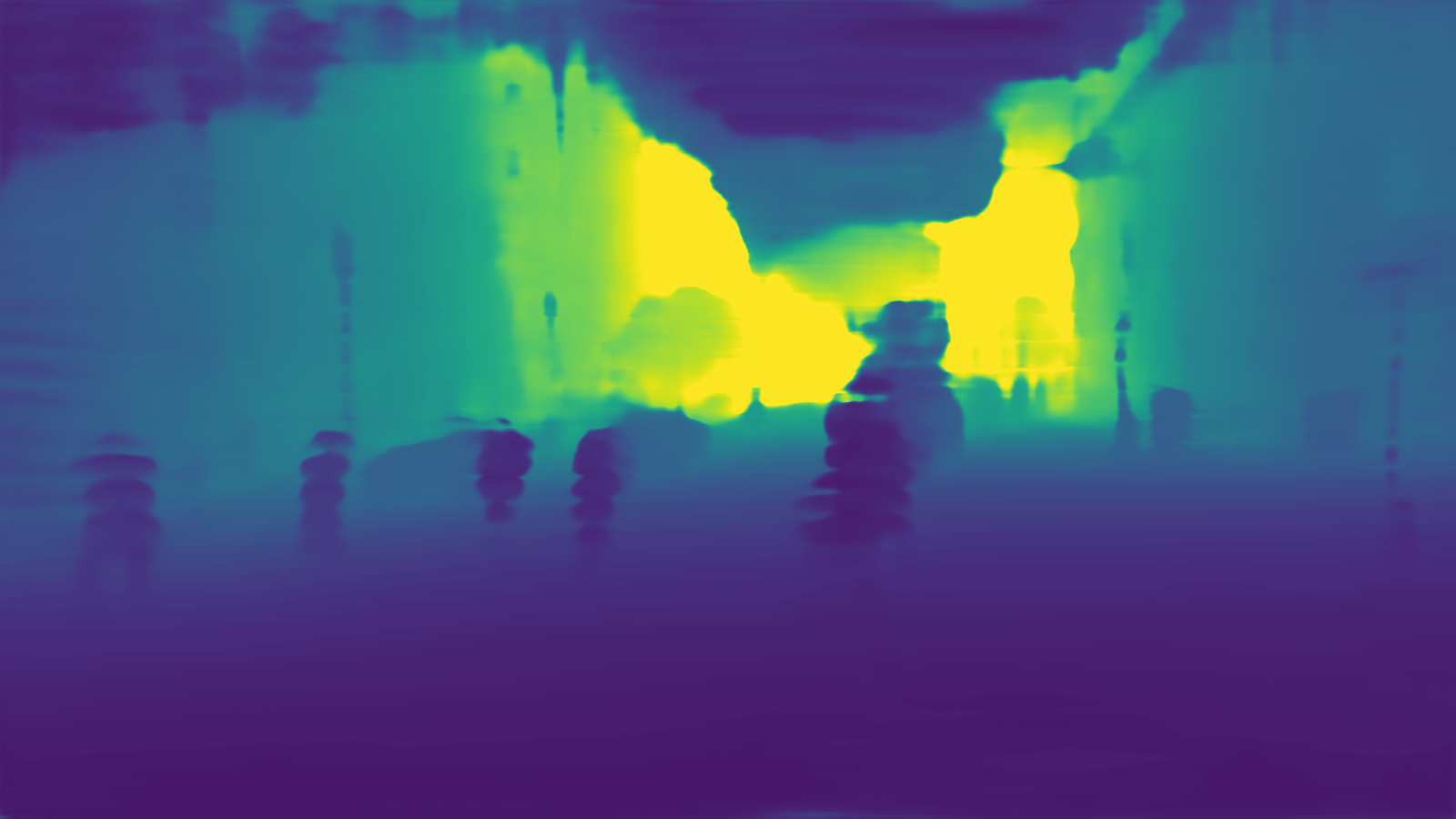}\label{fig:serial_design}}
\subfigure[Focal Length=5]{\includegraphics[width=0.23\textwidth]{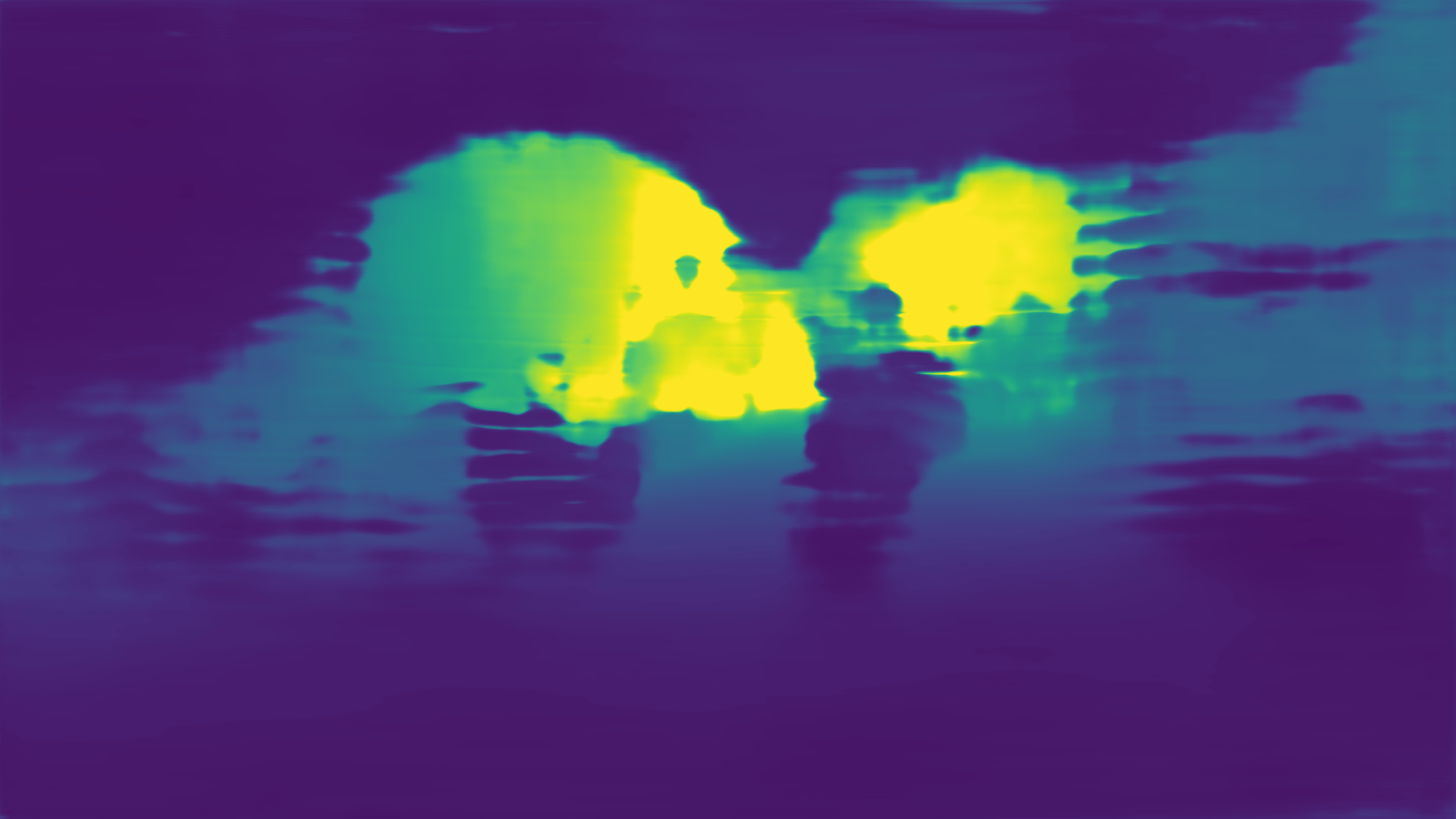}\label{fig:parallel_design}}
\subfigure[Focal Length=10]{\includegraphics[width=0.23\textwidth]{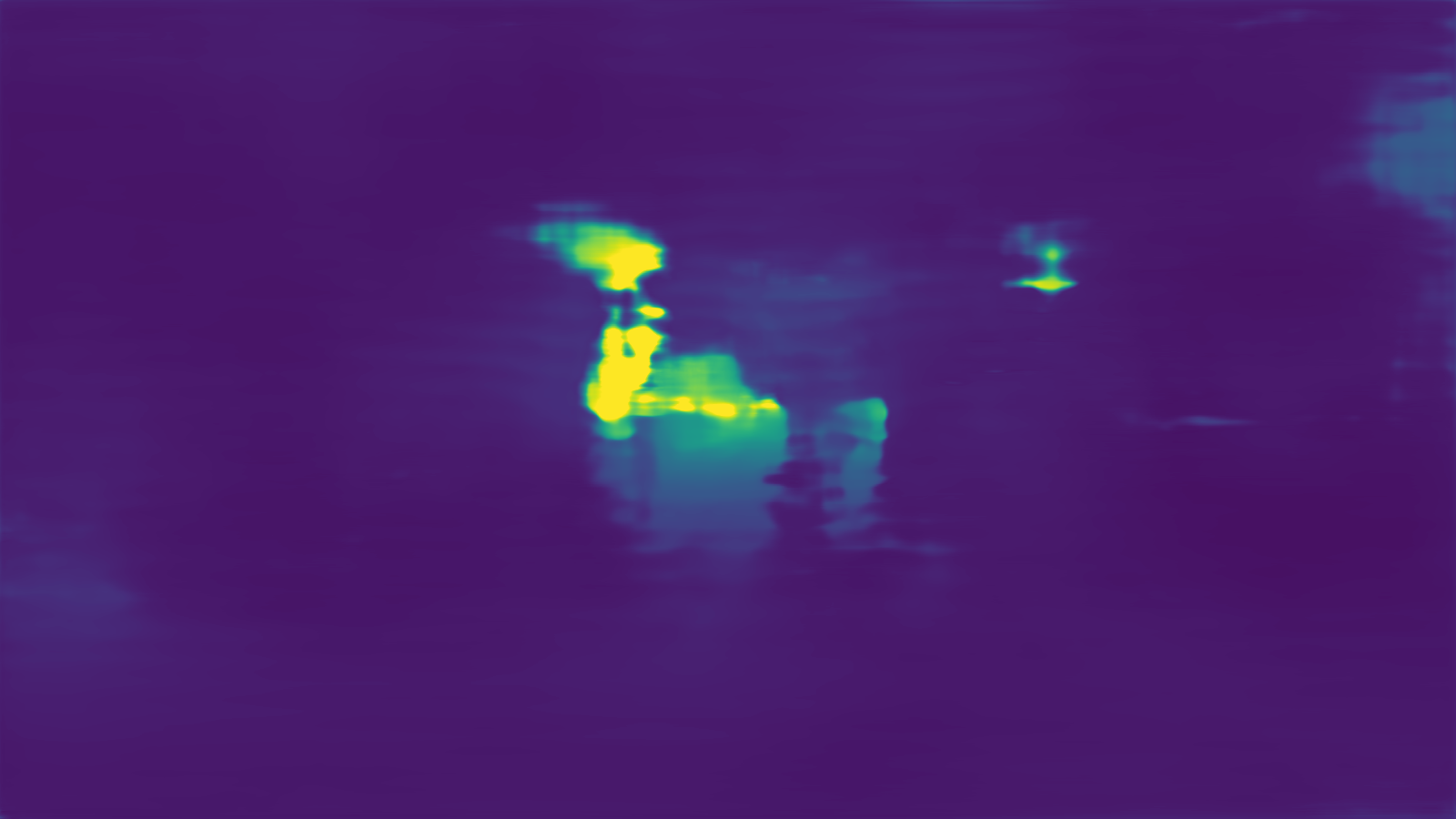}\label{fig:parallel_design}}
\subfigure[Focal Length=15]{\includegraphics[width=0.23\textwidth]{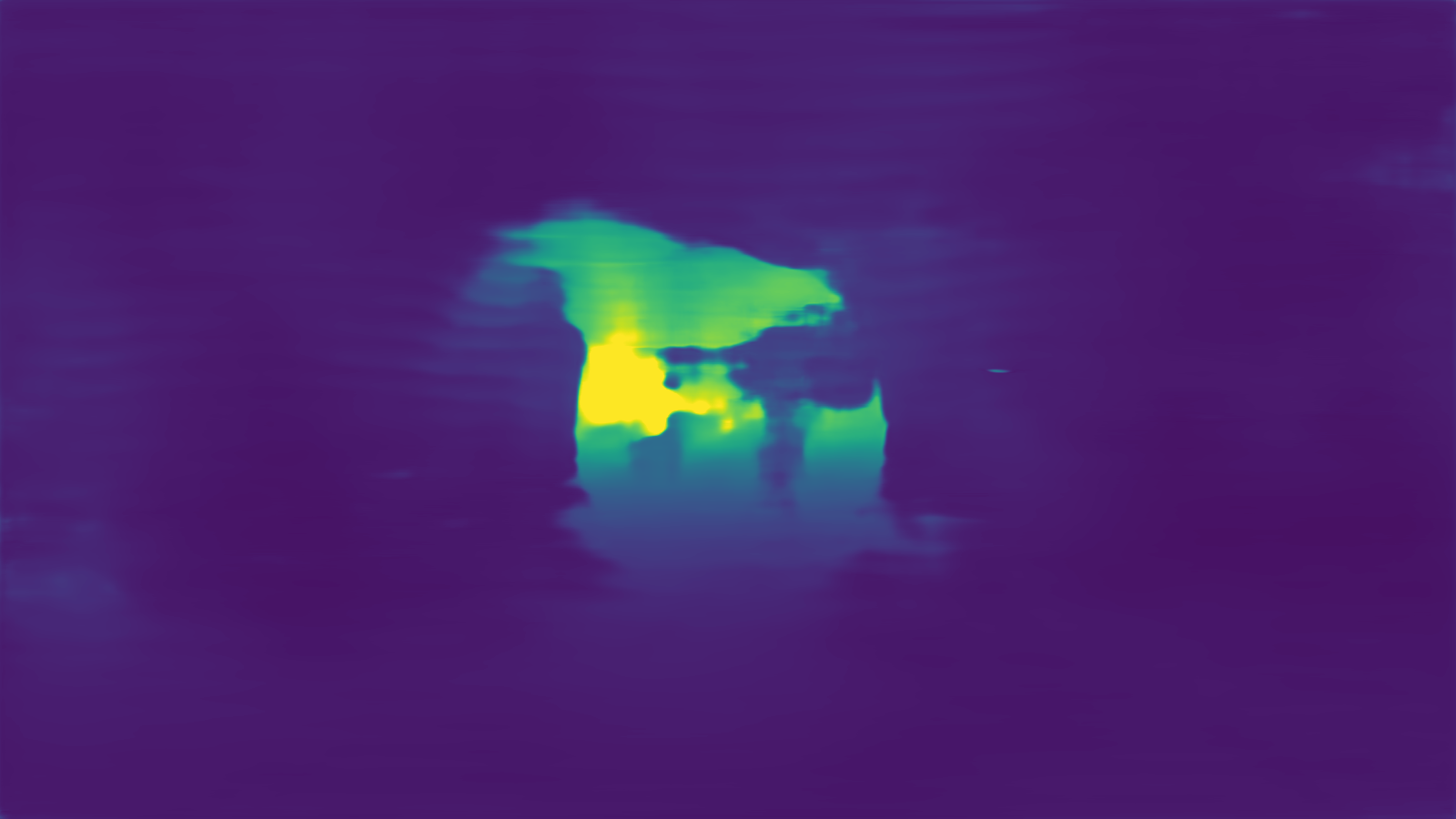}\label{fig:parallel_design}}

\subfigure[No attack]{\includegraphics[width=0.23\textwidth]{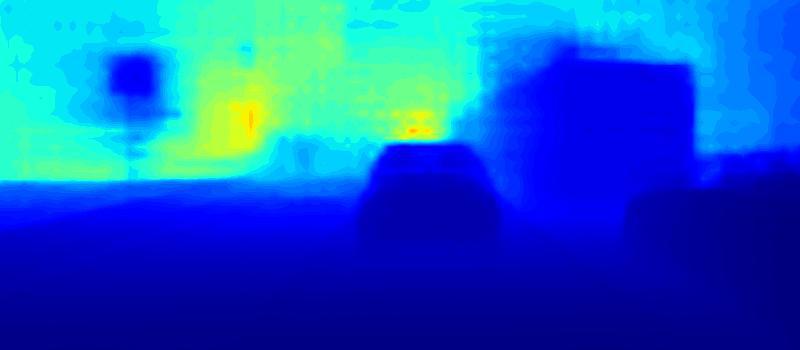}\label{fig:serial_design}}
\subfigure[Focal Length=5]{\includegraphics[width=0.23\textwidth]{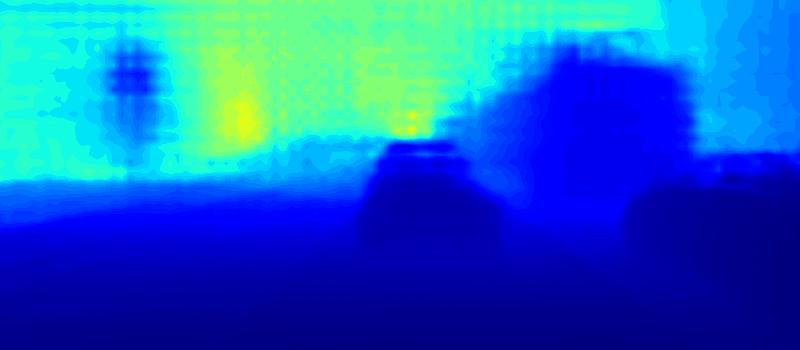}\label{fig:parallel_design}}
\subfigure[Focal Length=10]{\includegraphics[width=0.23\textwidth]{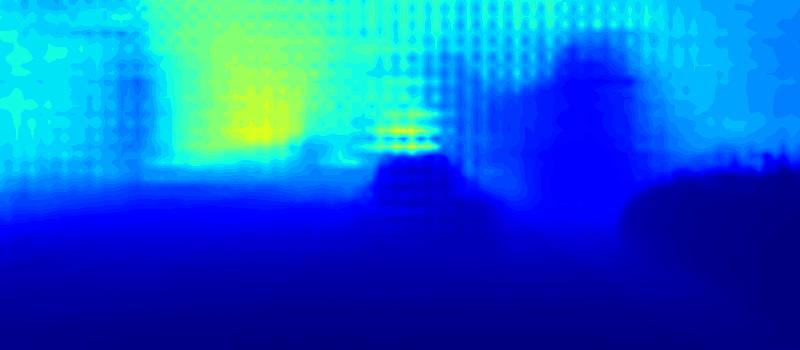}\label{fig:parallel_design}}
\subfigure[Focal Length=15]{\includegraphics[width=0.23\textwidth]{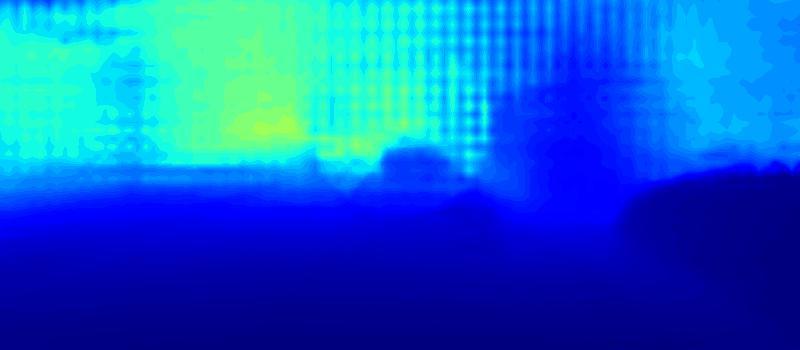}\label{fig:parallel_design}}

\caption{Examples of LensAttack on various object detection and depth estimation sensor fusion algorithms. The first to the last rows correspond to the object detection algorithms RADIANT~\cite{long2023radiant} and CRN~\cite{kim2023crn}, and the depth estimation algorithms, RC-PDA~\cite{long2021radar}, Cam\&mmWave~\cite{singh2023depth}, and DORN~\cite{lo2021depth}.}
\label{fig:Lens_attack}
\end{figure*}

The visualization results for the object detection and depth estimation algorithms under the above three attacks are presented in Figs.~\ref{fig:gaussian_attack}, \ref{fig:Vertical_attack}, and \ref{fig:Lens_attack}. As the blur kernel size or the Focal length increases, object detection performance degrades, causing incorrect bounding-box placement or failure to detect desired objects. Similarly, depth estimation performance also degrades, causing higher depth error in the estimation results.

\begin{figure*}[t]
\centering

\subfigure[No attack]{\includegraphics[width=0.23\textwidth,trim=20 130 80 130, clip]{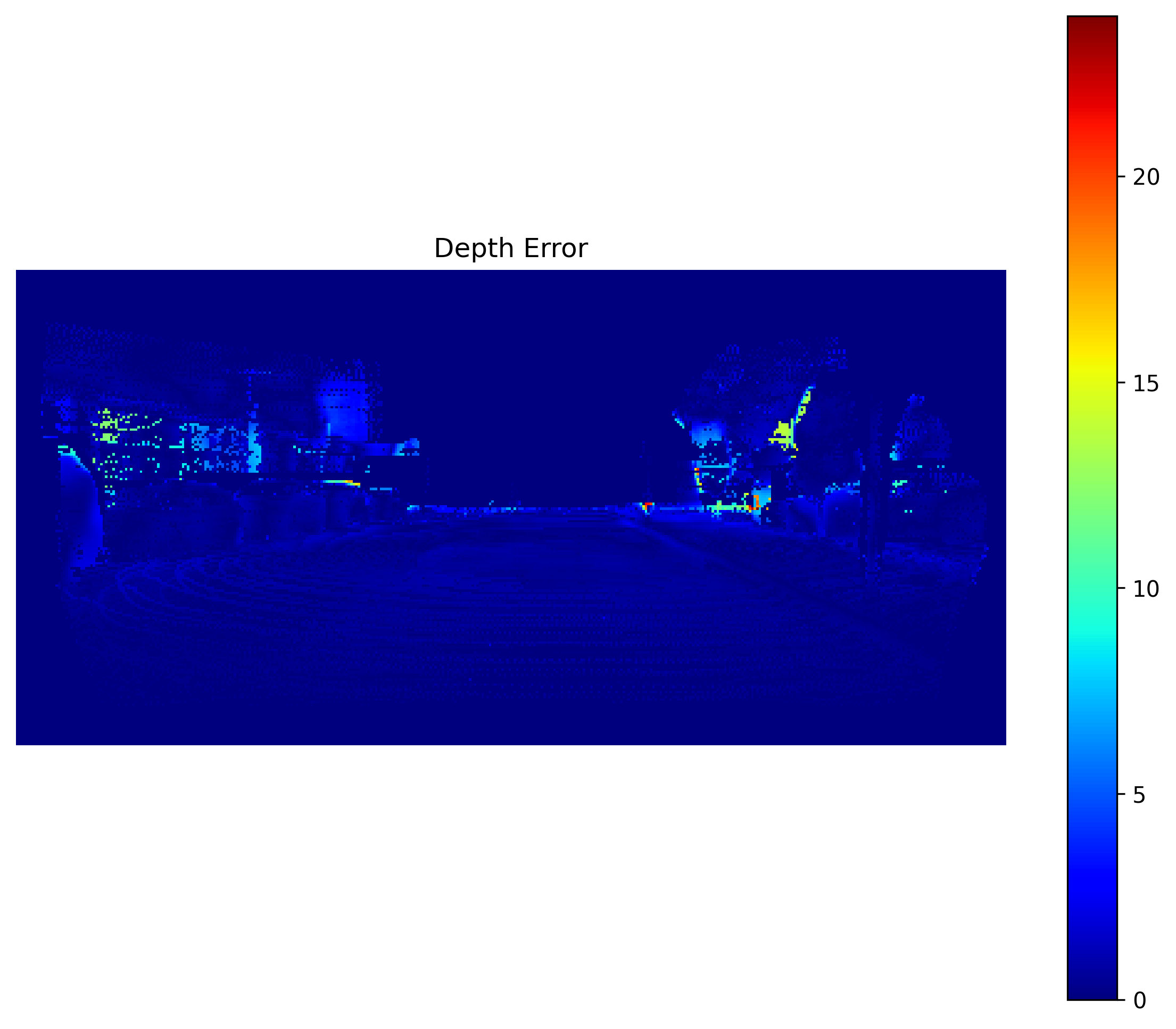}\label{fig:serial_design}}
\subfigure[Kernel=15]{\includegraphics[width=0.23\textwidth,trim=20 130 80 130, clip]{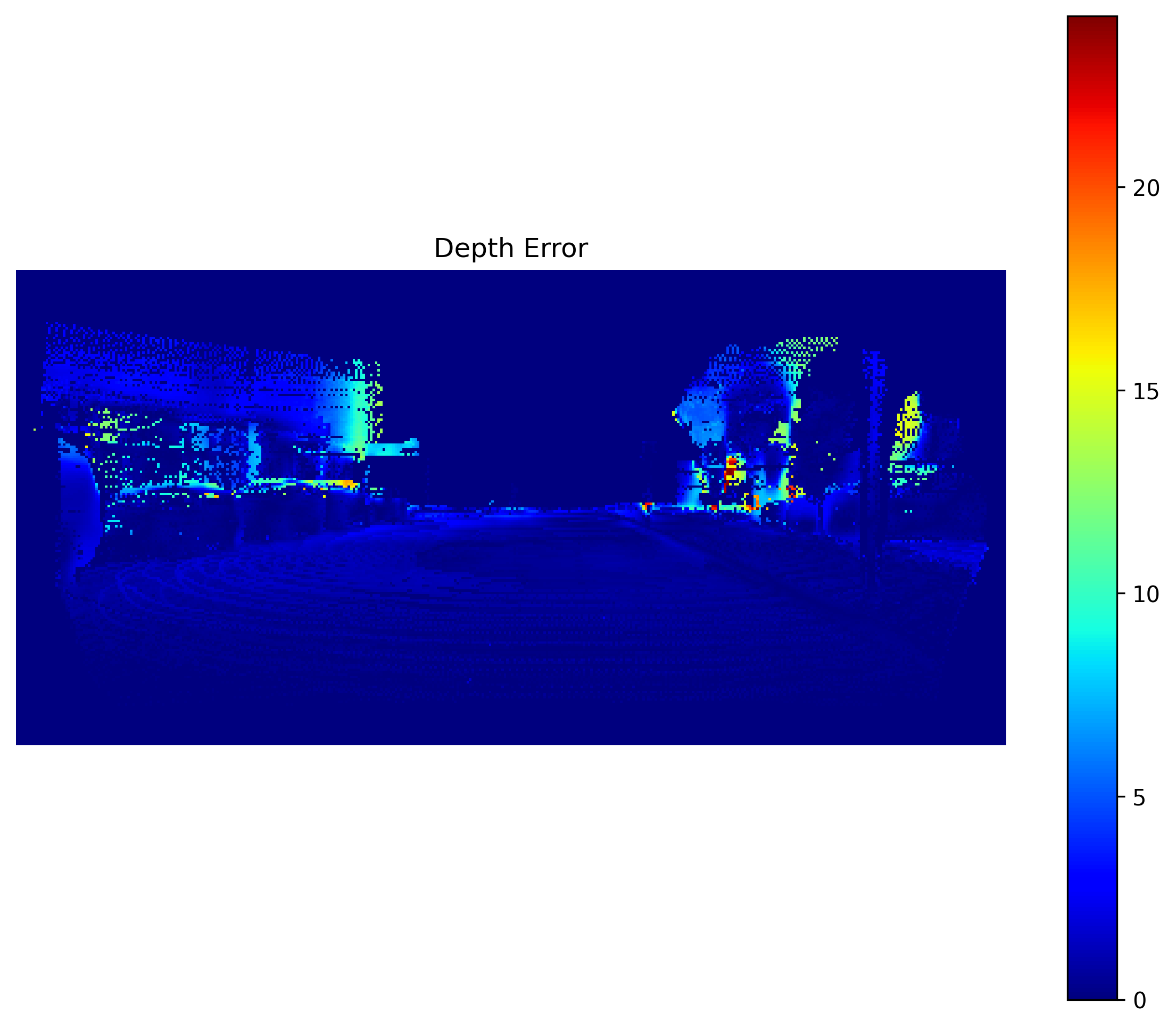}\label{fig:parallel_design}}
\subfigure[Kernel=45]{\includegraphics[width=0.23\textwidth,trim=20 130 80 130, clip]{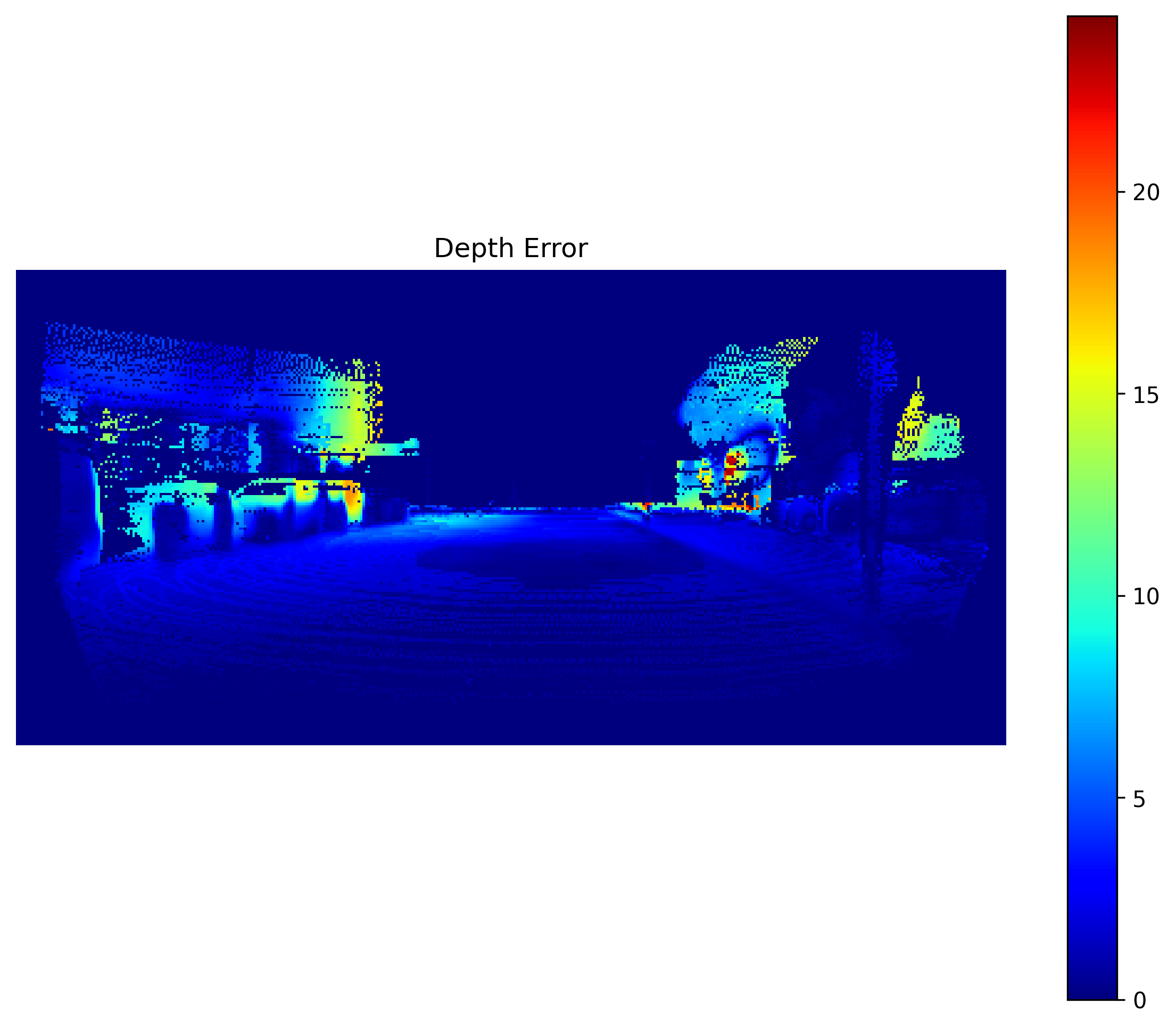}\label{fig:parallel_design}}
\subfigure[Kernel=75]{\includegraphics[width=0.23\textwidth,trim=20 130 80 130, clip]{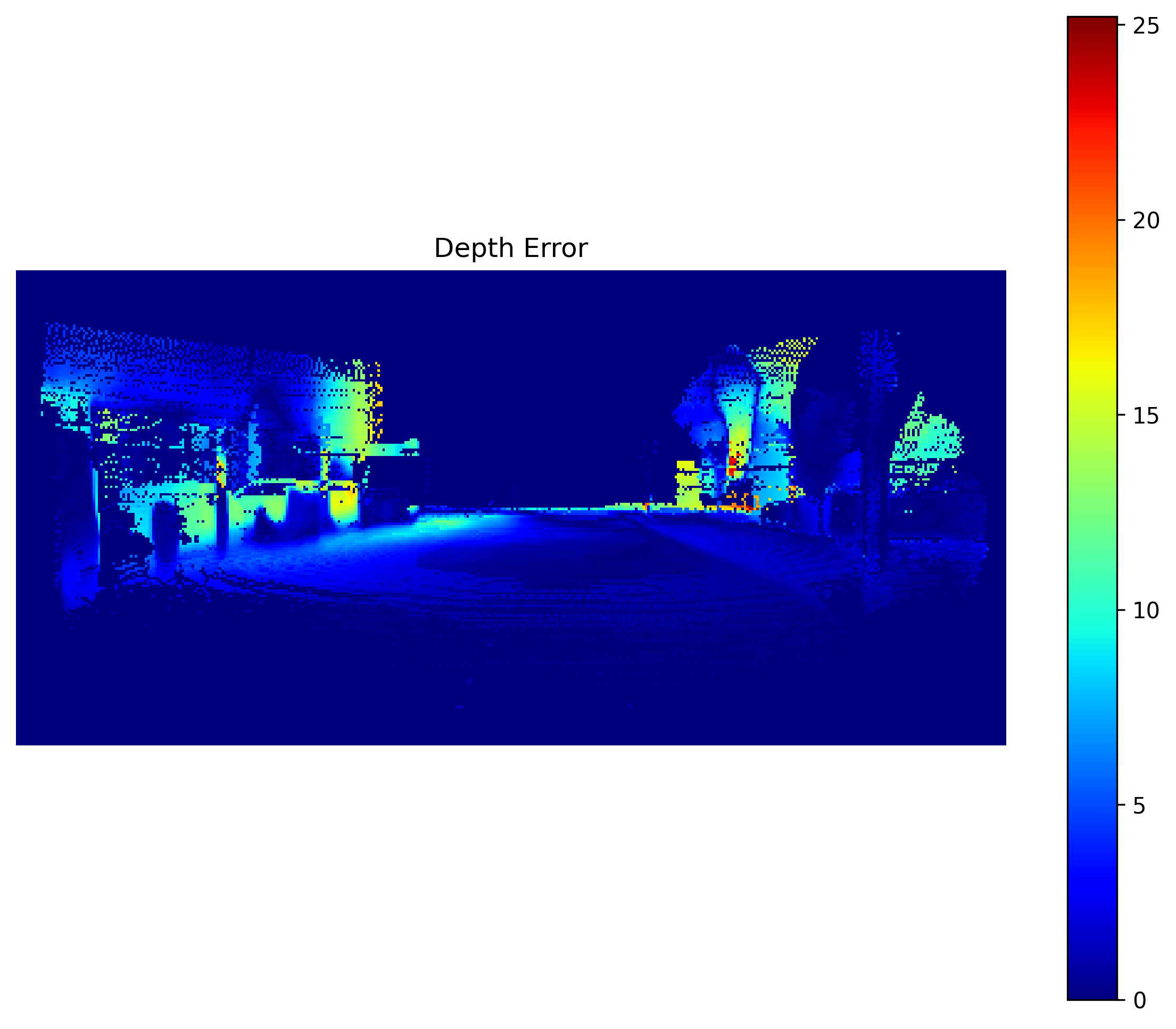}\label{fig:parallel_design}}

\subfigure[No attack]{\includegraphics[width=0.23\textwidth,trim=20 130 80 130, clip]{figure/3_rc-pda_deptherror/NoAttack_DE.png}\label{fig:serial_design}}
\subfigure[Kernel=20]{\includegraphics[width=0.23\textwidth,trim=20 130 80 130, clip]{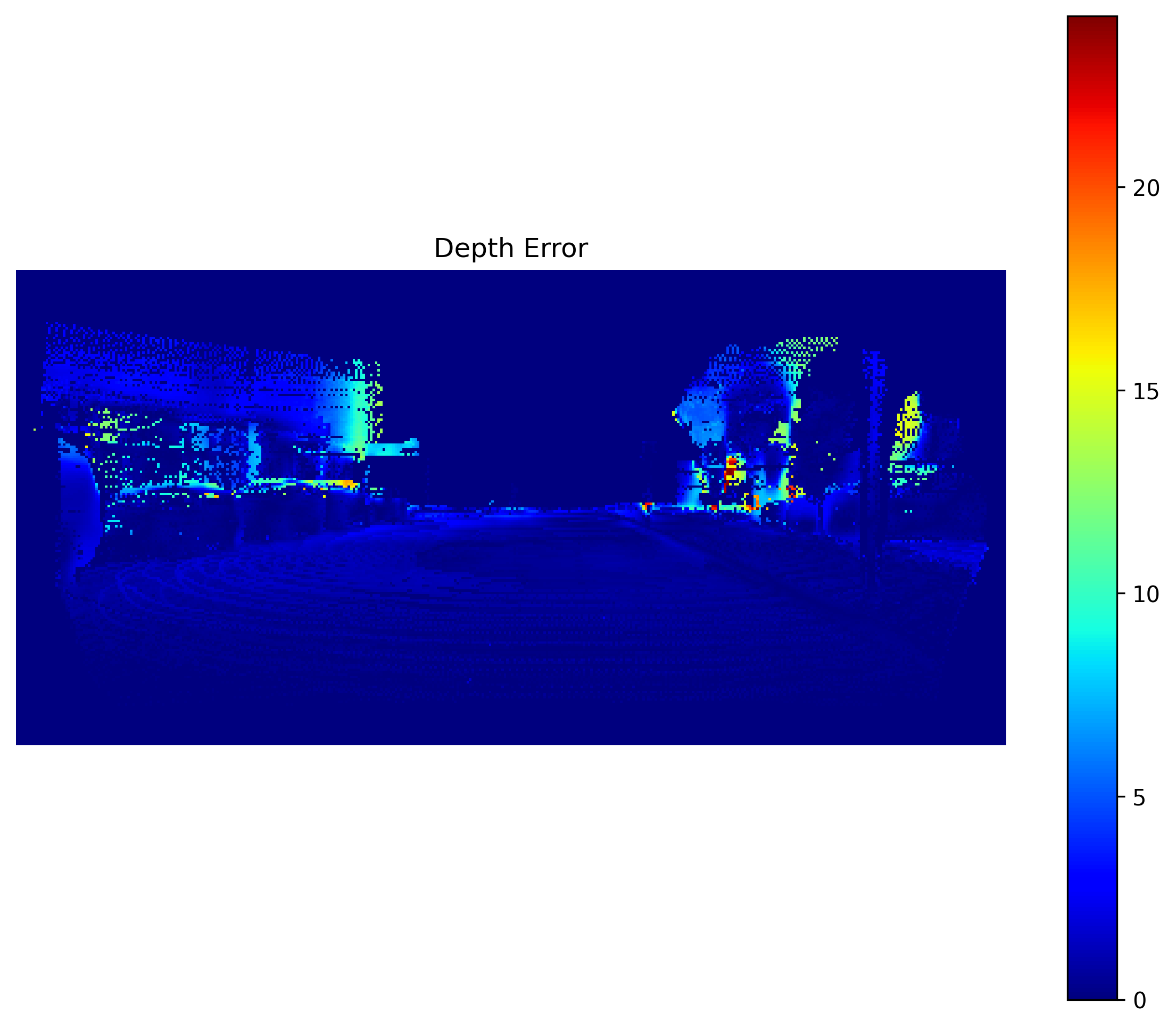}\label{fig:parallel_design}}
\subfigure[Kernel=50]{\includegraphics[width=0.23\textwidth,trim=20 130 80 130, clip]{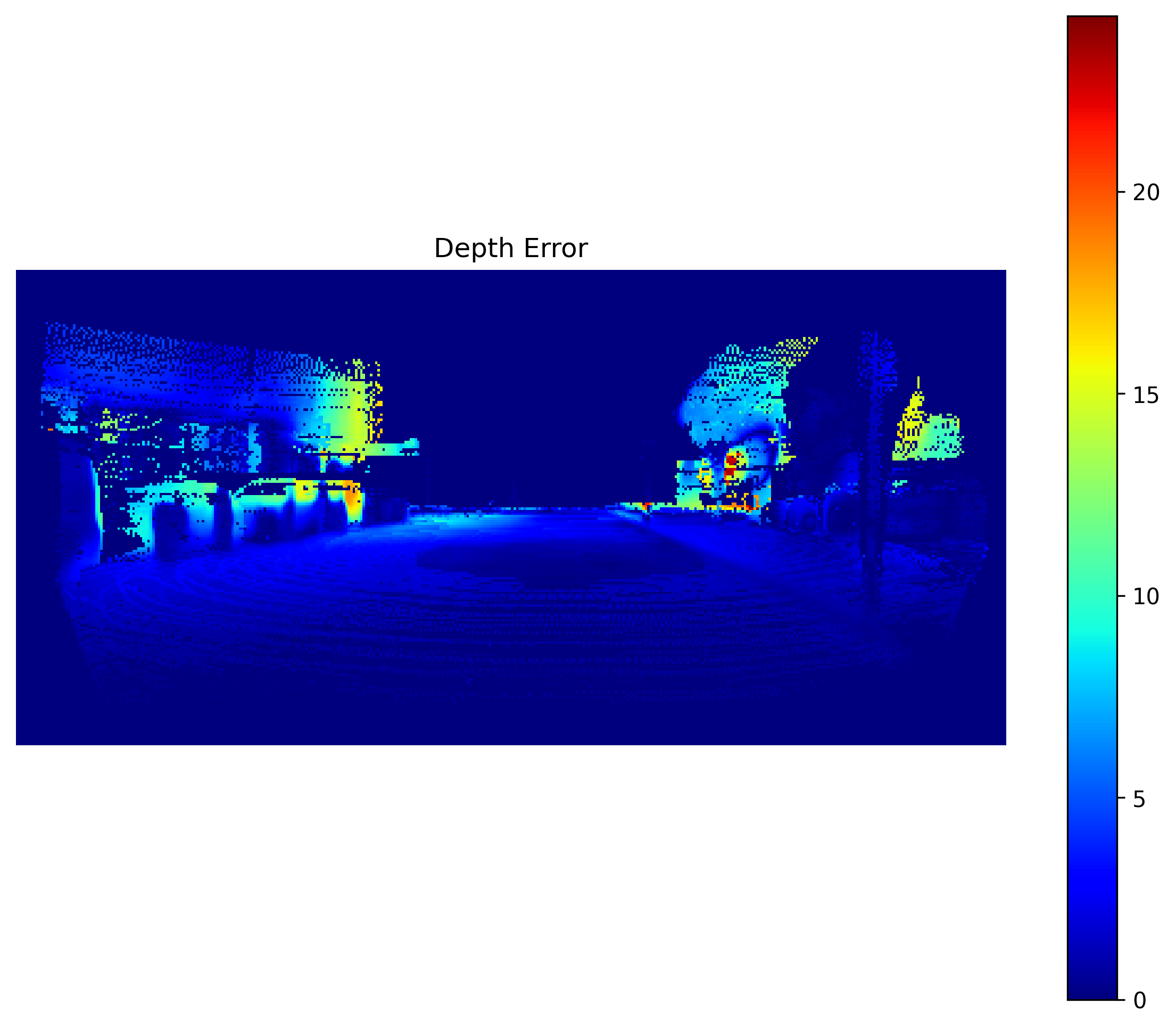}\label{fig:parallel_design}}
\subfigure[Kernel=80]{\includegraphics[width=0.23\textwidth,trim=20 130 80 130, clip]{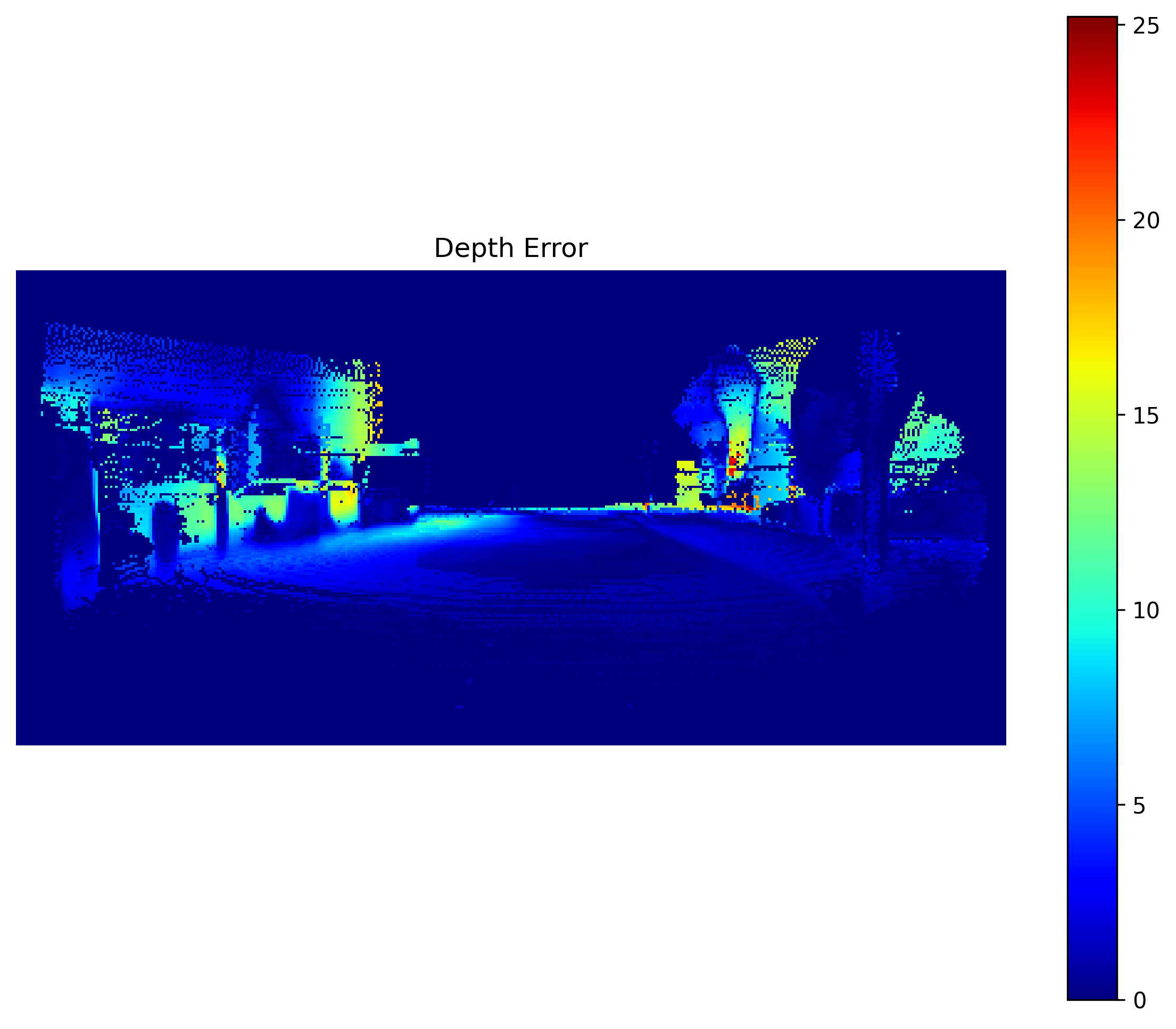}\label{fig:parallel_design}}

\subfigure[No attack]{\includegraphics[width=0.23\textwidth,trim=20 130 80 130, clip]{figure/3_rc-pda_deptherror/NoAttack_DE.png}\label{fig:serial_design}}
\subfigure[Focal Length=5]{\includegraphics[width=0.23\textwidth,trim=20 130 80 130, clip]{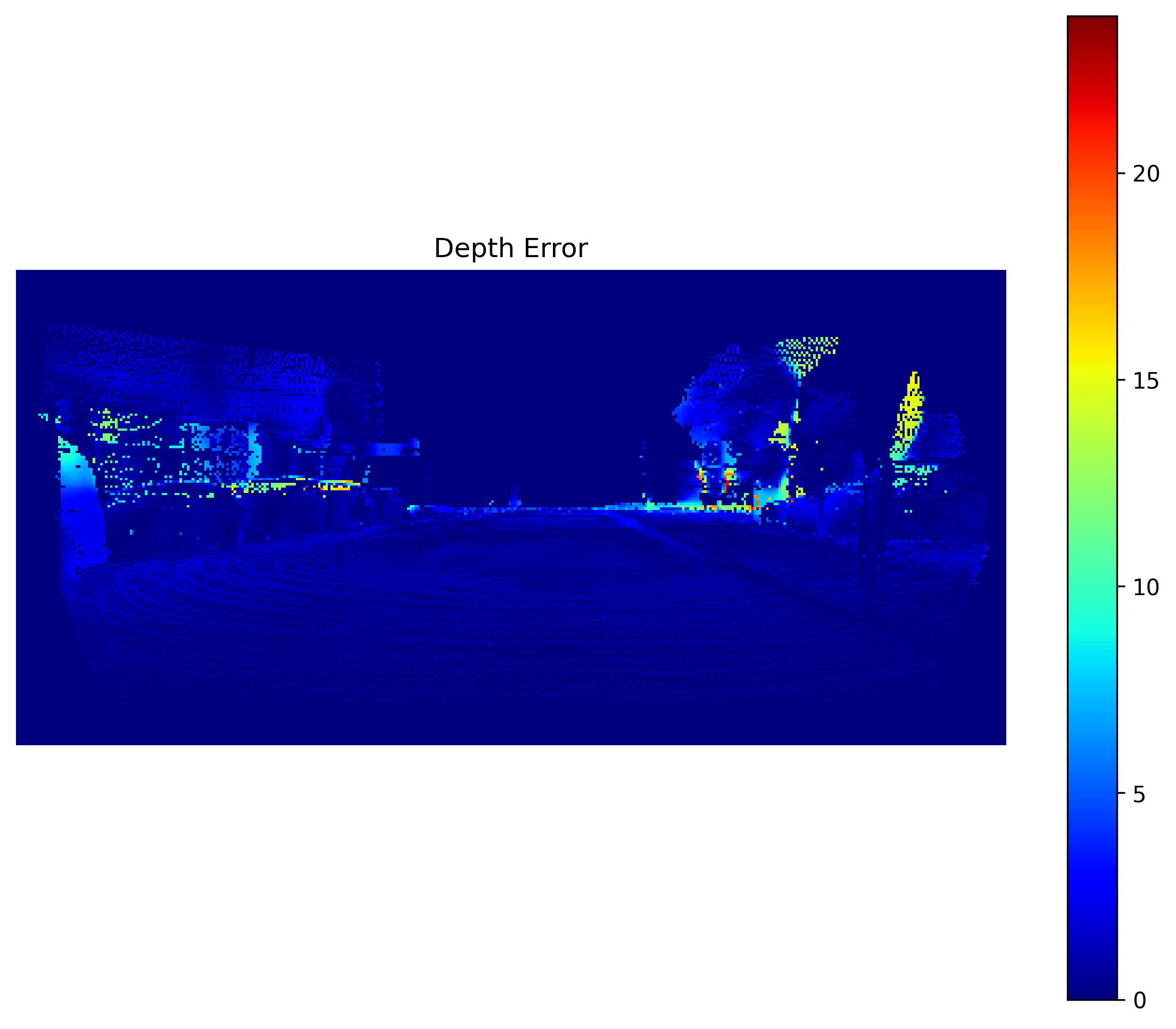}\label{fig:parallel_design}}
\subfigure[Focal Length=10]{\includegraphics[width=0.23\textwidth,trim=20 130 80 130, clip]{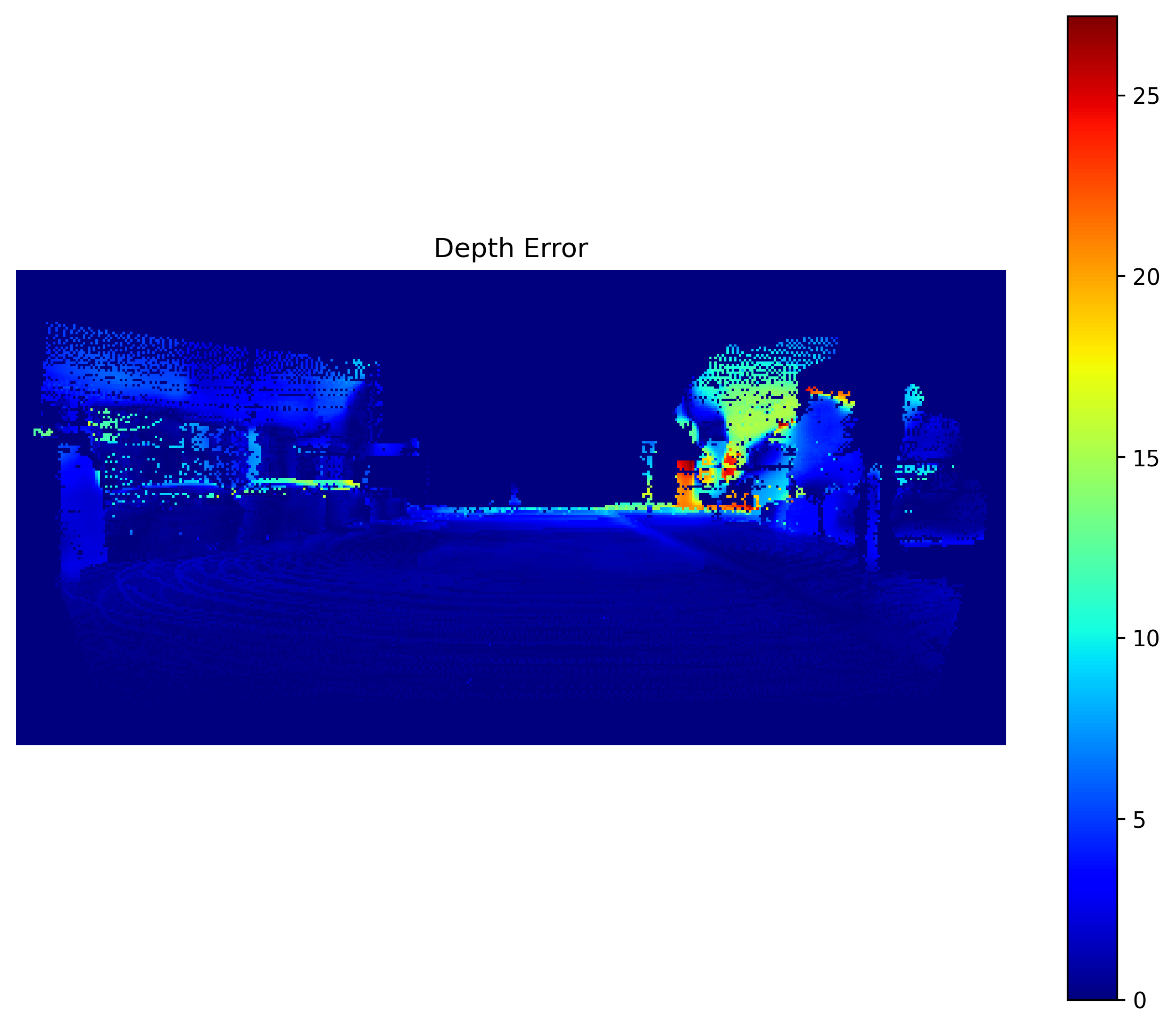}\label{fig:parallel_design}}
\subfigure[Focal Length=15]{\includegraphics[width=0.23\textwidth,trim=20 130 80 130, clip]{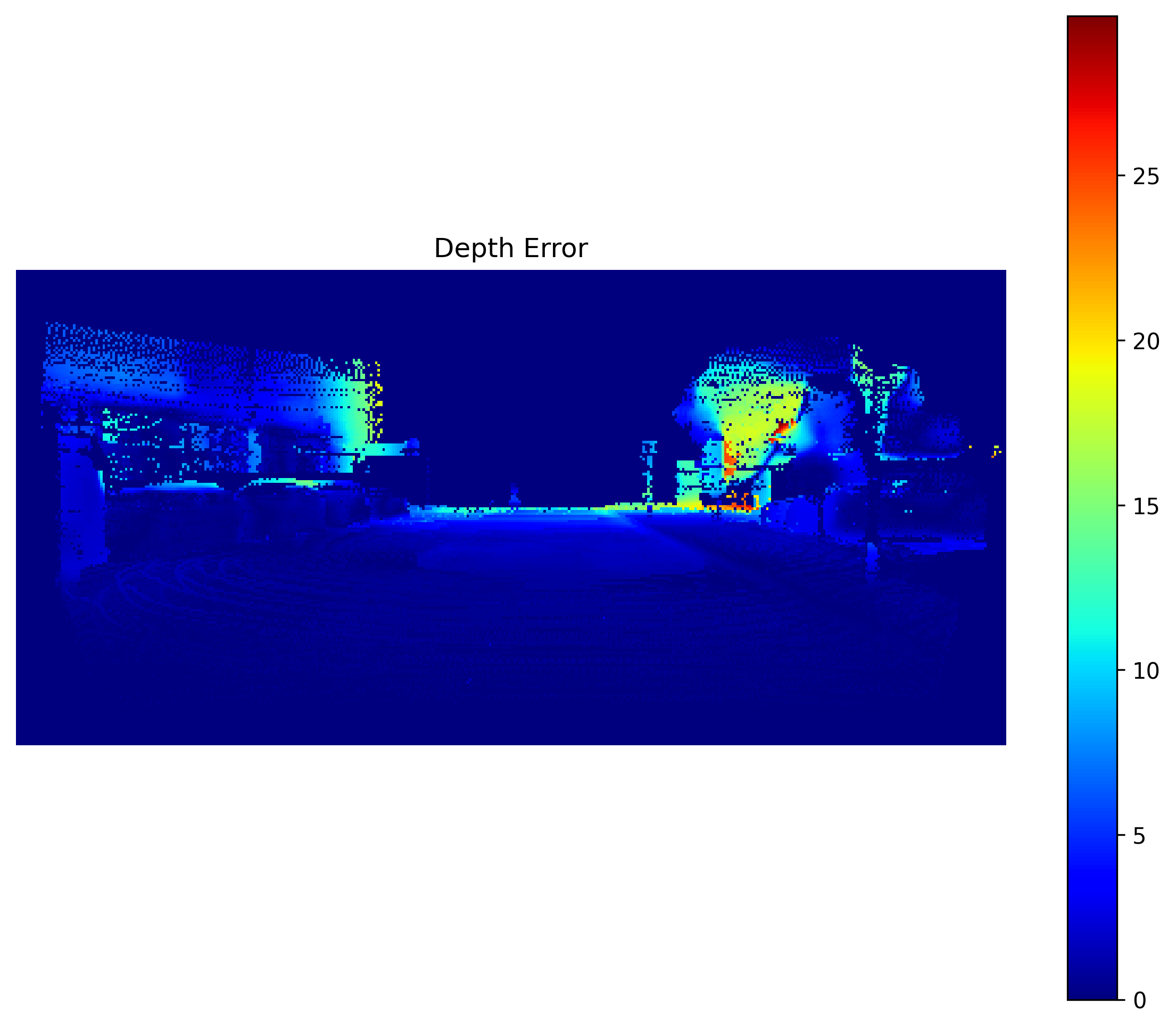}\label{fig:parallel_design}}

\caption{Depth error visualization of RC-PDA  under Gaussian Blur attack, Vertical Blur attack, and LensAttack with increasing adversarial strength.}
\label{fig:Depth_error_visualization_rcpda}
\end{figure*}

To better visualize the depth estimation performance, we further show the depth error visualization results for RC-PDA in Fig.~\ref{fig:Depth_error_visualization_rcpda}. 
The results clearly show that increasing adversarial strength results in greater depth errors within RC-PDA.

\subsection{Findings and Analysis}

Based on the above results, we present two findings: 

\cbox{Finding 1: Both DE and OD sensor fusion algorithms are affected by camera attacks, and the impacts are dramatically increased as the adversarial strength increases. }  

Two factors underlie this finding. First, a camera attack transforms
$I_t$ into $\widetilde{I}_t$, which in turn changes the camera
representation from

\begin{equation}
    F_t^c = \mathcal{E}_c(I_t)
\end{equation}
to
\begin{equation}
    \widetilde{F}_t^c
    =
    \mathcal{E}_c(\widetilde{I}_t).
\end{equation}

As the attack strength $\lambda_c$ increases, the discrepancy
between $F_t^c$ and $\widetilde{F}_t^c$ can propagate through the
subsequent fusion operation $\mathcal{F}(\cdot)$, ultimately
affecting $\hat{Y}_t^{\mathrm{OD}}$ or $\hat{D}_t$. This is
consistent with the known sensitivity of neural perception models
to image-space perturbations~\cite{goodfellow2014explaining, zhou2024optical, zhou2024transient}.

Second, camera images provide dense appearance and semantic
information, whereas the radar measurements in nuScenes are
comparatively sparse. Consequently, the evaluated fusion
architectures often rely more strongly on $F_t^c$ than on $F_t^r$.
A large change in the camera representation can therefore dominate
the fused representation $F_t^f$, explaining the pronounced
performance degradation observed under camera attacks.

\cbox{Finding 2: Moderate adversarial strength is sufficient to cause severe failures in both DE and OD sensor fusion algorithms.} 
Evidence for this finding comes from experiments on multiple camera–radar sensor fusion architectures, each assessed under three different levels of adversarial strengths.
As shown in Table~\ref{tab:vertical_blur}, for object detection algorithms, even the lowest adversarial strength (vertical motion blur with kernel size 20) reduces  performance by approximately 20\%. Increasing the attack to the medium and high settings further degrades performance, with drops of up to 25\%. For depth estimation algorithms, the impact is far more severe: the lowest-strength attack increases error by 251.2\%, and the medium and high settings raise it to about 391\%; notably, the medium level already induces errors comparable to the strongest attack. These results indicate that adversarial perturbations need not be extreme to cause substantial failures in camera–radar sensor fusion, and moderate distortions can produce degradations akin to those at maximum strength, underscoring critical vulnerabilities for real-world AD deployments.


\begin{table}[]
\centering
\caption{The performance of different camera-radar sensor fusion models under the LensAttack with various adversarial strengths. OD Metrics: mAP/NDS. DE Metrics: MAE/RMSE.}
\label{tab:lensattack_blur}
\resizebox{0.6\columnwidth}{!}{%
\begin{tabular}{|l|c|llll|}
\hline
\multicolumn{1}{|c|}{\multirow{2}{*}{Model}} & \multirow{2}{*}{Type} & \multicolumn{4}{c|}{Adversarial Strength (Focal Length)}                                                                                                                                                                                                                                                                                     \\ \cline{3-6} 
\multicolumn{1}{|c|}{}                       &                       & \multicolumn{1}{c|}{No Attack}                                                         & \multicolumn{1}{c|}{5}                                                                 & \multicolumn{1}{c|}{10}                                                                & \multicolumn{1}{c|}{15}                                           \\ \hline
RADIANT                                      & OD                    & \multicolumn{1}{l|}{0.363/0.340}                                                       & \multicolumn{1}{l|}{0.311/0.310}                                                       & \multicolumn{1}{l|}{0.276/0.293}                                                       & 0.270/0.291                                                       \\ \hline
CRN                                          & OD                    & \multicolumn{1}{l|}{0.473/0.562}                                                       & \multicolumn{1}{l|}{0.434/0.537}                                                       & \multicolumn{1}{l|}{0.390/0.508}                                                       & 0.366/0.495                                                       \\ \hline
RC-PDA                                       & DE                    & \multicolumn{1}{l|}{\begin{tabular}[c]{@{}l@{}}0.831/2.018\\ 1.279/2.682\end{tabular}} & \multicolumn{1}{l|}{\begin{tabular}[c]{@{}l@{}}1.099/2.485\\ 1.611/3.243\end{tabular}} & \multicolumn{1}{l|}{\begin{tabular}[c]{@{}l@{}}1.467/3.111\\ 2.042/3.871\end{tabular}} & \begin{tabular}[c]{@{}l@{}}1.750/3.520\\ 2.191/3.961\end{tabular} \\ \hline
Cam\&mmWave                                  & DE                    & \multicolumn{1}{l|}{2.037/4.548}                                                       & \multicolumn{1}{l|}{3.836/7.354}                                                       & \multicolumn{1}{l|}{6.515/11.652}                                                      & 7.738/13.318                                                      \\ \hline
DORN                                  & DE                    & \multicolumn{1}{l|}{2.384/5.264}                                                       & \multicolumn{1}{l|}{3.165/6.426}                                                       & \multicolumn{1}{l|}{3.756/7.462}                                                       & 4.199/8.356                                                       \\ \hline
\end{tabular}%
}
\end{table}

\begin{table}[h]
\centering
\caption{The performance of different camera-radar sensor fusion models under the Median blur attack with various adversarial strengths. OD Metrics: mAP/NDS. DE Metrics: MAE/RMSE.}
\label{tab:median_blur}
\resizebox{0.6\columnwidth}{!}{%
\begin{tabular}{|l|c|llll|}
\hline
\multicolumn{1}{|c|}{\multirow{2}{*}{Model}} & \multirow{2}{*}{Type} & \multicolumn{4}{c|}{Adversarial Strength (Kernel Size)}                                                                                                                                                                                                                                                                                      \\ \cline{3-6} 
\multicolumn{1}{|c|}{}                       &                       & \multicolumn{1}{c|}{No Attack}                                                         & \multicolumn{1}{c|}{15}                                                                & \multicolumn{1}{c|}{45}                                                                & \multicolumn{1}{c|}{75}                                           \\ \hline
RADIANT                                      & OD                    & \multicolumn{1}{l|}{0.363/0.340}                                                       & \multicolumn{1}{l|}{0.291/0.302}                                                       & \multicolumn{1}{l|}{0.272/ 0.293}                                                      & 0.271/0.292                                                       \\ \hline
CRN                                          & OD                    & \multicolumn{1}{l|}{0.473/0.562}                                                       & \multicolumn{1}{l|}{0.411/0.5242}                                                      & \multicolumn{1}{l|}{0.353/0.490}                                                       & 0.353/0.490                                                       \\ \hline
RC-PDA                                       & DE                    & \multicolumn{1}{l|}{\begin{tabular}[c]{@{}l@{}}0.831/2.018\\ 1.279/2.682\end{tabular}} & \multicolumn{1}{l|}{\begin{tabular}[c]{@{}l@{}}1.145/2.607\\ 1.645/3.317\end{tabular}} & \multicolumn{1}{l|}{\begin{tabular}[c]{@{}l@{}}1.792/3.633\\ 2.458/4.304\end{tabular}} & \begin{tabular}[c]{@{}l@{}}2.428/4.373\\ 2.973/4.808\end{tabular} \\ \hline
Cam\&mmWave                                  & DE                    & \multicolumn{1}{l|}{2.037/4.548}                                                       & \multicolumn{1}{l|}{5.516/9.864}                                                       & \multicolumn{1}{l|}{10.786/17.013}                                                     & 10.137/16.222                                                     \\ \hline
DORN                                  & DE                    & \multicolumn{1}{l|}{2.384/5.264}                                                       & \multicolumn{1}{l|}{3.331/6.794}                                                       & \multicolumn{1}{l|}{4.377/8.547}                                                       & 5.043/9.592                                                       \\ \hline
\end{tabular}%
}
\end{table}

\begin{table}[h]
\centering
\caption{The performance of different camera-radar sensor fusion models under the Bilateral blur attack with various adversarial strengths. OD Metrics: mAP/NDS. DE Metrics: MAE/RMSE.}
\label{tab:bilateral_blur}
\resizebox{0.6\columnwidth}{!}{%
\begin{tabular}{|l|c|llll|}
\hline
\multicolumn{1}{|c|}{\multirow{2}{*}{Model}} & \multirow{2}{*}{Type} & \multicolumn{4}{c|}{Adversarial Strength (Kernel, Color, Space)}                                                                                                                                                                                                                                                                             \\ \cline{3-6} 
\multicolumn{1}{|c|}{}                       &                       & \multicolumn{1}{l|}{No Attack}                                                         & \multicolumn{1}{c|}{{[}15,75,75{]}}                                                    & \multicolumn{1}{c|}{{[}45,150,150{]}}                                                  & \multicolumn{1}{c|}{{[}75,250,250{]}}                             \\ \hline
RADIANT                                      & OD                    & \multicolumn{1}{l|}{0.363/0.340}                                                       & \multicolumn{1}{l|}{0.338/0.324}                                                       & \multicolumn{1}{l|}{0.303/0.307}                                                       & 0.285/0.298                                                       \\ \hline
CRN                                          & OD                    & \multicolumn{1}{l|}{0.473/0.562}                                                       & \multicolumn{1}{l|}{0.444/0.542}                                                       & \multicolumn{1}{l|}{0.3995/0.5156}                                                     & 0.3725/0.5006                                                     \\ \hline
RC-PDA                                       & DE                    & \multicolumn{1}{l|}{\begin{tabular}[c]{@{}l@{}}0.831/2.018\\ 1.279/2.682\end{tabular}} & \multicolumn{1}{l|}{\begin{tabular}[c]{@{}l@{}}0.948/2.204\\ 1.335/2.747\end{tabular}} & \multicolumn{1}{l|}{\begin{tabular}[c]{@{}l@{}}1.457/3.181\\ 1.935/3.710\end{tabular}} & \begin{tabular}[c]{@{}l@{}}1.987/3.842\\ 2.564/4.317\end{tabular} \\ \hline
Cam\&mmWave                                  & DE                    & \multicolumn{1}{l|}{2.037/4.548}                                                       & \multicolumn{1}{l|}{3.051/6.112}                                                       & \multicolumn{1}{l|}{4.815/8.917}                                                       & 5.797/10.248                                                      \\ \hline
DORN                                  & DE                    & \multicolumn{1}{l|}{2.384/5.264}                                                       & \multicolumn{1}{l|}{2.928/6.101}                                                       & \multicolumn{1}{l|}{3.502/7.188}                                                       & 4.167/8.471                                                       \\ \hline
\end{tabular}%
}
\end{table}

\begin{table}[h]
\centering
\caption{The performance of different camera-radar sensor fusion models under the Horizontal blur attack with various adversarial strengths. OD Metrics: mAP/NDS. DE Metrics: MAE/RMSE.}
\label{tab:Horizontal_blur}
\resizebox{0.6\columnwidth}{!}{%
\begin{tabular}{|l|c|llll|}
\hline
\multicolumn{1}{|c|}{\multirow{2}{*}{Model}} & \multirow{2}{*}{Type} & \multicolumn{4}{c|}{Adversarial Strength (Kernel Size)}                                                                                                                                                                                                                                                                                      \\ \cline{3-6} 
\multicolumn{1}{|c|}{}                       &                       & \multicolumn{1}{l|}{No Attack}                                                         & \multicolumn{1}{c|}{20}                                                                & \multicolumn{1}{c|}{50}                                                                & \multicolumn{1}{c|}{80}                                           \\ \hline
RADIANT                                      & OD                    & \multicolumn{1}{l|}{0.363/0.340}                                                       & \multicolumn{1}{l|}{0.310/0.310}                                                       & \multicolumn{1}{l|}{0.276/0.294}                                                       & 0.271/0.292                                                       \\ \hline
CRN                                          & OD                    & \multicolumn{1}{l|}{0.473/0.562}                                                       & \multicolumn{1}{l|}{0.4055/0.5181}                                                     & \multicolumn{1}{l|}{0.356/0.490}                                                       & 0.3528/0.4893                                                     \\ \hline
RC-PDA                                       & DE                    & \multicolumn{1}{l|}{\begin{tabular}[c]{@{}l@{}}0.831/2.018\\ 1.279/2.682\end{tabular}} & \multicolumn{1}{l|}{\begin{tabular}[c]{@{}l@{}}1.197/2.730\\ 1.779/3.546\end{tabular}} & \multicolumn{1}{l|}{\begin{tabular}[c]{@{}l@{}}1.629/3.481\\ 2.160/3.973\end{tabular}} & \begin{tabular}[c]{@{}l@{}}1.901/3.822\\ 2.490/4.300\end{tabular} \\ \hline
Cam\&mmWave                                  & DE                    & \multicolumn{1}{l|}{2.037/4.548}                                                       & \multicolumn{1}{l|}{5.015/9.122}                                                       & \multicolumn{1}{l|}{8.494/14.180}                                                      & 9.109/14.893                                                      \\ \hline
DORN                                  & DE                    & \multicolumn{1}{l|}{2.384/5.264}                                                       & \multicolumn{1}{l|}{3.516/7.364}                                                       & \multicolumn{1}{l|}{4.624/8.898}                                                       & 5.361/9.747                                                       \\ \hline
\end{tabular}%
}
\end{table}

\begin{table}[h]
\centering
\caption{The performance of different camera-radar sensor fusion models under the Radial blur attack with various adversarial strengths. OD Metrics: mAP/NDS. DE Metrics: MAE/RMSE.}
\label{tab:Radial_blur}
\resizebox{0.6\columnwidth}{!}{%
\begin{tabular}{|l|c|llll|}
\hline
\multicolumn{1}{|c|}{\multirow{2}{*}{Model}} & \multirow{2}{*}{Type} & \multicolumn{4}{c|}{Adversarial Strength (Angle, Scale)}                                                                                                                                                                                                                                                                                     \\ \cline{3-6} 
\multicolumn{1}{|c|}{}                       &                       & \multicolumn{1}{l|}{No Attack}                                                         & \multicolumn{1}{c|}{{[}1, 5{]}}                                                         & \multicolumn{1}{c|}{{[}1, 10{]}}                                                        & \multicolumn{1}{c|}{{[}1, 15{]}}                                  \\ \hline
RADIANT                                      & OD                    & \multicolumn{1}{l|}{0.363/0.340}                                                       & \multicolumn{1}{l|}{0.276/0.294}                                                       & \multicolumn{1}{l|}{0.271/0.292}                                                       & 0.271/0.292                                                       \\ \hline
CRN                                          & OD                    & \multicolumn{1}{l|}{0.473/0.562}                                                       & \multicolumn{1}{l|}{0.370/0.497}                                                       & \multicolumn{1}{l|}{0.354/0.4905}                                                      & 0.354/0.490                                                       \\ \hline
RC-PDA                                       & DE                    & \multicolumn{1}{l|}{\begin{tabular}[c]{@{}l@{}}0.831/2.018\\ 1.279/2.682\end{tabular}} & \multicolumn{1}{l|}{\begin{tabular}[c]{@{}l@{}}1.670/2.532\\ 3.424/4.345\end{tabular}} & \multicolumn{1}{l|}{\begin{tabular}[c]{@{}l@{}}2.259/4.127\\ 3.043/4.899\end{tabular}} & \begin{tabular}[c]{@{}l@{}}2.851/4.826\\ 3.541/5.594\end{tabular} \\ \hline
Cam\&mmWave                                  & DE                    & \multicolumn{1}{l|}{2.037/4.548}                                                       & \multicolumn{1}{l|}{8.398/13.934}                                                      & \multicolumn{1}{l|}{9.691/15.474}                                                      & 9.858/15.616                                                      \\ \hline
DORN                                  & DE                    & \multicolumn{1}{l|}{2.384/5.264}                                                       & \multicolumn{1}{l|}{5.001/9.573}                                                       & \multicolumn{1}{l|}{5.708/10.272}                                                      & 6.051/10.694                                                      \\ \hline
\end{tabular}%
}
\end{table}

\begin{table}[h]
\centering
\caption{The performance of different camera-radar sensor fusion models under the Rotational blur attack with various adversarial strengths. OD Metrics: mAP/NDS. DE Metrics: MAE/RMSE.}
\label{tab:Rotational_blur}
\resizebox{0.6\columnwidth}{!}{%
\begin{tabular}{|l|c|llll|}
\hline
\multicolumn{1}{|c|}{\multirow{2}{*}{Model}} & \multirow{2}{*}{Type} & \multicolumn{4}{c|}{Adversarial Strength (Angle)}                                                                                                                                                                                                                                                                                            \\ \cline{3-6} 
\multicolumn{1}{|c|}{}                       &                       & \multicolumn{1}{l|}{No Attack}                                                         & \multicolumn{1}{c|}{10}                                                                & \multicolumn{1}{c|}{15}                                                                & \multicolumn{1}{c|}{20}                                           \\ \hline
RADIANT                                      & OD                    & \multicolumn{1}{l|}{0.363/0.340}                                                       & \multicolumn{1}{l|}{0.279/0.295}                                                       & \multicolumn{1}{l|}{0.273/0.293}                                                       & 0.271/0.292                                                       \\ \hline
CRN                                          & OD                    & \multicolumn{1}{l|}{0.473/0.562}                                                       & \multicolumn{1}{l|}{0.376/0.502}                                                       & \multicolumn{1}{l|}{0.3644/0.4957}                                                     & 0.3575/0.4924                                                     \\ \hline
RC-PDA                                       & DE                    & \multicolumn{1}{l|}{\begin{tabular}[c]{@{}l@{}}0.831/2.018\\ 1.279/2.682\end{tabular}} & \multicolumn{1}{l|}{\begin{tabular}[c]{@{}l@{}}1.874/3.862\\ 2.854/4.618\end{tabular}} & \multicolumn{1}{l|}{\begin{tabular}[c]{@{}l@{}}2.110/4.075\\ 3.208/4.944\end{tabular}} & \begin{tabular}[c]{@{}l@{}}2.415/4.578\\ 3.521/5.397\end{tabular} \\ \hline
Cam\&mmWave                                  & DE                    & \multicolumn{1}{l|}{2.037/4.548}                                                       & \multicolumn{1}{l|}{7.063/12.621}                                                      & \multicolumn{1}{l|}{8.259/14.128}                                                      & 9.031/15.038                                                      \\ \hline
DORN                                  & DE                    & \multicolumn{1}{l|}{2.384/5.264}                                                       & \multicolumn{1}{l|}{8.733/15.545}                                                      & \multicolumn{1}{l|}{10.095/17.200}                                                     & 10.799/17.991                                                     \\ \hline
\end{tabular}%
}
\end{table}
\section{Robustness Evaluation under Radar Attacks}
In this section, we evaluate the impact of radar sensor attacks on the performance of SOTA camera-radar sensor fusion algorithms.

\subsection{Attacks on Radar Sensors}
\label{sec:radar_attacks}

We next evaluate attacks that corrupt the radar observation $P_t$.
Following Eq.~\eqref{eq:radar_attack_general}, an attacked radar
point cloud is represented as

\begin{equation}
    \widetilde{P}_t
    =
    \mathcal{A}_r(P_t;\lambda_r),
\end{equation}
where $\lambda_r$ denotes the attack severity. In contrast to the
camera experiments, the camera observation $I_t$ remains unchanged,
and the fusion model receives

\begin{equation}
    (I_t,\widetilde{P}_t)
    =
    \left(
    I_t,\mathcal{A}_r(P_t;\lambda_r)
    \right).
    \label{eq:radar_attack_input}
\end{equation}

We consider three radar corruption mechanisms: depth estimation
error, ghost-target injection, and target disappearance.

\textbf{Depth Estimation Error.}
This attack perturbs the depth associated with the radar
measurements. The attack strength $\lambda_r$ is expressed as the
relative depth-error percentage with respect to the original radar
input. Larger values of $\lambda_r$ therefore represent stronger
corruption of radar-derived depth.

\textbf{Ghost Target.}
Ghost-target attacks inject additional spurious radar points into
$P_t$. Here, $\lambda_r$ denotes the percentage of added points
relative to the original radar point cloud. Increasing
$\lambda_r$ therefore increases the number of false radar
measurements presented to the fusion model.

\textbf{Target Disappearance.}
Target-disappearance attacks remove a subset of the original radar
points. In this case, $\lambda_r$ denotes the percentage of removed
radar points. A larger $\lambda_r$ corresponds to more severe loss
of radar information.



\subsection{Evaluation Results}
All evaluation results are derived using pre-trained models of the respective algorithms, assessed on the nuScenes test dataset under varying levels of radar-based adversarial attacks.
For depth estimation algorithms, two metrics are employed: Mean Absolute Error (MAE) and Root Mean Square Error (RMSE). 
In the case of object detection algorithms, two metrics are employed: Mean Average Precision (MAP) and Mean Average Translation Error (MATE). 

\textbf{Depth Estimation Error.}
Table \ref{tab:estimation_error} presents the performance of SOTA sensor fusion algorithms when subjected to radar attacks that introduce depth estimation errors. 
These errors are expressed as percentages relative to the orginal radar input.
For DE algorithms, performance degrades as the magnitude of depth estimation error increases. 
For OD algorithms, they exhibit only minor performance degradation under the same radar attacks, suggesting a higher degree of robustness.

\begin{figure*}[t]
\centering
\subfigure[No attack]{\includegraphics[width=0.23\textwidth]{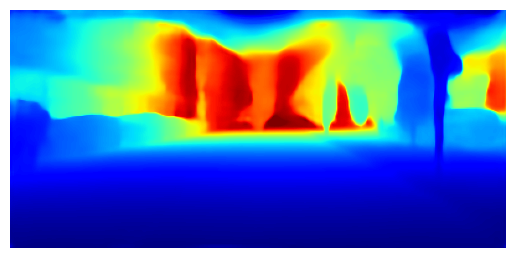}\label{fig:radar_depth_0_depth}}
\subfigure[50\% Estimation error]{\includegraphics[width=0.23\textwidth]{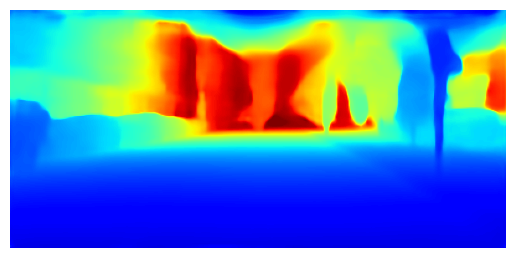}\label{fig:radar_depth_10_depth}}
\subfigure[100\% Estimation error]{\includegraphics[width=0.23\textwidth]{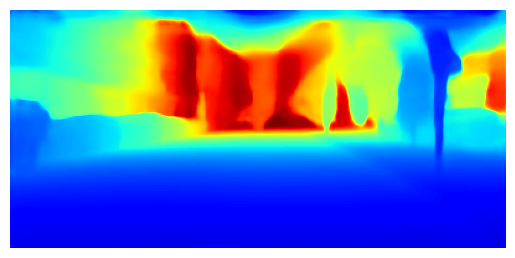}\label{fig:radar_depth_20_depth}}
\subfigure[200\% Estimation error]{\includegraphics[width=0.23\textwidth]{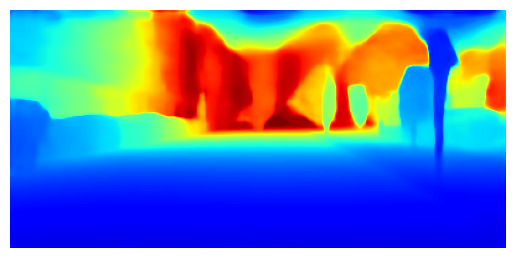}\label{fig:radar_depth_30_depth}}

\subfigure[No attack]{\includegraphics[width=0.23\textwidth]{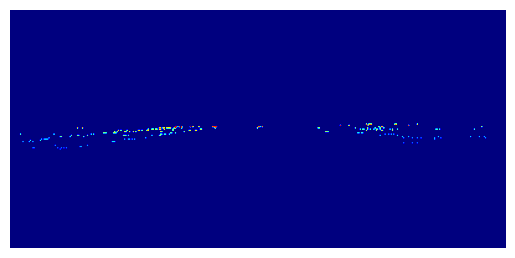}\label{fig:radar_depth_0_pcd}}
\subfigure[50\% Estimation error]{\includegraphics[width=0.23\textwidth]{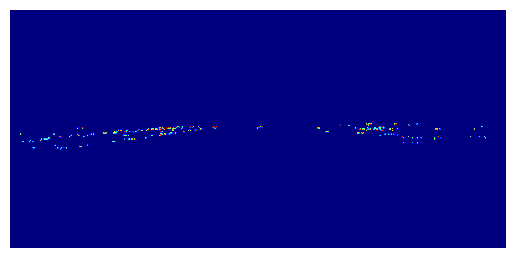}\label{fig:radar_depth_10_pcd}}
\subfigure[100\% Estimation error]{\includegraphics[width=0.23\textwidth]{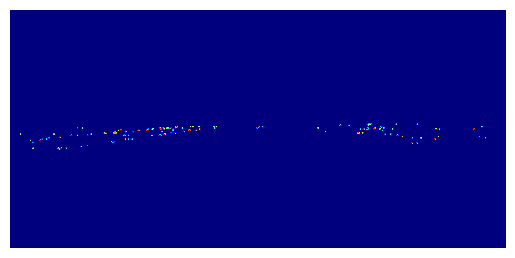}\label{fig:radar_depth_20_pcd}}
\subfigure[200\% Estimation error] {\includegraphics[width=0.23\textwidth]{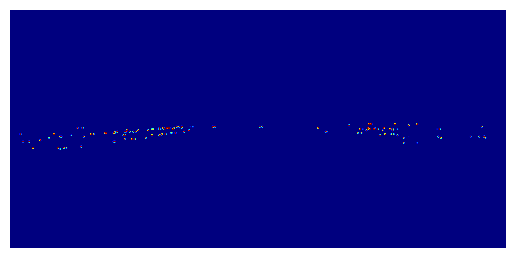}\label{fig:radar_depth_30_pcd}}

\caption{Examples of depth estimation error in RC-PDA \cite{long2021radar}. 
The first row illustrates depth estimation errors, while the second row shows the corresponding attack strength visualizations
}
\label{fig:radar_depth_error}
\end{figure*}

Figure~\ref{fig:radar_depth_error} shows the visualized results of the DE algorithm (RC-PDA) and different attack strengths applied to both DE and OD algorithms.
For the DE algorithm, performance degradation is observable, though less pronounced than under camera-based attacks.
In contrast, the OD algorithm demonstrates strong resilience: the location and size of bounding boxes remain largely unaffected even under stronger radar attacks, highlighting the robustness of OD methods against radar estimation errors.

\textbf{Target Disappearance.}
Table \ref{tab:target_disappearing} presents the evaluation results of SOTA sensor fusion algorithms under radar sensor attacks that cause varying levels of target disappearance. 
The attack level is quantified as the percentage of missing radar points relative to the original radar point cloud: higher percentages indicate more severe attacks and greater point loss.
For DE algorithms, as more radar points are removed due to the attack, the performance of SOTA sensor fusion algorithms degrades, but the decrease is small even with 70\% points disappeared.
For OD algorithms, they exhibit greater robustness to missing radar data compared to depth estimation algorithms.

\begin{figure*}[t]
\centering
\subfigure[No attack]{\includegraphics[width=0.23\textwidth]{figures_radar/depth_error/0_depth.png}\label{fig:radar_vanish_0_depth}}
\subfigure[30\% Target disappearance]{\includegraphics[width=0.23\textwidth]{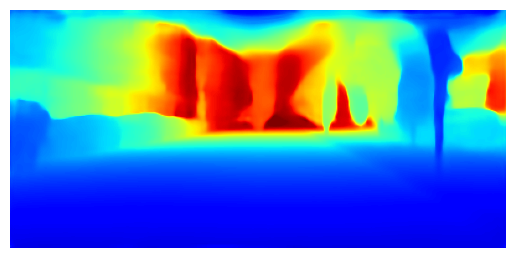}\label{fig:radar_vanish_10_depth}}
\subfigure[50\% Target disappearance]{\includegraphics[width=0.23\textwidth]{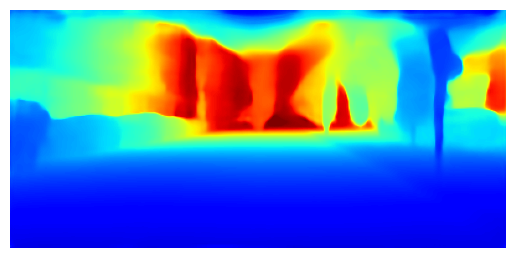}\label{fig:radar_vanish_20_depth}}
\subfigure[70\% Target disappearance]{\includegraphics[width=0.23\textwidth]{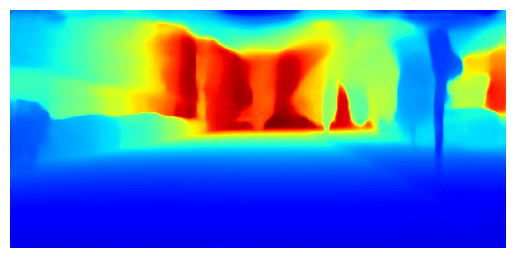}\label{fig:radar_vanish_30_depth}}

\subfigure[No attack]{\includegraphics[width=0.23\textwidth]{figures_radar/depth_error/0_radar.png}\label{fig:radar_vanish_0_pcd}}
\subfigure[30\% Target disappearance]{\includegraphics[width=0.23\textwidth]{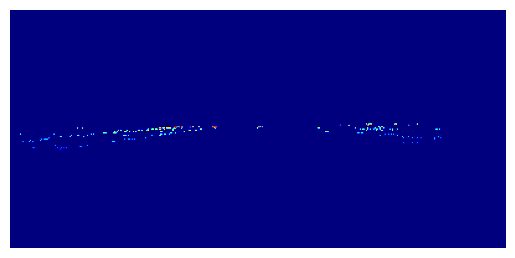}\label{fig:radar_vanish_10_pcd}}
\subfigure[50\% Target disappearance]{\includegraphics[width=0.23\textwidth]{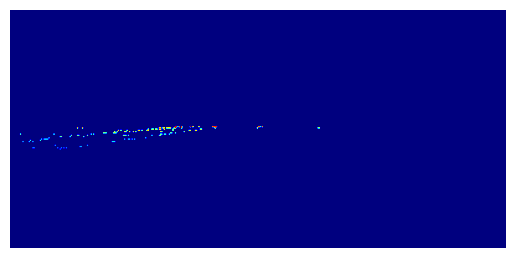}\label{fig:radar_vanish_20_pcd}}
\subfigure[70\% Target disappearance]{\includegraphics[width=0.23\textwidth]{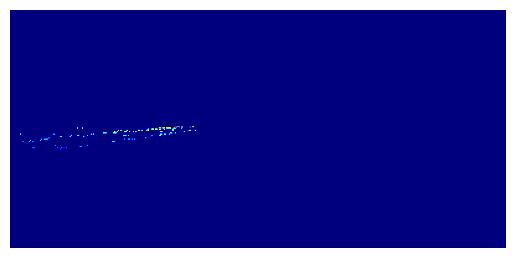}\label{fig:radar_vanish_30_pcd}}

\caption{Examples of target disappearance in RC-PDA \cite{long2021radar}. The first row illustrates depth estimation errors, while the second row shows the corresponding attack strength visualizations.}
\label{fig:radar_target_vanish}
\end{figure*}

Fig.~\ref{fig:radar_target_vanish} presents the visualized results of the DE algorithm (RC-PDA) and varying adversarial strengths on target disappearance.
In short, the DE algorithm experiences a slight degradation in the estimated depth with the increase in attack level. Moreover, the OD algorithm demonstrates greater robustness against radar target disappearance, with the detection performance remaining unaffected.

\textbf{Ghost Target.}
Table \ref{tab:ghost_target} presents the evaluation results of SOTA sensor fusion algorithms under radar sensor attacks that cause varying levels of ghost targets.
The attack level is quantified as the percentage of additional ghost radar points relative to the original radar point cloud: higher percentages indicate more severe attacks and a greater number of ghost points.
For DE algorithms, as the magnitude of estimation error increases, the performance of sensor fusion algorithms deteriorates slightly. 
For OD algorithms, they exhibit greater robustness to injected/spurious radar data compared to depth estimation algorithms.

\begin{figure*}[t]
\centering
\subfigure[No attack]{\includegraphics[width=0.23\textwidth]{figures_radar/depth_error/0_depth.png}\label{fig:radar_ghost_0_depth}}
\subfigure[10\% Ghost Targets]{\includegraphics[width=0.23\textwidth]{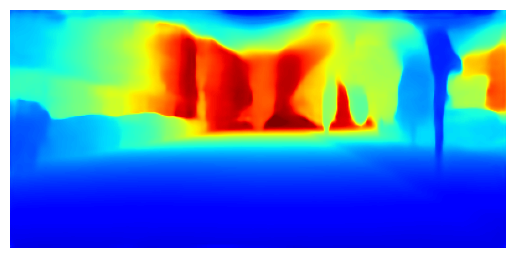}\label{fig:radar_ghost_10_depth}}
\subfigure[30\% Ghost Targets]{\includegraphics[width=0.23\textwidth]{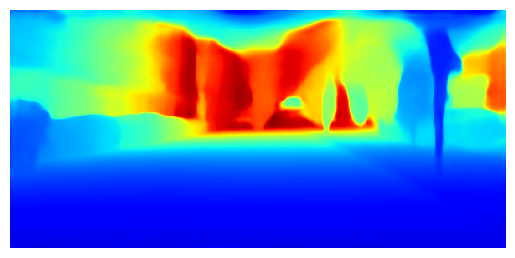}\label{fig:radar_ghost_20_depth}}
\subfigure[50\% Ghost Targets]{\includegraphics[width=0.23\textwidth]{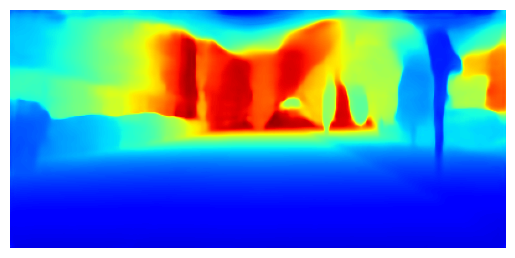}\label{fig:radar_ghost_30_depth}}

\subfigure[No attack]{\includegraphics[width=0.23\textwidth]{figures_radar/depth_error/0_radar.png}\label{fig:radar_ghost_0_pcd}}
\subfigure[10\% Ghost Targets]{\includegraphics[width=0.23\textwidth]{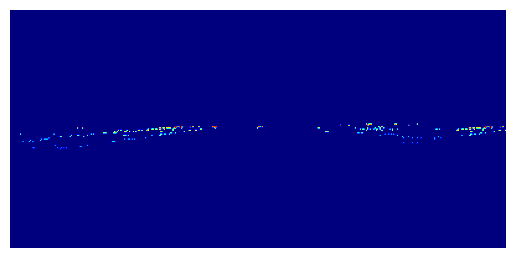}\label{fig:radar_ghost_10_pcd}}
\subfigure[30\% Ghost Targets]{\includegraphics[width=0.23\textwidth]{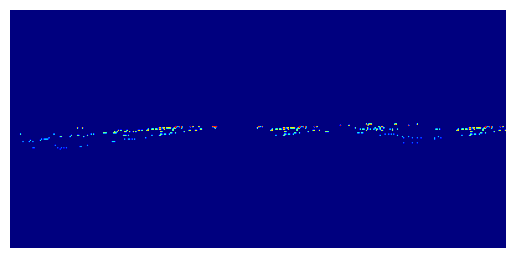}\label{fig:radar_ghost_20_pcd}}
\subfigure[50\% Ghost Targets]{\includegraphics[width=0.23\textwidth]{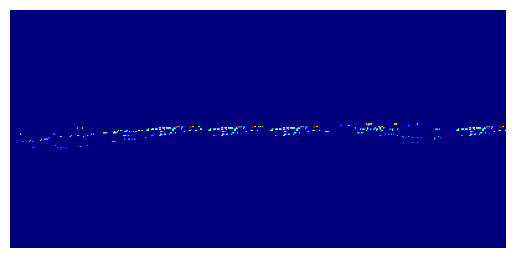}\label{fig:radar_ghost_30_pcd}}

\caption{Examples of ghost target in RC-PDA \cite{long2021radar}.
The first row illustrates depth estimation errors, while the second row shows the corresponding attack strength visualizations.}
\label{fig:radar_ghost_target}
\end{figure*}

Fig.~\ref{fig:radar_ghost_target} presents the visualized results of the DE algorithm (RC-PDA) and the corresponding attack strength visualizations.
We still observe that for the DE algorithm, higher adversarial levels cause a slight decline in depth estimation performance. In contrast, for the OD algorithm, the position and dimensions of the bounding boxes remain stable even under stronger attacks, highlighting the resilience of object detection methods to radar ghost targets.

\subsection{Findings and Analysis}

Based on the above results, we summarize the following key findings:

\cbox{Finding 3: DE sensor fusion algorithms are affected by radar attacks, but the impact of which is limited compared to that of the camera-based attacks.}

\cbox{Finding 4: OD sensor fusion algorithms are hardly affected by radar attacks and demonstrate comparatively robust under the evaluated radar perturbations.}  

We explain these findings by the way radar information is used in
the evaluated fusion architectures. Under radar attack, the radar
representation changes from

\begin{equation}
    F_t^r = \mathcal{E}_r(P_t)
\end{equation}
to
\begin{equation}
    \widetilde{F}_t^r
    =
    \mathcal{E}_r(\widetilde{P}_t).
\end{equation}

However, both the DE and OD models considered in our study are
strongly rooted in camera-based perception, with radar information
serving primarily as supplementary geometric evidence. Therefore,
even when $\widetilde{F}_t^r$ becomes less reliable, the models can
retain substantial predictive capability through the unchanged
camera representation $F_t^c$.

In addition, existing architectures extensively filter and weight
radar measurements before incorporating them into the fused
representation $F_t^f$. For DE, only a subset of radar measurements
with sufficiently strong association confidence may contribute to
the corresponding image regions. For OD, camera-derived object
hypotheses are commonly generated first, after which only radar
measurements associated with those hypotheses are retained.
Consequently, many corrupted measurements introduced by
$\mathcal{A}_r(\cdot;\lambda_r)$ are discarded before they can
substantially influence $F_t^f$.

Furthermore, retained radar measurements are commonly assigned
learned confidence weights. Low-confidence measurements therefore
contribute less to the final prediction, and some architectures can
effectively revert toward camera-dominant prediction when reliable
radar support is unavailable. This implicit reliability mechanism
helps explain why changes from $F_t^r$ to
$\widetilde{F}_t^r$ produce considerably smaller degradation than
the corresponding camera-side changes from $F_t^c$ to
$\widetilde{F}_t^c$.

Finally, the task outputs differ in their dependence on radar.
Depth-estimation models directly use radar-derived geometric
information to produce $\hat{D}_t$, making them more sensitive to
radar corruption. In the evaluated object-detection architectures,
several output components are primarily determined by camera
features, while radar contributes more selectively to geometric
estimation. This further limits the impact of radar attacks on
$\hat{Y}_t^{\mathrm{OD}}$.

\begin{table}[]
\centering
\caption{The performance of different camera-radar sensor fusion models under the radar attacks with various levels of depth estimation error. OD Metrics: MATE/MAP. DE Metrics: MAE/RMSE.}
\label{tab:estimation_error}
\resizebox{0.6\columnwidth}{!}{%
\begin{tabular}{|l|l|llll|}
\hline
\textbf{Model} & \textbf{Type} & \multicolumn{4}{c|}{\textbf{Depth Estimation Error}}                                                                                                                                                                                                                                                                                           \\ \cline{3-6} 
                                &                                & \multicolumn{1}{l|}{No Attack}                                                         & \multicolumn{1}{c|}{50\%}                                                       & \multicolumn{1}{c|}{100\%}                                                       & \multicolumn{1}{c|}{200\%}                                   \\ \hline
RADIANT                         & OD                             & \multicolumn{1}{l|}{0.609/0.380}                                                       & \multicolumn{1}{l|}{0.611/0.377}                                                       & \multicolumn{1}{l|}{0.612/0.373}                                                       & 0.617/0.365                                                       \\ \hline
CRN                             & OD                             & \multicolumn{1}{l|}{0.416/0.575}                                                       & \multicolumn{1}{l|}{0.424/0.569}                                                      & \multicolumn{1}{l|}{0.438/0.551}                                                       & 0.453/0.530                                                      \\ \hline
RC-PDA                          & DE                             & \multicolumn{1}{l|}{\begin{tabular}[c]{@{}l@{}}0.884/2.183\end{tabular}} & \multicolumn{1}{l|}{\begin{tabular}[c]{@{}l@{}}0.889/2.188\end{tabular}} & \multicolumn{1}{l|}{\begin{tabular}[c]{@{}l@{}}0.926/2.241\end{tabular}} & \begin{tabular}[c]
{@{}l@{}}1.069/2.524\end{tabular} \\ \hline
Cam\&mmWave                     & DE                             & \multicolumn{1}{l|}{2.037/4.548}                                                       & \multicolumn{1}{l|}{2.045/4.561}                                                      & \multicolumn{1}{l|}{2.102/4.646}                                                     & 2.307/4.901                                                    \\ \hline
DORN                     & DE                             & \multicolumn{1}{l|}{2.258/4.913}                                                       & \multicolumn{1}{l|}{2.265/4.919}                                                        & \multicolumn{1}{l|}{2.305/4.964}                                                      & 2.637/5.228                                                     \\ \hline
\end{tabular}%
}
\end{table}

\begin{table}[]
\centering
\caption{The performance of different camera-radar sensor fusion models under the radar attacks with various levels of target disappearing. OD Metrics: MATE/MAP. DE Metrics: MAE/RMSE.}
\label{tab:target_disappearing}
\resizebox{0.6\columnwidth}{!}{%
\begin{tabular}{|l|l|llll|}
\hline
\textbf{Model} & \textbf{Type} & \multicolumn{4}{c|}{\textbf{Target Disappearance}}                                                                                                                                                                                                                                                                                           \\ \cline{3-6} 
                                &                                & \multicolumn{1}{l|}{No Attack}                                                         & \multicolumn{1}{c|}{30\%}                                                       & \multicolumn{1}{c|}{50\%}                                                       & \multicolumn{1}{c|}{70\%}                                   \\ \hline
RADIANT                         & OD                             & \multicolumn{1}{l|}{0.609/0.380}                                                       & \multicolumn{1}{l|}{0.609/0.375}                                                       & \multicolumn{1}{l|}{0.611/0.371}                                                       & 0.613/0.370                                                       \\ \hline
CRN                             & OD                             & \multicolumn{1}{l|}{0.416/0.575}                                                       & \multicolumn{1}{l|}{0.418/0.572}                                                      & \multicolumn{1}{l|}{0.421/0.566}                                                       & 0.426/0.559                                                      \\ \hline
RC-PDA                          & DE                             & \multicolumn{1}{l|}{\begin{tabular}[c]{@{}l@{}}0.884/2.183\end{tabular}} & \multicolumn{1}{l|}{\begin{tabular}[c]{@{}l@{}}0.887/2.186\end{tabular}} & \multicolumn{1}{l|}{\begin{tabular}[c]{@{}l@{}}0.896/2.198\end{tabular}} & \begin{tabular}[c]
{@{}l@{}}0.920/2.231\end{tabular} \\ \hline
Cam\&mmWave                     & DE                             & \multicolumn{1}{l|}{2.037/4.548}                                                       & \multicolumn{1}{l|}{2.045/4.559}                                                      & \multicolumn{1}{l|}{2.060/4.576}                                                     & 2.091/4.611                                                    \\ \hline
DORN                     & DE                             & \multicolumn{1}{l|}{2.258/4.913}                                                       & \multicolumn{1}{l|}{2.263/4.919}                                                        & \multicolumn{1}{l|}{2.275/4.935}                                                      & 2.307/4.981                                                    \\ \hline
\end{tabular}%
}
\end{table}

\begin{table}[]
\centering
\caption{The performance of different camera-radar sensor fusion models under the radar attacks with various levels of ghost target. OD Metrics: MATE/MAP. DE Metrics: MAE/RMSE.}
\label{tab:ghost_target}
\resizebox{0.6\columnwidth}{!}{%
\begin{tabular}{|l|l|llll|}
\hline
\textbf{Model} & \textbf{Type} & \multicolumn{4}{c|}{\textbf{Ghost Target}}                                                                                                                                                                                                                                                                                           \\ \cline{3-6} 
                                &                                & \multicolumn{1}{l|}{No Attack}                                                         & \multicolumn{1}{c|}{10\%}                                                       & \multicolumn{1}{c|}{30\%}                                                       & \multicolumn{1}{c|}{50\%}                                   \\ \hline
RADIANT                         & OD                             & \multicolumn{1}{l|}{0.609/0.380}                                                       & \multicolumn{1}{l|}{0.604/0.386}                                                & \multicolumn{1}{l|}{0.607/0.386}                                                       & 0.615/0.372                                                       \\ \hline
CRN                             & OD                             & \multicolumn{1}{l|}{0.416/0.575}                                                       & \multicolumn{1}{l|}{0.413/0.582}                                                      & \multicolumn{1}{l|}{0.424/0.561}                                                       & 0.445/0.537                                                      \\ \hline
RC-PDA                          & DE                             & \multicolumn{1}{l|}{\begin{tabular}[c]{@{}l@{}}0.884/2.183\end{tabular}} & \multicolumn{1}{l|}{\begin{tabular}[c]{@{}l@{}}0.887/2.187\end{tabular}} & \multicolumn{1}{l|}{\begin{tabular}[c]{@{}l@{}}0.899/2.194\end{tabular}} & \begin{tabular}[c]
{@{}l@{}}0.906/2.196\end{tabular} \\ \hline
Cam\&mmWave                     & DE                             & \multicolumn{1}{l|}{2.037/4.548}                                                       & \multicolumn{1}{l|}{2.042/4.556}                                                     & \multicolumn{1}{l|}{2.059/4.565}                                                     & 2.072/4.581                                                    \\ \hline
DORN                     & DE                             & \multicolumn{1}{l|}{2.258/4.913}                                                       & \multicolumn{1}{l|}{2.263/4.922}                                                       & \multicolumn{1}{l|}{2.278/4.940}                                                      & 2.291/4.957                                                     \\ \hline
\end{tabular}%
}
\end{table}

\section{Future Research Directions}
\label{sec:future}

Our evaluation reveals that robustness does not arise automatically
from combining camera and radar measurements. In particular, the
evaluated fusion architectures remain strongly dependent on camera
features, causing camera perturbations to propagate through the
fusion pipeline and substantially degrade both object detection and
depth estimation. In contrast, radar attacks generally induce
smaller performance degradation because radar measurements are
sparse and are frequently filtered, associated, and
confidence-weighted before fusion. These observations suggest that
robust camera--radar fusion should explicitly reason about
\emph{sensor reliability} rather than assuming that both sensing
modalities remain equally trustworthy at inference time.

\begin{figure*}[t]
    \centering
    \includegraphics[width=\textwidth]
    {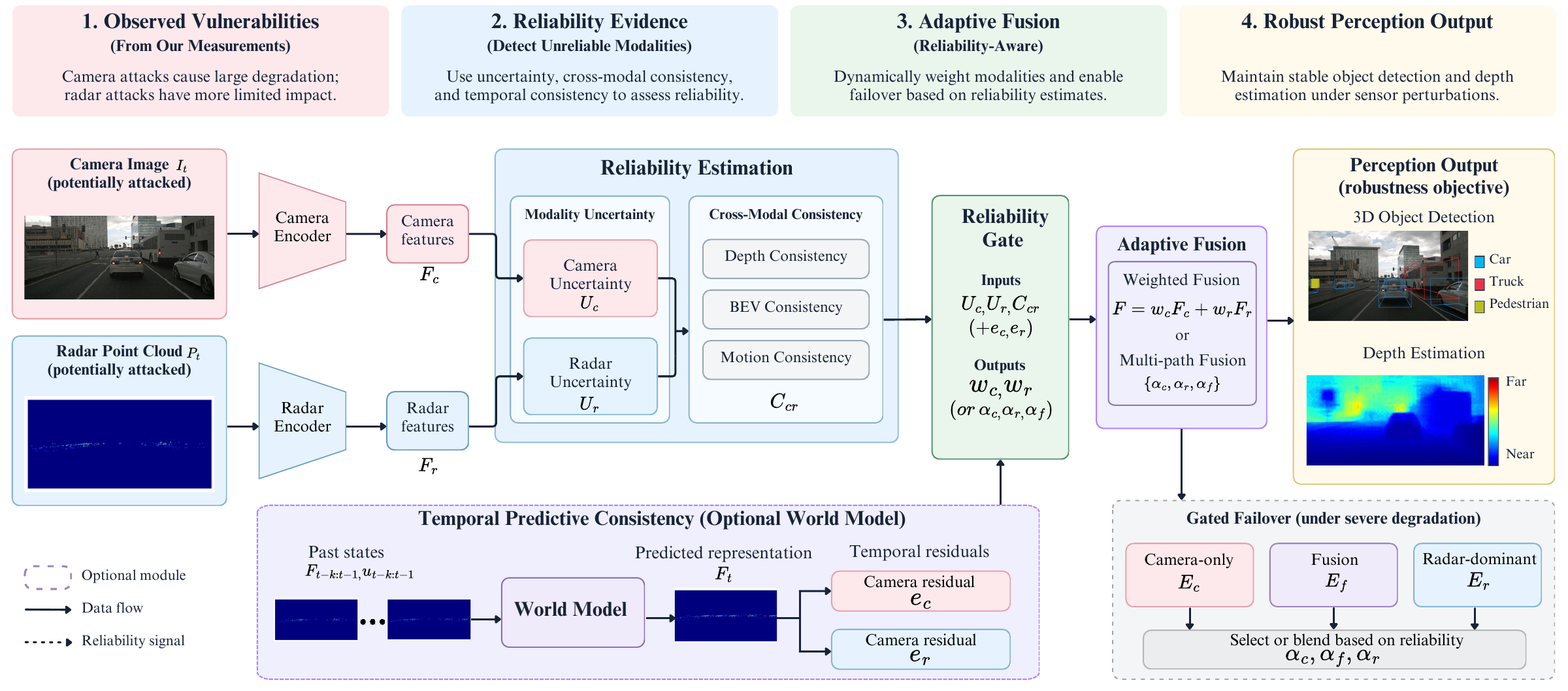}
    \caption{
    Measurement-guided design framework for future robust
    camera--radar fusion. The core design path uses
    modality-specific reliability and cross-modal consistency to
    guide adaptive fusion and, under severe degradation, gated
    failover. Temporal predictive consistency is shown as an
    optional extension that may provide additional evidence when
    cross-modal disagreement alone cannot identify the unreliable
    modality. The framework summarizes design implications derived
    from our measurements and is not an evaluated defense method in
    this work.
    }
    \label{fig:defense_framework}
\end{figure*}

Table~\ref{tab:defense_implications} summarizes how the major
empirical findings of this work translate into corresponding design
implications. Based on these observations, we outline a
measurement-guided design framework for future robust
camera--radar fusion systems, illustrated in
Fig.~\ref{fig:defense_framework}. The central design principle is
to use modality-specific reliability and cross-modal consistency to
determine how strongly each modality should influence the fused
prediction. Temporal predictive consistency is further considered
as an optional source of reliability evidence when cross-modal
disagreement alone is ambiguous.

Importantly, the framework in this section is intended to organize
the design implications derived from our measurement results rather
than introduce an evaluated defense method. The formulations below
illustrate possible realizations of these principles; their
implementation and systematic evaluation are left to future work.

\begin{table*}[t]
\centering
\caption{From empirical findings to design implications for robust
camera--radar fusion.}
\label{tab:defense_implications}
\renewcommand{\arraystretch}{1.25}
\resizebox{\textwidth}{!}{
\begin{tabular}{
|p{0.23\textwidth}
|p{0.32\textwidth}
|p{0.36\textwidth}|
}
\hline

\textbf{Measurement Finding}
&
\textbf{Observed Cause}
&
\textbf{Design Implication}
\\
\hline

Camera attacks substantially degrade both object detection and
depth estimation.
&
Current fusion models rely heavily on dense camera features, while
radar generally provides supplementary geometric information.
&
Explicitly estimate camera reliability and dynamically reduce its
contribution when camera observations become unreliable.
\\
\hline

Moderate camera perturbations can already cause severe performance
degradation.
&
Fixed fusion behavior cannot respond to gradual sensor-quality
degradation before severe perception failure occurs.
&
Reliability estimation should operate continuously and support
progressive reweighting rather than only binary attack detection.
\\
\hline

Radar attacks have a comparatively smaller effect, particularly for
object detection.
&
Radar measurements are sparse and are often filtered,
confidence-weighted, or discarded when inconsistent with camera
predictions.
&
Generalize filtering and uncertainty-aware weighting principles to
both sensing modalities rather than treating camera information as
implicitly reliable.
\\
\hline

Camera and radar provide complementary geometric and semantic
information.
&
Corruption of one modality can create cross-modal disagreement in
depth, object location, motion, or BEV occupancy.
&
Explicitly model camera--radar consistency and use disagreement as
a runtime reliability signal.
\\
\hline

Cross-modal disagreement indicates abnormal sensing but does not
always reveal which modality is responsible.
&
Either camera corruption or radar corruption can create
camera--radar inconsistency.
&
Temporal predictive consistency can provide an optional additional
cue for identifying which modality deviates from expected scene
evolution.
\\
\hline

\end{tabular}
}
\end{table*}

\subsection{Cross-Modal Reliability Estimation}
\label{sec:future_reliability}

A primary design implication of our measurements is that the fusion
process should explicitly estimate the reliability of each sensing
modality. Let $F_t^c$ and $F_t^r$ denote the camera and radar
feature representations defined in Section~\ref{sec:system_overview}.
Instead of assuming that both representations remain equally
reliable, a future fusion system could estimate modality-specific
uncertainty as

\begin{equation}
    U_t^c = \mathcal{U}_c(F_t^c),
    \qquad
    U_t^r = \mathcal{U}_r(F_t^r),
\end{equation}
where $U_t^c$ and $U_t^r$ represent the estimated uncertainty of
the camera and radar streams. Camera uncertainty may reflect
prediction entropy, depth uncertainty, or feature-quality
statistics, while radar uncertainty may incorporate point
confidence, density, association ambiguity, or measurement quality.
The particular uncertainty estimator would depend on the underlying
fusion architecture.

Modality-specific uncertainty alone, however, may not be sufficient.
An attacked sensor can remain highly confident while producing an
incorrect output. Camera and radar provide partially independent
physical observations of the same environment, allowing their
agreement to serve as an additional reliability signal.

For example, consider a radar measurement projected to image
location $i$ at time $t$. Let $d_{t,i}^c$ and $d_{t,i}^r$ denote
the corresponding camera- and radar-derived depth estimates. One
possible normalized disagreement measure is

\begin{equation}
    e_{t,i}^{\mathrm{depth}}
    =
    \frac{
        \left| d_{t,i}^c - d_{t,i}^r \right|
    }{
        \sigma_{t,i}^c + \sigma_{t,i}^r + \epsilon
    },
\end{equation}
where $\sigma_{t,i}^c$ and $\sigma_{t,i}^r$ characterize the
uncertainty of the two estimates and $\epsilon$ prevents numerical
instability. A large disagreement would indicate that the two
modalities provide incompatible geometric evidence.

Cross-modal consistency can also be examined beyond individual
depth measurements. In a shared BEV representation, camera- and
radar-derived occupancy or object hypotheses could be compared to
measure spatial consistency. Likewise, Doppler measurements could
be compared with camera-derived motion cues. These complementary
signals can be represented generally as

\begin{equation}
    C_t^{cr}
    =
    \phi
    \left(
        C_t^{\mathrm{depth}},
        C_t^{\mathrm{BEV}},
        C_t^{\mathrm{motion}}
    \right),
\end{equation}
where $\phi(\cdot)$ denotes an aggregation function and
$C_t^{cr}$ represents camera--radar consistency.

The key principle is that reliability estimation need not identify
which specific attack has occurred. Instead, it can determine
whether the current sensing evidence remains mutually consistent.
Such an attack-agnostic formulation could potentially generalize
better to sensor perturbations that were not explicitly included
during training.

\subsection{Optional Temporal Predictive Consistency}
\label{sec:future_temporal}

Cross-modal disagreement can indicate that at least one sensing
modality has become unreliable, but disagreement alone does not
necessarily identify which modality is responsible. For example,
either camera distortion or radar corruption can produce a large
camera--radar consistency residual. Temporal predictive consistency
provides one possible additional source of evidence for resolving
this ambiguity.

As illustrated by the dashed branch in
Fig.~\ref{fig:defense_framework}, a lightweight temporal predictive
model could use previous fused scene representations and ego-motion
to estimate an expected current representation,

\begin{equation}
    \widehat{F}_t^f
    =
    \mathcal{W}
    \left(
        F_{t-k:t-1}^f,
        u_{t-k:t-1}
    \right),
\end{equation}
where $\mathcal{W}$ denotes a temporal predictive model and
$u_{t-k:t-1}$ contains ego-motion information. Depending on the
fusion architecture, $\widehat{F}_t^f$ could represent an expected
BEV feature map, depth representation, object-level state, or other
compact scene representation.

The current evidence from each modality could then be compared
with the predicted scene representation,

\begin{equation}
    e_t^c
    =
    D_c(F_t^c,\widehat{F}_t^f),
    \qquad
    e_t^r
    =
    D_r(F_t^r,\widehat{F}_t^f),
\end{equation}
where $D_c(\cdot,\cdot)$ and $D_r(\cdot,\cdot)$ denote
architecture-dependent comparison functions after the necessary
projection or alignment into a compatible representation.

These residuals could complement cross-modal consistency. For
example, when camera and radar disagree, a relatively large
$e_t^c$ together with a smaller $e_t^r$ would provide additional
evidence that the camera stream has deviated from the expected
scene evolution. Conversely, a large $e_t^r$ with a stable
$e_t^c$ would provide evidence of anomalous radar measurements.

We emphasize that temporal predictive consistency is an
\emph{optional extension} rather than a requirement of the core
reliability-aware fusion principle. It becomes particularly useful
when cross-modal disagreement is detected but the unreliable
modality cannot be identified from instantaneous observations
alone.

\subsection{Reliability-Aware Adaptive Fusion}
\label{sec:future_fusion}

Once modality reliability and cross-modal consistency have been
estimated, these signals could be used to adapt the fusion process.
For example, a lightweight reliability gate could estimate the
relative contributions of the two sensing modalities,

\begin{equation}
    [w_t^c,w_t^r]
    =
    \mathrm{Softmax}
    \left(
        G_{\theta}
        (U_t^c,U_t^r,C_t^{cr},e_t^c,e_t^r)
    \right),
\end{equation}
where $G_{\theta}$ denotes a possible learnable reliability
estimator and

\begin{equation}
    w_t^c+w_t^r=1.
\end{equation}

When temporal predictive consistency is not used, the gate could
operate only on $U_t^c$, $U_t^r$, and $C_t^{cr}$.

Because the evaluated architectures perform fusion at different
network stages and their modality-specific representations need not
share the same feature space, reliability-aware fusion should not
necessarily be implemented as direct feature addition. A general
formulation is
\begin{equation}
    F_t^{\mathrm{robust}}
    =
    \mathcal{F}_{\mathrm{adapt}}
    \left(
        F_t^c,F_t^r;
        w_t^c,w_t^r
    \right),
\end{equation}
where $\mathcal{F}_{\mathrm{adapt}}$ represents an
architecture-specific adaptive fusion operation. For architectures
in which the modality representations are aligned in a common
feature space, a weighted combination is one possible realization,

\begin{equation}
    F_t^{\mathrm{robust}}
    =
    w_t^c F_t^c + w_t^r F_t^r.
\end{equation}

Under benign conditions, such a mechanism could retain the fusion
behavior that provides strong clean-data performance. When camera
uncertainty or camera--radar disagreement increases, the system
could reduce the camera contribution and rely more heavily on
radar. Similarly, unreliable radar measurements could be
downweighted when radar uncertainty becomes large.

Reliability estimation may also be performed spatially rather than
using a single scalar for the entire frame. Several perturbations,
particularly localized optical distortions, may affect only part of
the sensing field. A spatially varying reliability representation,

\begin{equation}
    w_t^c(p), \qquad w_t^r(p),
\end{equation}
where $p$ indexes an image region, object proposal, or BEV cell,
could preserve reliable information in unaffected regions while
reducing the influence of locally inconsistent observations.

\subsection{Gated Failover under Severe Degradation}
\label{sec:future_failover}

Adaptive weighting may be sufficient under mild or moderate sensor
degradation, but severe attacks may make one sensing stream
effectively unusable. In this case, a robust system should degrade
gracefully rather than continue forcing unreliable information
through the normal fusion pathway.

One possible realization is to maintain multiple prediction
pathways representing camera-only, radar-dominant, and
camera--radar fused operation, where such pathways can be supported
by the underlying architecture. Let these candidate pathways be
denoted as

\begin{equation}
    E_c,\qquad E_r,\qquad E_f.
\end{equation}

A reliability module could estimate corresponding pathway weights,

\begin{equation}
    [\alpha_t^c,\alpha_t^r,\alpha_t^f]
    =
    \mathrm{Softmax}
    \left(
        G_{\theta}
        (U_t^c,U_t^r,C_t^{cr},e_t^c,e_t^r)
    \right),
\end{equation}
subject to
\begin{equation}
    \alpha_t^c+\alpha_t^r+\alpha_t^f=1.
\end{equation}

Under normal conditions, the fused pathway could remain dominant.
When camera reliability becomes sufficiently low, the system could
shift toward a radar-dominant pathway if one is available.
Conversely, when radar observations become unreliable, the system
could rely more strongly on the camera-only pathway.

The specific implementation could use either soft combination or
hard pathway selection depending on whether the candidate pathways
produce compatible output representations. The broader design
principle is to provide a controlled fallback mechanism rather than
continue using a severely degraded sensing modality. This
generalizes the filtering, confidence weighting, and fallback
behaviors that already help limit the impact of corrupted radar
measurements in several existing fusion architectures.

\subsection{Reliability-Aware Training and Evaluation}
\label{sec:future_training}

A future reliability estimator should ideally learn whether a
sensing modality is trustworthy rather than classify a particular
attack type. One possible training strategy would expose the fusion
model to clean inputs, camera-corrupted inputs, radar-corrupted
inputs, and modality-dropout conditions.

For each training sample at time $t$, let $L_t^c$, $L_t^r$, and
$L_t^f$ denote the downstream task losses obtained from the
camera-only, radar-dominant, and fused pathways, respectively.
These losses could provide an oracle indication of which pathway
would have produced the most reliable prediction for the current
sample. One possible soft reliability target is

\begin{equation}
    \alpha_{t,k}^{*}
    =
    \frac{
        \exp(-L_t^k/\tau)
    }{
        \sum_{j\in\{c,r,f\}}
        \exp(-L_t^j/\tau)
    },
    \qquad
    k\in\{c,r,f\},
\end{equation}
where $\tau$ controls the sharpness of the target distribution.
Such oracle targets could be used during training to supervise a
reliability estimator using only signals available at inference
time.

This formulation is only one possible training strategy rather than
a required component of the framework. Other uncertainty
calibration, corruption-robust training, or modality-dropout
strategies could also instantiate the same design principle.

Future evaluation of such defenses should emphasize three
properties suggested by our measurements. First, robustness should
be measured across a range of perturbation strengths, particularly
moderate attack severities, because substantial degradation can
occur before the strongest attack level is reached. Second, the
reliability mechanism should be evaluated on perturbation types not
included during training to determine whether it captures general
sensor reliability rather than attack-specific signatures. Third,
evaluation should report not only downstream perception accuracy but
also failover behavior, including whether the system correctly
reduces its dependence on an unreliable modality while retaining
useful information from the remaining sensor.

\subsection{Discussion}
\label{sec:future_discussion}

The framework above should be viewed as a synthesis of design
principles derived from the failure modes observed in our
measurement study. The strong sensitivity to camera attacks
suggests that camera features should not be treated as implicitly
reliable, while the comparatively limited impact of the evaluated
radar attacks demonstrates the benefit of filtering, confidence
weighting, and selective use of radar measurements. These findings
directly motivate the core components of the framework:
modality-specific reliability estimation, cross-modal consistency
checking, and adaptive fusion or failover.

The framework intentionally separates these core design principles
from optional extensions. Cross-modal reliability estimation can be
implemented without a temporal predictive model, while temporal
consistency may provide additional evidence when instantaneous
camera--radar disagreement cannot identify the unreliable modality.
Likewise, the particular gating, fusion, and training formulations
presented above are illustrative realizations rather than
architectural requirements.

Accordingly, Fig.~\ref{fig:defense_framework} does not represent a
defense system evaluated in this work. Its purpose is to connect the
empirical findings of Sections~4 and~5 to concrete research
directions for future robust camera--radar fusion. Future studies
should investigate the trade-off between clean-data accuracy and
attack robustness, generalization to unseen sensor perturbations,
the reliability of modality selection, and the latency and
computational overhead introduced by runtime consistency
monitoring.

At a higher semantic level, multimodal foundation models may also
be explored in future work as an additional scene-level consistency
signal. However, such models are not part of the core framework
considered here, and their calibration, latency, and robustness
under adversarial sensing conditions would need to be established
before they could serve as trustworthy runtime monitors.
\section{Conclusion}
\label{sec:conclusion}

This paper presents a systematic robustness measurement study of
camera--radar sensor fusion under adversarial driving scenarios. We
develop a unified evaluation pipeline that applies eight
camera-image perturbations and three radar point-cloud perturbations
to five representative fusion architectures across two perception
tasks, 3D object detection and depth estimation. Our results show that multi-sensor fusion does not inherently
provide robustness against adversarial sensing conditions. More
importantly, we observe a clear asymmetry between the two sensing
modalities. Camera perturbations substantially degrade both object
detection and depth estimation, with moderate perturbation strengths
already causing severe performance loss. In contrast, the evaluated
radar perturbations have a smaller and task-dependent impact,
particularly for object detection. Our analysis shows that this
difference is closely related to the camera-dominant design of
current fusion architectures and to the filtering, association, and
confidence-weighting mechanisms that limit the influence of
unreliable radar measurements.

These findings indicate that robust camera--radar perception should
not assume that all sensing modalities remain equally reliable at
inference time. Instead, future fusion systems should explicitly
estimate modality reliability, monitor cross-modal consistency, and
adapt the fusion process when one sensing stream becomes unreliable.
Based on the observed failure modes, we outline measurement-guided
design directions including reliability-aware adaptive fusion,
cross-modal consistency checking, and gated failover. We hope these
findings provide a foundation for the development and evaluation of
more robust multi-sensor perception systems for autonomous driving.

\bibliographystyle{ACM-Reference-Format}
\bibliography{reference}

\end{document}